\documentclass[pra,twocolumn,amsmath,amssymb,superscriptaddress]{revtex4-1}
\usepackage{epsfig,amsmath}
\usepackage{subfigure}
\usepackage{graphicx}
\usepackage{dcolumn}
\usepackage{stmaryrd}
\usepackage{mathrsfs}
\usepackage{pifont}
\usepackage{amsthm}
\usepackage{amssymb}
\usepackage{bm}
\usepackage{latexsym}
\usepackage[colorlinks=true,linkcolor=blue,citecolor=blue]{hyperref}
\usepackage{color}
\usepackage{epstopdf}

\usepackage{booktabs}
\usepackage{threeparttable}
\usepackage{multirow}

\begin{document}

\title{Universal quantum control over bosonic network}

\author{Zhu-yao Jin}
\affiliation{School of Physics, Zhejiang University, Hangzhou 310027, Zhejiang, China}

\author{Jun Jing}
\email{Contact author: jingjun@zju.edu.cn}
\affiliation{School of Physics, Zhejiang University, Hangzhou 310027, Zhejiang, China}

\date{\today}

\begin{abstract}
Perfect transfer of {\em unknown} states across distinct nodes is a basic function in bosonic quantum networks. Here we develop a general theory to construct an $N$-node bosonic network governed by the time-dependent Hamiltonian, as the universal quantum control theory for continuous-variable systems. In particular, we can activate nonadiabatic passages superposed of initial and target modes by the commutation condition about the Hamiltonian's coefficient matrix and projection operator in the representation of time-independent ancillary modes, which serves as the necessary and sufficient condition to solve the time-dependent Schr\"odinger equation of the full Hamiltonian. To exemplify the versatility of our theory on the Heisenberg-picture passages, we perform arbitrary state exchange between two nodes, chiral entanglement transfer among three bosonic nodes, and chiral Fock-state transfer among three of four bosonic nodes. Our work provides a promising avenue toward the universal control of any pair of nodes or modes as well as the entire bosonic network.
\end{abstract}

\maketitle

\section{Introduction}

A quantum network~\cite{Kimble2008Quantum,Stephanie2018Quantum} exhibits fundamental advantages over its classical counterpart in specific applications, such as quantum key distribution~\cite{Xu2020Secure}, long-distance quantum computation~\cite{Duan2001Longdistance,Stannigel2010Optomechanical,Muralidharan2016Optimal,Wallnofer2020Machine}, distributed quantum computation~\cite{Lim2005Repeat,Cohen2018Deterministic,Cuomo2020Towards}, and quantum metrology~\cite{Glovannetti2006Quantum,Joo2011Quantum,Giovannetti2011Advances,Chin2012Quantum,Zhou2020Quantum}. It is typically composed of two or more quantum nodes, constituted of atoms~\cite{Reiserer2015Cavity,Liu2022Hybrid,Covey2023Quantum}, ions~\cite{Duan2010Quantum,Chen2023Scalable}, and bosonic modes~\cite{Briegel1998Quantum,Yung2005Perfect,Ma2021Quantum,Zhou2024Quantum}. State transfer and entanglement distribution in a quantum network can be mediated by the indirect connection between remote nodes, which is established via the mutual interactions between neighboring quantum nodes~\cite{Cirac1997Quantum,Chou2007Functional,Kimble2008Quantum}.

In comparison to the quantum networks constructed by the atomic or ionic nodes, the bosonic network is featured with versatile functions, including (1) the efficient simulation of boson sampling~\cite{Justin2013Boson,Matthew2013Photonic,
Tillmann2013Photonic,Carolan2014On,Spagnolo2014Experimental,Wang2019Boson,Arrazola2021Quantum}, which demonstrates a clear quantum advantage over the classical computer~\cite{Aaronson2013Computational}, (2) the feasibility of the universal quantum computer based on the Knill-Laflamme-Milburn schemes~\cite{Knill2001Scheme,Kok2007Linear}, and (3) the fault-tolerant quantum computation with the error-correction codes~\cite{Chuang1997Bosonic,Gottesman2001Encoding,
Mirrahimi2014Dynamically,Michael2016New}. Bosonic networks can be set up on many physical systems, including a cavity quantum electrodynamics (QED) systems~\cite{Kimble1998strong,Haroche2006Exploring}, circuit QED systems~\cite{Vermersch2017Quantum,Blais2021Circuit}, synthetic photonic lattices~\cite{Regensburger2012Parity,Celi2014Synthetic,Lustig2019Photonic,Ozawa2019Topological}, Bose-Einstein condensates~\cite{Morsch2006Dynamics}, optomechanical systems~\cite{Heinrich2011Collective,Stannigel2011Optomechanical,
Aspelmeyer2014Cavity,Zhang2015Synchronization,Peterson2017Demonstration,Xu2020Cavity}, circuit quantum acoustodynamics systems~\cite{von2024Engineering}, cavity magnomechanical systems~\cite{Zhang2016Cavity,Li2021Quantum,Shen2022Coherent,Shen2025Cavity}, and cavity magnonic systems~\cite{Zhang2014Strongly,Zhang2016Optomagnonic,Lachance2019Hybrid,Xu2021Coherent,Babak2022Cavity,
Xu2023Quantum,Zheng2023Tutorial}.

State transfer between bosonic modes or nodes has been explored in various control protocols, particularly those based on the quantum adiabatic theorem~\cite{Claridge2009high}. Under the adiabatic condition, the state transfer between two cavity modes in optomechanical systems~\cite{Dong2012Optomechanical,Wang2012Using,Tian2012Adiabatic,Fedoseev2021Stimulated} can be enabled by the mechanical dark mode even under the thermal noise. It mimics the stimulated Raman adiabatic passage in a discrete three-level system~\cite{Vitanov2017Stimulated}. Generalized from the three-mode systems~\cite{Dong2012Optomechanical,Wang2012Using,Tian2012Adiabatic}, a dark-mode theorem~\cite{Jian2023Dark} recently presented a comprehensive analysis over a bosonic network composed of two coupled subsystems, each of which consisted of multiple uncoupled modes. The adiabatic evolution along the dark mode is inherently susceptible to the environmental noise due to the long-time exposure. It will induce a significant decoherence of the quantum system~\cite{Jing2016Eigenstate}. Alternatively, a leakage-free path~\cite{Jing2022Onecomponent} enforced by control, such as the general dynamical-decoupling approach~\cite{Jing2013Nonperturbative,Jing2015Nonperturbative}, elevates the adiabatic condition of the system, where the quantum channel is realized through the time-dependent quantum eigenstate. Existing research suggests that fast and reliable state transfer among bosonic modes is usually constrained by the system size, such as the two-mode~\cite{Zhang2019Fast,Qi2022Accelerated,Chen2018Invariant,Li2021Quantum} and three-mode ~\cite{Dong2012Optomechanical,Wang2012Using,Tian2012Adiabatic} systems since it is hard to solve the quantum Langevin equation for larger system. Various approaches, such as the transitionless driving~\cite{Zhang2019Fast,Qi2022Accelerated}, the inverse engineering~\cite{Chen2018Invariant}, and the pulse optimization~\cite{Xiang2023Universal,Lu2023High} seem to inspire the accelerated adiabatic passage in the continuous-variable systems, yet are typically confined to the single-excitation subspace. In general, a universal theoretical framework, which is insensitive to system size, connection geometry (the presence or absence of the dark modes), and target states, is desired for the bosonic networks.

In this paper, we develop a universal theory to control a general $N$-node bosonic network, which extends the universal quantum control (UQC) theory for discrete-variable systems~\cite{Jin2025Universal,Jin2025Entangling,Jin2025ErrCorr,
Jin2025Rydberg,Jin2025NonHerm,Jin2025Majorana} to that for continuous-variable systems. Universal passages in the Heisenberg picture can be activated by the partial or complete commutation condition about the coefficient matrix of the time-dependent network Hamiltonian in the stationary representation, which is equivalent to solving the time-dependent Schr\"odinger equation for the entire network. A variety of quantum controls over the network can be performed through the activated passages. Our theory is justified by arbitrary state exchange between two bosonic modes and chiral entanglement state transfer across multiple modes.

The rest of this paper is structured as follows. In Sec.~\ref{general}, we introduce a general theoretical framework for solving the time-dependent Schr\"odinger equation about the bosonic network of arbitrary size. In the representation of stationary ancillary modes, the commutation condition about the coefficient matrix of the Hamiltonian and the projection operator of matrix bases gives rise to useful nonadiabatic passages. Section~\ref{BosonTwo} exemplifies the passage-construction protocol in a paradigmatic two-mode system with mutual state conversion. Section~\ref{BosonThree} presents the chiral entanglement transfer of three bosonic modes. Section~\ref{BosonNet} describes the protocol for a general $N$-node system and demonstrates the chiral Fock-state transfer across four bosonic modes. Our protocol is found to be robust against the unwanted coupling in the network. Discussion and conclusion can be found in Sec.~\ref{conclusion}. Appendix~\ref{proof} provides a detailed derivation about the ancillary-mode transformation for the time-dependent network Hamiltonian.

\section{General framework}\label{general}

Our study is conducted on a quantum network composed of $N$ bosonic nodes, addressed by their annihilation operators $a_1$, $a_2$, $\cdots$, $a_N$. The system dynamics can be described by the time-dependent Schr\"odinger equation as ($\hbar\equiv1$)
\begin{equation}\label{Sch}
i\frac{d}{dt}|\psi(t)\rangle=H(t)|\psi(t)\rangle,
\end{equation}
where $|\psi(t)\rangle$ is the pure-state solution and the time-dependent Hamiltonian $H(t)$ can be written as
\begin{equation}\label{Ham}
H(t)=\vec{a}^\dagger H^a(t)\vec{a}^T, \quad \vec{a}\equiv (a_1,a_2,\cdots,a_N),
\end{equation}
with the row operator-vector $\vec{a}^\dagger=(a_1^\dagger,a_2^\dagger,\cdots,a_N^\dagger)$ and the time-dependent $N\times N$ coefficient matrix $H^a(t)$. The superscript $T$ means the transposition from a row vector to a column vector. Solving the time-dependent Schr\"odinger equation~(\ref{Sch}) is fundamentally challenging for continuous-variable systems, due to their infinite-dimensional Hilbert space and the noncommutativity of Hamiltonian at distinct moments.

In the framework of UQC theory~\cite{Jin2025Universal,Jin2025Entangling,Jin2025ErrCorr,Jin2025Rydberg,Jin2025NonHerm}, we consider the dynamics of bosonic systems by a set of time-dependent ancillary basis operators $\mu_k(t)$'s, $1\leq k\leq N$, which are typically superposed of the bosonic nodes $a_j$'s in laboratory. They satisfy the canonical communication relation, i.e., $[\mu_j(t), \mu_k^\dagger(t)]=\delta_{jk}$.

The ancillary basis modes $\mu_k(t)$'s are connected to the bosonic modes $a_k$'s by an $N\times N$ unitary transformation matrix $\mathcal{M}^\dagger(t)$ as
\begin{equation}\label{TimeAnci}
\vec{\mu}_t^T=\mathcal{M}^\dagger(t)\vec{a}^T, \quad \vec{\mu}_t\equiv[\mu_1(t), \mu_2(t), \cdots, \mu_N(t)].
\end{equation}
The adjoint matrix $\mathcal{M}^\dagger(t)$ implies the geometric structure of the underlying manifold of $\mu_k(t)$'s, admitting a general representation of the form 
\begin{equation}\label{unitary}
\mathcal{M}^\dagger(t)=\left[\vec{M}_1(t), \vec{M}_2(t), \cdots, \vec{M}_N(t)\right]^T
\end{equation}
with the row vectors of $N$ dimensionality
\begin{equation}\label{unitarVec}
\begin{aligned}
\vec{M}_1(t)&=\left(\cos\theta_1e^{i\frac{\alpha_1}{2}},-\sin\theta_1e^{-i\frac{\alpha_1}{2}},0,\cdots,0\right),\\
\vec{M}_k(t)&=\left[\cos\theta_ke^{i\frac{\alpha_k}{2}}\vec{b}_{k-1}(t),-\sin\theta_ke^{-i\frac{\alpha_k}{2}},0,\cdots,0\right],\\
\vec{M}_{N-1}(t)&=\Big[\cos\theta_{N-1}e^{i\frac{\alpha_{N-1}}{2}}\vec{b}_{N-2}(t), \\ &-\sin\theta_{N-1}e^{-i\frac{\alpha_{N-1}}{2}}\Big],\\
\vec{M}_N(t)&=\vec{b}_{N-1}(t),
\end{aligned}
\end{equation}
where $k$ runs from $2$ to $N-2$. The bright vector $\vec{b}_k(t)$ is a row vector of $k+1$ dimensionality defined as
\begin{equation}\label{unitarVecBri}
\vec{b}_k(t)\equiv\left[\sin\theta_ke^{i\frac{\alpha_k}{2}}\vec{b}_{k-1}(t), \cos\theta_ke^{-i\frac{\alpha_k}{2}}\right],
\end{equation}
where $1\leq k\leq N-1$ and $\vec{b}_0(t)\equiv 1$. We can define the bright-mode operators by the inner product $b_k(t)=\vec{b}_k(t)\cdot\vec{a}_{k+1}^T$ with $\vec{a}_k^T=(a_1,a_2,\cdots,a_k)^T$ and $b_0(t)=a_1$, e.g., $b_1(t)=\sin\theta_1e^{i\frac{\alpha_1}{2}}a_1+\cos\theta_1e^{-i\frac{\alpha_1}{2}}a_2$. For the sake of readability, the time-dependence of the parameters $\theta_k(t)$ and $\alpha_k(t)$ are implicit in Eqs.~(\ref{unitarVec}) and (\ref{unitarVecBri}). They can be either time-dependent or time-independent.

Using Eq.~(\ref{TimeAnci}), the system Hamiltonian~(\ref{Ham}) can be rewritten as
\begin{equation}\label{HamTimeAnci}
H(t)=\vec{\mu}_t^\dagger H^\mu(t)\vec{\mu}_t^T,
\end{equation}
where $H^\mu(t)=\mathcal{M}^\dagger(t)H^a(t)\mathcal{M}(t)$ is the coefficient matrix for $H(t)$ in terms of the time-dependent ancillary modes $\mu_k(t)$'s. To solve the Schr\"odinger equation with the Hamiltonian in Eq.~(\ref{HamTimeAnci}), we have to find a rotation to a representation of time-independent or stationary ancillary modes, by which $\vec{\mu}_t\rightarrow\vec{\mu}_0$ with $\vec{\mu}_0=[\mu_1(0), \mu_2(0), \cdots, \mu_N(0)]$. In general, this rotation can be performed by $V_{N-1}^\dagger(t)\mu_k(t)V_{N-1}(t)\rightarrow\mu_k(0)$ with
\begin{equation}\label{unitaryV}
V_{N-1}(t)=V_{\alpha_1}V_{\theta_1}V_{\alpha_2}V_{\theta_2}\cdots V_{\alpha_{N-1}}V_{\theta_{N-1}}=\prod_{k=1}^{N-1}V_{\alpha_k}V_{\theta_k},
\end{equation}
where
\begin{equation}\label{uniVat}
\begin{aligned}
V_{\alpha_k}(t)&=e^{-i\frac{\delta\alpha_k}{2}\left[b_{k-1}^\dagger(0)b_{k-1}(0)-a_{k+1}^\dagger a_{k+1}\right]},\\
V_{\theta_k}(t)&=e^{-\delta\theta_k\left[e^{i\alpha_k(0)}a_{k+1}^\dagger b_{k-1}(0)-e^{-i\alpha_k(0)}b_{k-1}^\dagger(0)a_{k+1}\right]}
\end{aligned}
\end{equation}
with $\delta\alpha_k=\alpha_k(t)-\alpha_k(0)$ and $\delta\theta_k=\theta_k(t)-\theta_k(0)$. The detailed derivation can be found in appendix~\ref{proof}.

In the rotating frame with respect to $V_{N-1}(t)$, we have
\begin{equation}\label{HamTimeInAnci}
\begin{aligned}
H_{\rm rot}(t)&=V_{N-1}^\dagger(t)H(t)V_{N-1}(t)-iV_{N-1}^\dagger(t)\frac{dV_{N-1}(t)}{dt}\\
&=\vec{\mu}_0^\dagger\left[H^\mu(t)-\mathcal{A}(t)\right]\vec{\mu}_0^T,
\end{aligned}
\end{equation}
where the dynamical coefficient matrix $H^\mu(t)$ and the gauge potential $\mathcal{A}$~\cite{Michael2017Geometry,Claeys2019Floquet,Takahashi2024Shortcuts} are determined by the system Hamiltonian and the rotated representation, respectively. Then the time-dependent Schr\"odinger equation~(\ref{Sch}) can be written as
\begin{equation}\label{Schrot}
i\frac{d}{dt}|\psi(t)\rangle_{\rm rot}=H_{\rm rot}(t)|\psi(t)\rangle_{\rm rot}
\end{equation}
with the rotated pure-state solution $|\psi(t)\rangle_{\rm rot}=V_{N-1}^\dagger(t)|\psi(t)\rangle$. The time-evolution operator for Eq.~(\ref{Schrot}) can be written in the Dyson series~\cite{Dyson1949TheRadiation} as
\begin{equation}\label{UrotDyson}
\begin{aligned}
&U_{\rm rot}(t)=\hat{T}e^{-i\int_0^tH_{\rm rot}(t')dt'}\\
=&\sum_{n=0}^{\infty}(-i)^n\int_0^tdt_1\cdots\int_0^{t_{n-1}}dt_nH_{\rm rot}(t_1)\cdots H_{\rm rot}(t_n),
\end{aligned}
\end{equation}
where $\hat{T}$ is the time-ordered operator.

\emph{Main result.}--- Here we prove that the commutation condition:
\begin{equation}\label{diagEq}
\left[H^\mu(t)-\mathcal{A}(t), \Pi^k\right]=0,
\end{equation}
is a necessary and sufficient condition for partially and fully determining the time-evolution operator $U_{\rm rot}(t)$. Here $\Pi^k$, $1\leq k\leq N$, is a projection operator or an $N\times N$ coefficient matrix with the only nonzero element at the $k$th row and the $k$th column, i.e., $\Pi^k_{jl}=\delta_{jk}\delta_{lk}$.

\emph{Necessary condition.}--- $U_{\rm rot}(t)$ in Eq.~(\ref{UrotDyson}) can be explicitly obtained when $[H_{\rm rot}(t_j), H_{\rm rot}(t_k)]=0$ for arbitrary $t_j$ and $t_k$. It is equivalent to the condition that the coefficient matrix in Eq.~(\ref{HamTimeInAnci}) is diagonal at any moment, i.e., $H^\mu_{km}(t)-\mathcal{A}_{km}(t)=0$ for $k\ne m$. In this case, $H_{\rm rot}(t)$ in Eq.~(\ref{HamTimeInAnci}) can be reduced as
\begin{equation}\label{HamrotDia}
H_{\rm rot}(t)=\sum_{k=1}^{N'\leq N}\left[H^\mu_{kk}(t)-\mathcal{A}_{kk}(t)\right]\mu_k^\dagger(0)\mu_k(0).
\end{equation}
$N'<N$ means that the coefficient matrix is partially diagonal within the first $N'$ degrees of freedom in the vector $\vec{\mu}^\dagger_0$. It gives rise to
\begin{equation}\label{NeceDia}
\left[H^\mu(t)-\mathcal{A}(t)\right]\Pi^k=\Pi^k\left[H^\mu(t)-\mathcal{A}(t)\right]
\end{equation}
with $k$ running from $1$ to $N'$, which can be written as Eq.~(\ref{diagEq}) in a more compact form. $N'=N$ means that $H_{\rm rot}(t)$ as well as $U_{\rm rot}(t)$ can be fully diagonalized. If the diagonal element in the gauge potential $\mathcal{A}_{kk}$ vanishes for arbitrary $t$, then the relevant $\mu_k(t)=\mu_k(0)$ describes a decoupled dark mode due to the fact that $\mathcal{A}_{km}\equiv-i[\mu_k(t), d\mu_m^\dagger(t)/dt]$. Otherwise, the time-dependent ancillary operator $\mu_k(t)$ determined by $H^\mu(t)$ can activate a useful passage. In general, we have
\begin{equation}\label{Urot}
U_{\rm rot}(t)=\sum_{k=1}^{N'\le N}e^{-if_{kk}(t)}\mu_k^\dagger(0)\mu_k(0),
\end{equation}
where the global phase is
\begin{equation}\label{global}
f_{kk}(t)=\int_0^tdt'\left[H^\mu_{kk}(t')-\mathcal{A}_{kk}(t')\right].
\end{equation}

\emph{Sufficient condition.}--- If the coefficient matrix $H^\mu(t)-\mathcal{A}(t)$ for Hamiltonian satisfies the commutation condition in Eq.~(\ref{diagEq}) with $k$ running from $1$ to $N'$, then $H_{\rm rot}(t)$ in the relevant subspace takes the diagonal form in Eq.~(\ref{HamrotDia}). Consequently, the time-evolution operator $U_{\rm rot}(t)$ can be directly obtained as Eq.~(\ref{Urot}).

Using Eqs.~(\ref{unitaryV}) and (\ref{Urot}), together with the Heisenberg equation of motion, the dynamics of each ancillary operator $\mu_k(t)$, $1\le k\le N'$, is found to be
\begin{equation}\label{Oper}
V_{N-1}(t)U_{\rm rot}(t)\mu_k(0)U_{\rm rot}^\dagger(t)V_{N-1}^\dagger(t)=e^{-if_{kk}(t)}\mu_k(t).
\end{equation}
Equation (\ref{Oper}) indicates that if the system initial state resides in the mode $\mu_k(0)$, then later it will evolve to the mode $\mu_k(t)$, with an accumulated global phase $f_{kk}(t)$. In other words, the constraints by the commutation condition in Eq.~(\ref{diagEq}) for the time-dependent Hamiltonian $H(t)$ can activate the ancillary mode $\mu_k(t)$ as useful nonadiabatic passage. The bosonic modes appear in the activated passages can be populated through the time evolution under control. In what follows, the feasibility of our universal control theory is further verified by the two-mode, three-mode, and $N$-mode bosonic systems.

\section{State exchange between two bosonic modes}\label{BosonTwo}

In this section, our universal control theory in Sec.~\ref{general} is used to exchange arbitrary states, including the Fock state, the coherent state, the cat state, and the thermal state, of two bosonic modes. In this minimal network, the two modes $a_1$ and $a_2$ are coupled by the exchange interaction with a strength $J$ and a phase $\varphi$. The full Hamiltonian reads
\begin{equation}\label{HamTwo}
H(t)=\frac{1}{2}\left(\omega_1a_1^\dagger a_1+\omega_2a_2^\dagger a_2\right)+\left(Je^{i\varphi}a_1^\dagger a_2+{\rm H.c.}\right),
\end{equation}
where $\omega_1$ and $\omega_2$ are the frequencies of the bosonic modes $a_1$ and $a_2$, respectively. In the rotating frame with respect to $H_0=\omega_0(t)/2(a_1^\dagger a_1+a_2^\dagger a_2)$, the Hamiltonian~(\ref{HamTwo}) can be transformed as
\begin{equation}\label{HamTwoDetu}
H(t)=\frac{1}{2}\Delta(t)(a_1^\dagger a_1-a_2^\dagger a_2)+(Je^{i\varphi}a_1^\dagger a_2+{\rm H.c.}),
\end{equation}
where the detuning satisfies $\Delta(t)=\omega_1-\omega_0(t)=-\omega_2+\omega_0(t)$.

Using Eq.~(\ref{TimeAnci}), the dynamics of the two-mode system can be described by the ancillary modes:
\begin{equation}\label{AnciTwo}
[\mu_1(t), \mu_2(t)]^T=\mathcal{M}^\dagger(t)(a_1, a_2)^T
\end{equation}
with the unitary transformation matrix
\begin{equation}\label{Mtwo}
\mathcal{M}^\dagger(t)=\begin{pmatrix}\cos\theta_1(t)e^{i\frac{\alpha_1(t)}{2}}&-\sin\theta_1(t)e^{-i\frac{\alpha_1(t)}{2}}\\ \sin\theta_1(t)e^{i\frac{\alpha_1(t)}{2}}&\cos\theta_1(t)e^{-i\frac{\alpha_1(t)}{2}}\end{pmatrix},
\end{equation}
where the parameters $\theta_1(t)$ and $\alpha_1(t)$ are associated with the population and the relative phase of the modes $a_1$ and $a_2$, respectively. Then the dynamics under the system Hamiltonian~(\ref{HamTwoDetu}) can be obtained by the rotation to the time-independent representation of ancillary modes, i.e., $V_1^\dagger(t)\mu_1(t)V_1(t)\rightarrow\mu_1(0)$ and $V_1^\dagger(t)\mu_2(t)V_1(t)\rightarrow\mu_2(0)$. Using Eq.~(\ref{unitaryV}), we have
\begin{equation}\label{Vtwo}
V_1(t)=V_{\alpha_1}(t)V_{\theta_1}(t),
\end{equation}
where
\begin{equation}\label{VtVa}
\begin{aligned}
V_{\alpha_1}(t)&=e^{-i\frac{\alpha_1(t)-\alpha_1(0)}{2}(a_1^\dagger a_1-a_2^\dagger a_2)},\\
V_{\theta_1}(t)&=e^{-[\theta_1(t)-\theta_1(0)][e^{i\alpha_1(0)}a_2^\dagger a_1-e^{-i\alpha_1(0)}a_1^\dagger a_2]}.
\end{aligned}
\end{equation}
Substituting Eqs.~(\ref{HamTwoDetu}), (\ref{AnciTwo}), and (\ref{Vtwo}) to the commutation condition~(\ref{diagEq}), the coupling strength and the detuning can be expressed by the parameters of the ancillary modes in Eq.~(\ref{AnciTwo}):
\begin{equation}\label{conditiontwo}
\begin{aligned}
&J(t)=\frac{\dot{\theta}_1(t)}{\sin\left[\varphi+\alpha_1(t)\right]},\\
&\Delta(t)=\dot{\alpha}_1(t)-2J\cos\left[\varphi+\alpha_1(t)\right]\cot[2\theta_1(t)].
\end{aligned}
\end{equation}
The conditions in Eq.~(\ref{conditiontwo}) share the similar forms as those for a discrete two-dimensional system~\cite{Jin2025Universal,Jin2025Entangling,Jin2025ErrCorr,Jin2025Rydberg}. It is attributed to the fact that the manifold geometry of two bosonic modes is essentially the same as that of a two-level system in determining the gauge field or Berry connection.

According to Eq.~(\ref{Oper}), the ancillary modes $\mu_1(0)$ and $\mu_2(0)$ in the Heisenberg picture evolve with time as
\begin{equation}\label{evolveTwo}
\mu_1(0)\rightarrow e^{-if(t)}\mu_1(t),\quad \mu_2(0)\rightarrow e^{if(t)}\mu_2(t),
\end{equation}
where the mode-dependent global phase $f(t)$ satisfies
\begin{equation}\label{globaltwo}
\dot{f}(t)=-J\frac{\cos[\varphi+\alpha_1(t)]}{\sin2\theta_1(t)}
=-\dot{\theta}_1(t)\frac{\cot[\varphi+\alpha_1(t)]}{\sin2\theta_1(t)}.
\end{equation}
Equations~(\ref{AnciTwo}) and (\ref{evolveTwo}) show that the ancillary modes $\mu_1(t)$ and $\mu_2(t)$ can be used to implement the state transfer or exchange. Both initial and target states can be specified by the proper setting of the boundary conditions, e.g., $\theta_1(t)$ and $\alpha_1(t)$. For example, arbitrary state of the mode $a_1$ can be faithfully transferred to the mode $a_2$ via the evolution along the passage $\mu_1(t)$ when $t=\tau$, under the conditions of $\theta_1(0)=0$ and $\theta_1(\tau)=\pi/2$ with $\tau$ the evolution period. During the same period, the initial state of $a_2$ is transferred to $a_1$ via the passage $\mu_2(t)$.

To avoid the singularity of the parameters in laboratory, we take $\theta_1(t)$, $\alpha_1(t)$, and $f(t)$ as independent variables. Using Eq.~(\ref{globaltwo}), Eq.~(\ref{conditiontwo}) is equivalent to
\begin{subequations}\label{ConditionInv}
\begin{align}
&2\Delta(t)=\dot{\alpha}_1(t)+2\dot{f}(t)\cos2\theta_1(t), \label{Deltar} \\
&\dot{\alpha}_1(t)=-\frac{\ddot{\theta}_1\dot{f}\sin2\theta_1-\ddot{f}\dot{\theta}_1\sin2\theta_1
-2\ddot{f}\dot{\theta}_1^2\cos2\theta_1}{\dot{f}^2\sin^22\theta_1+\dot{\theta}_1^2},\label{dotalpha}\\
&J(t)=-\sqrt{\dot{\theta}_1(t)^2+\dot{f}(t)^2\sin^22\theta_1(t)}. \label{Om0}
\end{align}
\end{subequations}
It is straightforward to find that the detuning $\Delta(t)$ can be controlled by $\dot{f}(t)$. In particular, $\dot{f}(t)=0$ yields a resonant detuning $\Delta(t)=0$, which is the parallel transport condition in discrete-variable systems as a key element of conventional holonomic transformation~\cite{Sjoqvist2012Nonadiabatic}. In contrast, the setting $\dot{f}(t)\neq0$ gives rise to off-resonance $\Delta(t)\neq0$. Our passage-construction protocol applies to both resonant and off-resonant detuning regimes. In the latter case, the protocol can directly start from Eq.~(\ref{HamTwo}) rather than Eq.~(\ref{HamTwoDetu}) in the rotating frame.

In case of $\omega_1\neq\omega_2$, our protocol can be alternatively performed at the cost of the time modulation over the the driving phase $\varphi=\varphi(t)$ instead of the driving detuning $\Delta(t)$. Using the original Hamiltonian~(\ref{HamTwo}) and the commutation condition (\ref{diagEq}), the constraint condition in Eq.~(\ref{conditiontwo}) for $\Delta(t)$ is replaced with
\begin{equation}\label{conditionTimeIn}
\varphi(t)=-\alpha(t)-\arctan\left[\frac{4\dot{\theta}(t)\cos2\theta(t)}{\omega_1-\omega_2-2\dot{\alpha}(t)}\right].
\end{equation}
And the condition for the coupling strength $J$ remains invariant.

\begin{figure}[htbp]
\centering
\includegraphics[width=0.9\linewidth]{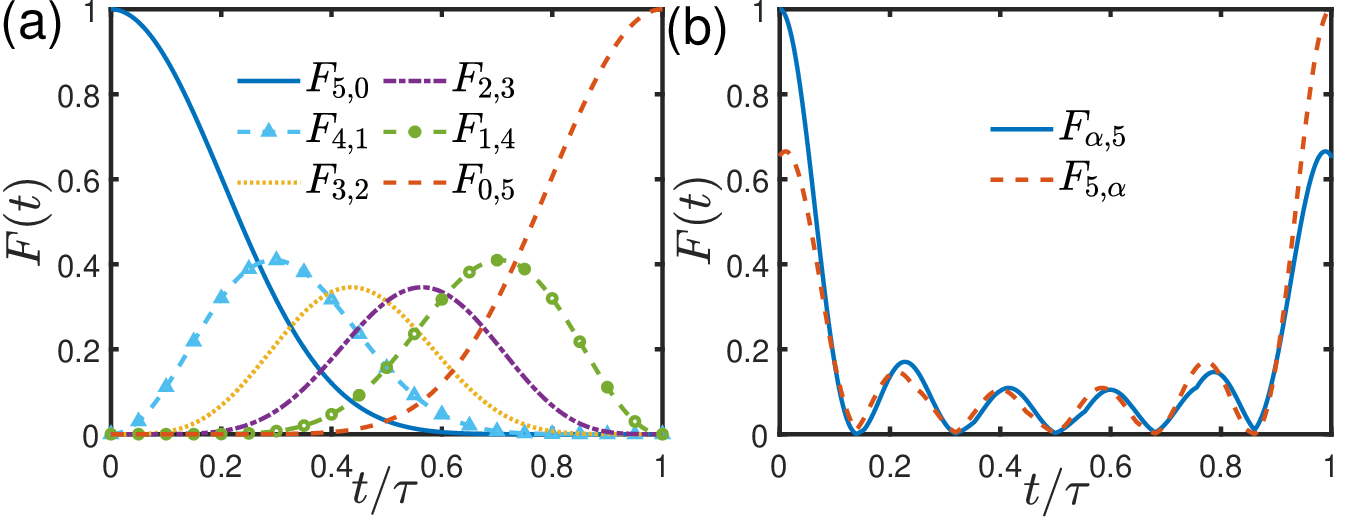}
\includegraphics[width=0.9\linewidth]{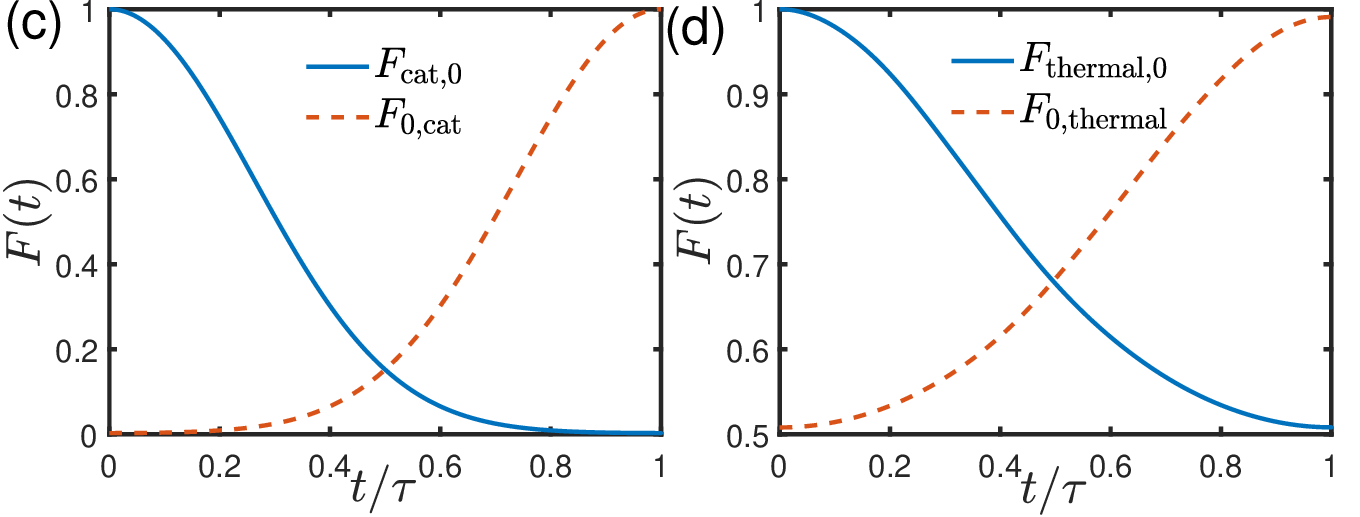}
\caption{Fidelity dynamics $F(t)$ for the state exchange in the two-mode system about (a) the Fock state $|5,0\rangle\rightarrow|0,5\rangle$, (b) the product of coherent state and Fock state $|\alpha,5\rangle\rightarrow|5,\alpha\rangle$ with $\alpha=5$, (c) the cat state $|{\rm cat},0\rangle\rightarrow|0,{\rm cat}\rangle$, where $|\rm cat\rangle=(|\alpha\rangle+|-\alpha\rangle)$ with $\alpha=5$, and (d) the thermal state $\rho_{\rm th}\otimes|0\rangle\langle0|\rightarrow|0\rangle\langle0|\otimes\rho_{\rm th}$, where $\rho_{\rm th}=\sum_np_n|n\rangle\langle n|$ with $p_n=(\bar{n}^n)/(1+\bar{n})^{n+1}$ and $\bar{n}=1$. The coupling strength $J$ and the detuning $\Delta(t)$ are set as Eq.~(\ref{ConditionInv}) with $\theta_1(t)=\pi t/(2\tau)$ and $f(t)=0$ in (a) and (b), or $f(t)=3\theta_1(t)$ in (c) and (d).}\label{PopuTwo}
\end{figure}

In Fig.~\ref{PopuTwo}, we demonstrate the performance of our protocol about the state exchange between the modes $a_1$ and $a_2$ by the fidelity $F=\langle\psi(t)|\rho|\psi(t)\rangle$, where $|\psi(t)\rangle$ is the pure-state solution of the time-dependent Schr\"odinger equation $id|\psi(t)\rangle/dt=H(t)|\psi(t)\rangle$ with the original Hamiltonian~(\ref{HamTwo}). Here $\rho$ can be the initial, intermediate, or target states. In Fig.~\ref{PopuTwo}(a), $a_1$ is initially prepared in the Fock state $|n=5\rangle$ and $a_2$ is prepared in the vacuum state $|0\rangle$. $\rho$ is then chosen such that both modes $a_1$ and $a_2$ are in Fock states. Consequently, the fidelity can be written as $F_{n_1,n_2}=|\langle n_1|\langle n_2|\psi(t)\rangle|^2$. During the time evolution, the nonadiabatic passage is described by the temporary occupations on the intermediate states: $P_{4,1}=0.41$ when $t=0.30\tau$, $P_{3,2}=0.35$ when $t=0.44\tau$, $P_{2,3}=0.35$ when $t=0.56\tau$, and $P_{1,4}=0.41$ when $t=0.71\tau$. The initial Fock state $|5,0\rangle$ completely becomes $|0,5\rangle$ when $t=\tau$. In Fig.~\ref{PopuTwo}(b), the initial state is a tensor product of a coherent state and a Fock state $|\alpha=5\rangle_1\otimes|n=5\rangle_2$. Then $\rho=|\alpha,5\rangle\langle\alpha,5|$ or $\rho=|5,\alpha\rangle\langle5,\alpha|$. It is found that in the end of the passage, the states of modes $a_1$ and $a_2$ are perfectly exchanged. And it is interesting to find that in between the beginning and the end of the passage, there exist $m=5$ fidelity peaks during the time evolution, the same as Fig.~\ref{PopuTwo}(a).

Similarly, in Fig.~\ref{PopuTwo}(c), the mode $a_1$ is initially prepared as a cat state $|\rm cat\rangle=(|\alpha\rangle+|-\alpha\rangle)/\mathcal{N}$ with $\alpha=5$ and $\mathcal{N}$ the normalization coefficient. Again it is confirmed that $a_2(\tau)=a_1(0)$ and $a_1(\tau)=a_2(0)$. Our protocol even applies to the mixed state. In Fig.~\ref{PopuTwo}(d), the fidelity is evaluated by $F={\rm Tr}[\rho(t)\rho_{\rm th}]$, where $\rho(t)$ is the solution to the von-Neumann equation driven by the original Hamiltonian~(\ref{HamTwo}) and $\rho_{\rm th}=\sum_np_n|n\rangle\langle n|$ with $p_n=(\bar{n}^n)/(1+\bar{n})^{n+1}$ with $\bar{n}=1$ the average occupation. A complete exchange is also observed for the thermal states $\rho_{\rm th}$ and $|0\rangle\langle0|$, as shown in Fig.~\ref{PopuTwo}(d).

\begin{figure}[htbp]
\centering
\includegraphics[width=0.9\linewidth]{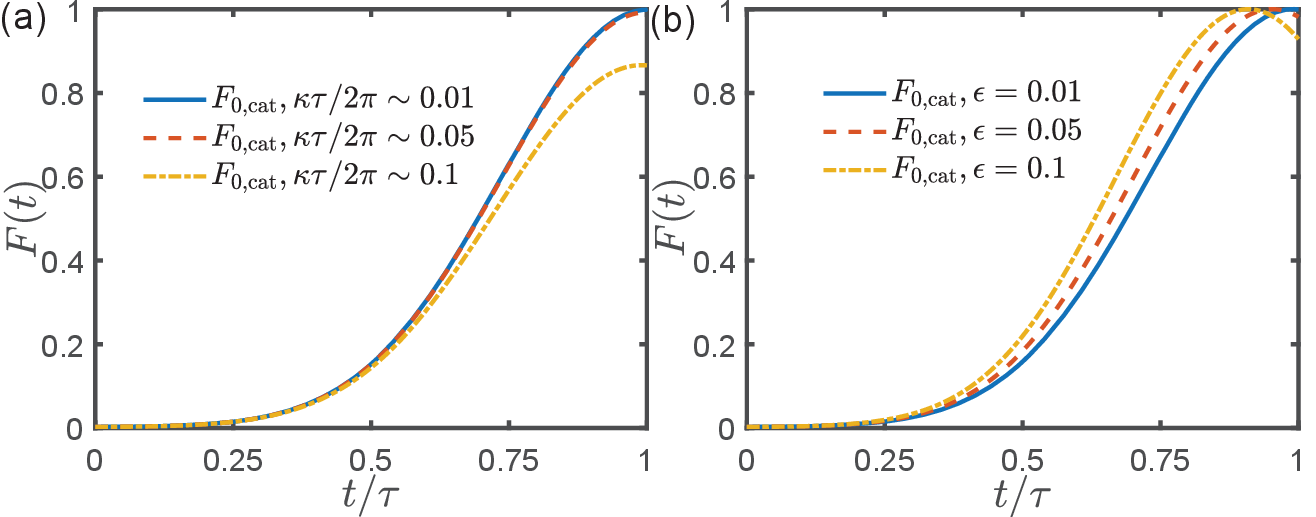}
\caption{Fidelity dynamics $F(t)$ for the target state $|0,\rm cat\rangle$ during the exchange of cat state $|\rm cat,0\rangle\rightarrow|0,\rm cat\rangle$: (a) for various environmental decay rates $\kappa$; and (b) for various derivation coefficients $\epsilon$ in the imperfect parameter setting $\theta_1(t)\rightarrow(1+\epsilon)\theta_1(t)$ and $\theta_1(t)=(\pi t)/(2\tau)$. The evolution period is set as $\tau=100$ ns and the coupling strength $J/2\pi\sim25$ MHz. The other parameters are the same as Fig.~\ref{PopuTwo}(c).}\label{Decay}
\end{figure}

The robustness of our protocol can be discussed against the environmental coupling or the parameter imperfection. The system dynamics in the presence of decoherence can be described by the master equation~\cite{Carmichael1999statistical},
\begin{equation}\label{masterequationGeneral}
\frac{\partial}{\partial t}\rho=-i[H(t), \rho]+\frac{\kappa}{2}\mathcal{L}_{a_1}[\rho]+\frac{\kappa}{2}\mathcal{L}_{a_2}[\rho],
\end{equation}
where $H(t)$ is the ideal Hamiltonian in Eq.~(\ref{HamTwo}) and the Lindblad superoperators are defined as $\mathcal{L}_o[\rho]=2o\rho o^\dagger-o^\dagger o\rho-\rho o^\dagger o$ with $o=a_1,a_2$ indicating the decay channels for the bosonic modes $a_1$ and $a_2$, respectively. The decay rates are set as the same value. In the state-of-art cavity magnonic systems~\cite{Lachance2019Hybrid,Babak2022Cavity}, the exchange coupling strength between the cavity and magnon modes can be as large as $J/2\pi\sim25$ MHz~\cite{Lachance2019Hybrid,Babak2022Cavity} and the decay rate is in the order of $\kappa/2\pi\sim1$ MHz. In Fig.~\ref{Decay}(a), we demonstrate the fidelity dynamics of the target state $|0,\rm cat\rangle$ during the cat state exchange of $|\rm cat,0\rangle\rightarrow|0,\rm cat\rangle$ between the bosonic modes $a_1$ and $a_2$. Our protocol is found to be robust against decoherence. In particular, we have $F_{0,\rm cat}(\tau)=0.993$ when $\kappa\tau/2\pi\sim0.05$ (close to the experimental conditions~\cite{Lachance2019Hybrid,Babak2022Cavity}) and $F_{0,\rm cat}(\tau)=0.867$ when $\kappa\tau/2\pi\sim0.1$. In Fig.~\ref{Decay}(b), we plot the same fidelity under the parametric fluctuation, i.e., $\theta_1(t)\rightarrow(1+\epsilon)\theta_1(t)$ with the deviation coefficient $\epsilon$, which affects both detuning and coupling strength due to Eqs.~(\ref{Deltar}) and (\ref{Om0}). It is found that our protocol is also insensitive to the imperfect parameter setting. In particular, we have $F_{0,\rm cat}(\tau)=0.999$ when $\epsilon=0.01$, $F_{0,\rm cat}(\tau)=0.982$ when $\epsilon=0.05$, and $F_{0,\rm cat}(\tau)=0.929$ when $\epsilon=0.1$.

\section{Chiral entanglement transfer among three modes}\label{BosonThree}

\begin{figure}[htbp]
\centering
\includegraphics[width=0.5\linewidth]{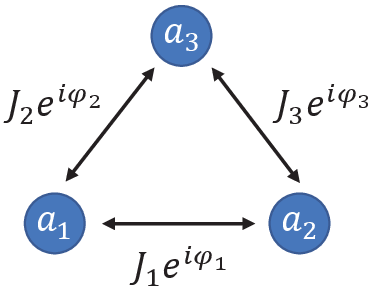}
\caption{Sketch of a tripartite system comprising three bosonic modes $a_1$, $a_2$, and $a_3$, which are coupled by the exchange interactions with the coupling strengths $J_1$, $J_2$, and $J_3$, and the phases $\varphi_1$, $\varphi_2$, and $\varphi_3$, respectively.}\label{modelThree}
\end{figure}

This section is devoted to the control over a time-dependent tripartite bosonic system in Fig.~\ref{modelThree}. Our target is to realize the chiral transfer of the two-body maximally entangled state in this ``triangle'' system. Consider three non-degenerate bosonic modes $a_1$, $a_2$, and $a_3$ with the frequencies $\omega_1$, $\omega_2$, and $\omega_3$, respectively. Each pair of the bosonic modes are coupled through an exchange interaction. In particular, the full Hamiltonian reads
\begin{equation}\label{HamThree}
\begin{aligned}
&H(t)=\frac{1}{2}\left(\omega_1a_1^\dagger a_1+\omega_2a_2^\dagger a_2+\omega_3a_3^\dagger a_3\right)\\
+&\left(J_1e^{i\varphi_1}a_1^\dagger a_2+J_2e^{i\varphi_2}a_1^\dagger a_3+J_3e^{i\varphi_3}a_2^\dagger a_3+{\rm H.c.}\right),
\end{aligned}
\end{equation}
where $J_j$ and $\varphi_j$ are coupling strength and phase, respectively. In the rotating frame with respect to $H_0=\omega_0(t)/2(a_1^\dagger a_1+a_2^\dagger a_2+a_3^\dagger a_3)$, we have
\begin{equation}\label{HamThreeDetu}
\begin{aligned}
&H(t)=\frac{1}{2}\left[\Delta_1(t)a_1^\dagger a_1+\Delta_2(t)a_2^\dagger a_2+\Delta_3(t)a_3^\dagger a_3\right]\\
+&\left(J_1e^{i\varphi_1}a_1^\dagger a_2+J_2e^{i\varphi_2}a_1^\dagger a_3+J_3e^{i\varphi_3}a_2^\dagger a_3+{\rm H.c.}\right)
\end{aligned}
\end{equation}
under the conditions of $\Delta_j(t)=\omega_j-\omega_0(t)$ with $j=1,2,3$.

Similar to Eqs.~(\ref{TimeAnci}) and (\ref{AnciTwo}), the dynamics of an arbitrary three-mode system can be described in the ancillary representation, in which the ancillary modes can be alternatively chosen as
\begin{equation}\label{AnciThree}
\left[\mu_1(t), \mu_2(t), \mu_3(t)\right]^T=\mathcal{M}^\dagger(t)(a_1, a_2, a_3)^T,
\end{equation}
where $\mathcal{M}^\dagger(t)$ is a $3\times3$ unitary transformation matrix
\begin{equation}\label{Mthree}
\mathcal{M}^\dagger(t)=\begin{pmatrix}\vec{u}_1(t)&0\\ \cos\theta_2(t)e^{i\frac{\alpha_2(t)}{2}}\vec{b}_1(t)&-\sin\theta_2(t)e^{-i\frac{\alpha_2(t)}{2}}\\
\sin\theta_2(t)e^{i\frac{\alpha_2(t)}{2}}\vec{b}_1(t)&\cos\theta_2(t)e^{-i\frac{\alpha_2(t)}{2}}\end{pmatrix}
\end{equation}
with $\vec{u}_1(t)=[\cos\theta_1(t)e^{i\alpha_1(t)/2}, -\sin\theta_1(t)e^{-i\alpha_1(t)/2}]$ and $\vec{b}_1(t)=[\sin\theta_1(t)e^{i\alpha_1(t)/2}, \cos\theta_1(t)e^{-i\alpha_1(t)/2}]$ as two row vectors. Again, the parameters $\theta_1(t)$ and $\theta_2(t)$ are associated with the populations of the bosonic modes $a_1$, $a_2$, and $a_3$, and $\alpha_1(t)$ and $\alpha_2(t)$ are associated with their relative phases.

Using Eq.~(\ref{unitaryV}), the system dynamics can be solved in the stationary ancillary representation, by which $V_2^\dagger(t)\mu_k(t)V_2(t)\rightarrow\mu_k(0)$ with $k=1,2,3$. Specifically, the unitary transformation $V_2(t)$ can be chosen as
\begin{equation}\label{Vthree}
V_2(t)=V_{\alpha_1}(t)V_{\theta_1}(t)V_{\alpha_2}(t)V_{\theta_2}(t),
\end{equation}
where $V_{\alpha_1}(t)$ and $V_{\theta_1}(t)$ have been given by Eq.~(\ref{VtVa}) and
\begin{equation}\label{VpVbVtVa}
\begin{aligned}
V_{\alpha_2}(t)&=e^{-i\frac{\alpha_2(t)-\alpha_2(0)}{2}\left[b_1^\dagger(0)b_1(0)-a_3^\dagger a_3\right]},\\
V_{\theta_2}(t)&=e^{-[\theta_2(t)-\theta_2(0)][e^{i\alpha_2(0)}a_3^\dagger b_1(0)-e^{-i\alpha_2(0)}b_1^\dagger(0)a_3]}.
\end{aligned}
\end{equation}

Plugging Eqs.~(\ref{HamThreeDetu}), (\ref{AnciThree}), and (\ref{Vthree}) into the commutation condition~(\ref{diagEq}), we have
\begin{equation}\label{detuning}
\begin{aligned}
\Delta_1(t)&=-\Delta(t)\sin^2\theta_1(t)-\Delta_a(t),\\
\Delta_2(t)&=-\Delta(t)\cos^2\theta_1(t)+\Delta_a(t),\\
\Delta_3(t)&=\Delta(t)
\end{aligned}
\end{equation}
with the scaling detunings $\Delta(t)$ and $\Delta_a(t)$, and
\begin{equation}\label{CouplingStrength}
\begin{aligned}
J_1e^{i\varphi_1}&=J_a-\frac{1}{2}\Delta(t)\sin\theta_1(t)\cos\theta_1(t)e^{-i\alpha_1(t)},\\
J_2e^{i\varphi_2}&=J\sin\theta_1(t)e^{-i\frac{\alpha_1(t)}{2}},\\
J_3e^{i\varphi_3}&=J\cos\theta_1(t)e^{i\frac{\alpha_1(t)}{2}}
\end{aligned}
\end{equation}
with the scaling coupling strengths $J_a$ and $J$. The scaling parameters are determined by
\begin{equation}\label{ConditionMu0}
\begin{aligned}
&\Delta_a(t)=\dot{\alpha}_1(t)+2J_a\cot2\theta_1(t)\cos\alpha_1(t),\\
&\Delta(t)=\dot{\alpha}_2(t)+2J\cot2\theta_2(t)\cos\alpha_2(t)+\frac{J_a\cos\alpha_1(t)}{\sin2\theta_1(t)},\\
&J_a(t)=-\frac{\dot{\theta}_1(t)}{\sin\alpha_1(t)},\\
&J(t)=-\frac{\dot{\theta_2}(t)}{\sin\alpha_2(t)}.
\end{aligned}
\end{equation}

Under the constraint conditions in Eqs.~(\ref{detuning}), (\ref{CouplingStrength}), and (\ref{ConditionMu0}), the ancillary modes $\mu_k(t)$'s in Eq.~(\ref{AnciThree}) can be activated as useful nonadiabatic passages. The ancillary modes $\mu_k(0)$'s evolve with time in accordance to Eq.~(\ref{Oper}) as
\begin{equation}\label{evolveThree}
\mu_k(0)\rightarrow e^{if_k(t)}\mu_k(t), \quad k=1,2,3,
\end{equation}
where the global phases can be expressed as
\begin{equation}\label{globalphase}
\begin{aligned}
&\dot{f}_1(t)=J_a\frac{\cos\alpha_1(t)}{\sin2\theta_1(t)}=-\dot{\theta}_1(t)\frac{\cot\alpha_1(t)}{\sin2\theta_1(t)},\\
&\dot{f}_2(t)=\dot{f}(t)-\frac{1}{2}\dot{f}_1(t), \\
&\dot{f}_3(t)=-\dot{f}(t)-\frac{1}{2}\dot{f}_1(t)
\end{aligned}
\end{equation}
with
\begin{equation}\label{globalphasef}
\dot{f}(t)=J\frac{\cos\alpha_2(t)}{\sin2\theta_2(t)}=-\dot{\theta}_2(t)\frac{\cot\alpha_2(t)}{\sin2\theta_2(t)}.
\end{equation}
During the practical control, one can choose $\theta_1(t)$, $\theta_2(t)$, $f_1(t)$, and $f(t)$ as independent variables to avoid the singularity of the experimental parameters. Specifically, using Eq.~(\ref{globalphasef}), Eq.~(\ref{ConditionMu0}) can be rewritten as
\begin{equation}\label{Mu0Inverse}
\begin{aligned}
&\Delta_a(t)=\dot{\alpha}_1(t)+2\dot{f}_1(t)\cos2\theta_1(t),\\
&\dot{\alpha}_1(t)=-\frac{\ddot{\theta}_1\dot{f}_1\sin2\theta_1-\ddot{f}_1\dot{\theta}_1\sin2\theta_1
-2\dot{f}_1\dot{\theta}_1^2\cos2\theta_1}{\dot{f}_1^2\sin^22\theta_1+\dot{\theta}_1^2},\\
&J_a(t)=-\sqrt{\dot{\theta}_1(t)^2+\dot{f}_1(t)^2\sin^22\theta_1(t)},
\end{aligned}
\end{equation}
and
\begin{equation}\label{Mu12Inverse}
\begin{aligned}
&\Delta(t)=\dot{\alpha}_2(t)+2\dot{f}(t)\cos2\theta_2(t)+\dot{f}_1(t),\\
&\dot{\alpha}_2(t)=-\frac{\ddot{\theta}_2\dot{f}\sin2\theta_2-\ddot{f}\dot{\theta}_2\sin2\theta_2
-2\dot{f}\dot{\theta}_2^2\cos2\theta_2}{\dot{f}^2\sin^22\theta_2+\dot{\theta}_2^2},\\
&J(t)=-\sqrt{\dot{\theta}_2(t)^2+\dot{f}(t)^2\sin^22\theta_2(t)}.
\end{aligned}
\end{equation}

We assume that the initial state of the entire system reads $|\psi(0)\rangle=|\phi(2)\rangle_{13}\otimes|0\rangle_2$, where
\begin{equation}\label{NOONdef}
|\phi(N)\rangle_{jk}\equiv\frac{1}{\sqrt{2}}(|N0\rangle+|0N\rangle)_{jk}.
\end{equation}
In other words, when $t=0$, the first and third modes are prepared in a maximally entangled state, i.e., the NOON state~\cite{Lee2002Quantum,Pezze2018Quantum} $|\phi(N=2)\rangle_{13}$ and the second mode is in the vacuum state $|0\rangle_2$. Our target is to realize a chiral transfer of the NOON state along the triangle network by appropriately setting the boundary conditions of $\theta_1(t)$ and $\theta_2(t)$ for the passages $\mu_k(t)$ in Eq.~(\ref{AnciThree}). In the counterclockwise direction, the transfer can be divided into three stages of equal period: (i) $|\psi(0)\rangle\rightarrow|\phi(2)\rangle_{12}\otimes|0\rangle_3$ when $t\in[0, \tau]$, (ii) $|\phi(2)\rangle_{12}\otimes|0\rangle_3\rightarrow|\phi(2)\rangle_{23}\otimes|0\rangle_1$ when $t\in[\tau, 2\tau]$, and (iii) $|\phi(2)\rangle_{23}\otimes|0\rangle_1\rightarrow|\psi(0)\rangle$ when $t\in[2\tau, 3\tau]$, where $\tau$ is the period of each stage.

\begin{figure}[htbp]
\centering
\includegraphics[width=0.85\linewidth]{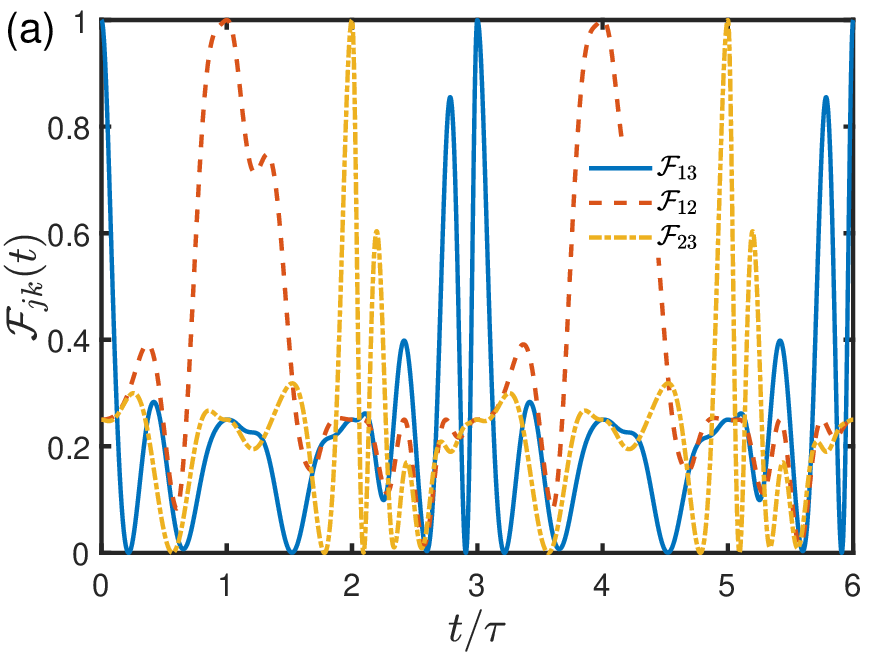}
\includegraphics[width=0.85\linewidth]{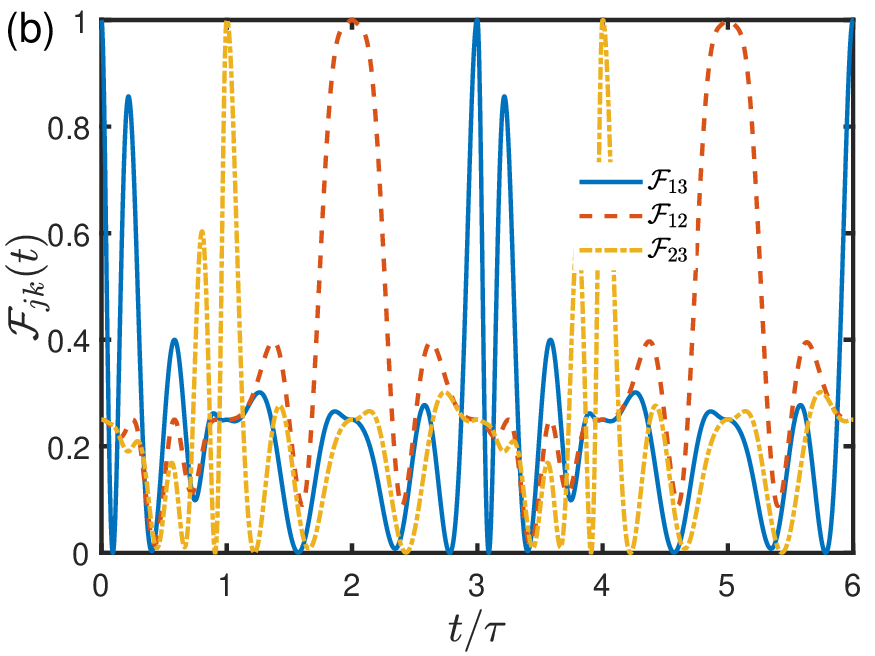}
\caption{Fidelity dynamics $\mathcal{F}_{jk}(t)$ about the chiral transfer of the NOON state in the three-mode system of a triangular configuration (see Fig.~\ref{modelThree}) along (a) the counterclockwise direction and (b) the clockwise direction. Under the conditions in Eqs.~(\ref{detuning}) and (\ref{CouplingStrength}), the parameters $\Delta_a(t)$, $\Delta(t)$, $J_a(t)$, and $J(t)$ are set according to Eqs.~(\ref{Mu0Inverse}) and (\ref{Mu12Inverse}) with $f_1(t)=0$ and $f(t)=3\theta_2(t)$. In (a) $\theta_1(t)$ and $\theta_2(t)$ are set by Eq.~(\ref{parathree}) and in (b) $\theta_1(t)$ and $\theta_2(t)$ are set by Eq.~(\ref{parathreeCount}).}\label{Chiral}
\end{figure}

In particular, during Stage (i), under the boundary conditions of $\theta_1(0)=0$, $\theta_2(0)=0$, $\theta_1(\tau)=\pi/2$, and $\theta_2(\tau)=\pi/2$, we have $\mu_1(0)=a_1\rightarrow\mu_1(\tau)=a_2$, $\mu_2(0)=a_2\rightarrow\mu_2(\tau)=a_3$, and $\mu_3(0)=a_3\rightarrow\mu_1(\tau)=a_1$ according to Eq.~(\ref{AnciThree}). It indicates that the states of modes $a_1$, $a_2$, and $a_3$ at $t=0$, are transferred to $a_2$, $a_3$, and $a_1$, respectively, at $t=\tau$. Then the bases in $|\psi(0)\rangle$ are transformed as
\begin{equation}\label{StageoneTrans}
\begin{cases}
  |2\rangle_1|0\rangle_2|0\rangle_3\rightarrow|0\rangle_1|2\rangle_2|0\rangle_3, \\
  |0\rangle_1|0\rangle_2|2\rangle_3\rightarrow|2\rangle_1|0\rangle_2|0\rangle_3.
\end{cases}
\end{equation}
Similar to the process by Eq.~(\ref{StageoneTrans}), the boundary conditions for Stages (ii) and (iii) are set as $\theta_1(\tau+0^+)=0$, $\theta_2(\tau+0^+)=0$, $\theta_1(2\tau)=\pi/2$, $\theta_2(2\tau)=\pi/2$, $\theta_1(2\tau+0^+)=\pi$, $\theta_2(2\tau+0^+)=\pi/2$, $\theta_1(3\tau)=\pi/2$, and $\theta_2(3\tau)=\pi$. In general, the parameters $\theta_1(t)$ and $\theta_2(t)$ for the $k$th loop of the counterclockwise NOON-state transfer can be set as
\begin{equation}\label{parathree}
\begin{aligned}
\theta_1(t)&=\frac{\pi[t-3(k-1)\tau]}{2\tau}, \quad \theta_2(t)=\theta_1(t),\\
\theta_1(t)&=\frac{\pi[t-3(k-1)\tau]}{2\tau}-\frac{\pi}{2}, \quad \theta_2(t)=\theta_1(t),\\
\theta_1(t)&=\frac{\pi[t-3(k-1)\tau]}{2\tau}+\pi, \quad \theta_2(t)=\theta_1(t)-\frac{\pi}{2}
\end{aligned}
\end{equation}
for the three stages, respectively.

As for the clockwise NOON-state transfer, $\theta_1(t)$ and $\theta_2(t)$ for the three stages can be respectively set as
\begin{equation}\label{parathreeCount}
\begin{aligned}
\theta_1(t)&=\frac{\pi[t-3(k-1)\tau]}{2\tau}+\frac{\pi}{2}, \quad \theta_2(t)=\theta_1(t)-\frac{\pi}{2},\\
\theta_1(t)&=\frac{\pi[t-3(k-1)\tau]}{2\tau}, \quad \theta_2(t)=\theta_1(t)+\frac{\pi}{2},\\
\theta_1(t)&=\frac{\pi[t-3(k-1)\tau]}{2\tau}+\frac{\pi}{2}, \quad \theta_2(t)=\theta_1(t)+\frac{\pi}{2},
\end{aligned}
\end{equation}
in the $k$th loop, $k\geq1$.

The performance of our protocol can be evaluated by the dynamics of the state or entanglement fidelity $\mathcal{F}_{jk}=|\langle \psi(t)|\phi(2)\rangle_{jk}|0\rangle_{l\neq j,k}|^2$ with respect to the target state $|\phi(2)\rangle_{jk}$ defined in Eq.~(\ref{NOONdef}), where $|\psi(t)\rangle$ is the pure-state solution of the Schr\"odinger equation with the original Hamiltonian~(\ref{HamThree}). In Fig.~\ref{Chiral}(a), it is found that the perfect counterclockwise entanglement transfer can be achieved as $\mathcal{F}_{13}(0)=1$ when $t=0$, $\mathcal{F}_{12}(\tau)=1$ when $t=\tau$, $\mathcal{F}_{23}(2\tau)=1$ when $t=2\tau$ and $\mathcal{F}_{13}(3\tau)=1$ when $t=3\tau$. During the period $t\in[3\tau, 6\tau]$, the second loop perfectly repeats the first one. In Fig.~\ref{Chiral}(b), the NOON state is perfectly transferred in a clockwise manner. Specifically, we have $\mathcal{F}_{13}(0)=1$ when $t=0$, $\mathcal{F}_{23}(\tau)=1$ when $t=\tau$, $\mathcal{F}_{12}(2\tau)=1$ when $t=2\tau$, and $\mathcal{F}_{13}(3\tau)=1$ when $t=3\tau$. The behavior of the second loop is also identical to that of the first loop. Note the chiral transfer of continuous-variable system in the current protocol is realized through nonadiabatic control rather than the Floquet driving~\cite{Qi2022Floquet} that was featured with a fixed ratio of the driving intensity and frequency and the uniformly distributed local phases. Thus, the nonadiabatic passage is much flexible in parametric setting than the Floquet driving, e.g., Eqs.~(\ref{parathree}) and (\ref{parathreeCount}) can be replaced with any functions holding the same boundary conditions. Also, the current protocol enables the chiral transfer of selected nodes in the entire network, without eliminating the irrelevant couplings or connections.

\section{Chiral state transfer in network}\label{BosonNet}

\begin{figure}[htbp]
\centering
\includegraphics[width=0.8\linewidth]{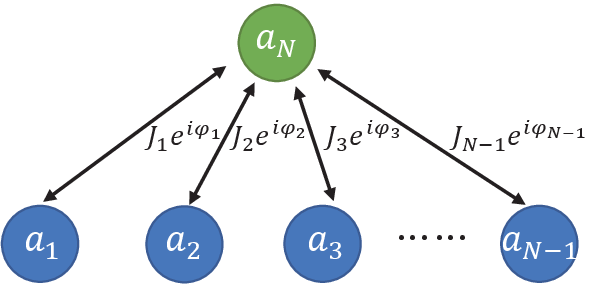}
\caption{Sketch of the bosonic network comprising $N$ bosonic modes, in which a central bosonic mode $a_N$ is coupled to the other uncoupled bosonic modes $a_n$ with $1\leq n\leq N-1$ via the exchange interaction that is characterized by the coupling strength $J_n$ and the phase $\varphi_n$.}\label{modelN}
\end{figure}

In this section, our control protocol is applied to a central-configuration bosonic network as shown in Fig.~\ref{modelN} which consists of $N$ bosonic modes. The central node $a_N$ is coupled to the other $a_n$'s, $1\leq n\leq N-1$, via the exchange interaction that is characterized by the coupling strength $J_n$ and the phase $\varphi_n$. Then the full Hamiltonian can be written as
\begin{equation}\label{HamN}
H(t)=\sum_{n=1}^N\frac{1}{2}\omega_na_n^\dagger a_n+\sum_{n=1}^{N-1}\left(J_ne^{i\varphi_n}a_N^\dagger a_n+{\rm H.c.}\right),
\end{equation}
where $\omega_n$ is the mode frequency. In the rotating frame with respect to $H_0=\omega_0(t)/2\sum_{n=1}^Na_n^\dagger a_n$, the Hamiltonian can be transformed as
\begin{equation}\label{HamNdetu}
H(t)=\sum_{n=1}^N\frac{1}{2}\Delta_n(t)a_n^\dagger a_n+\sum_{n=1}^{N-1}\left(J_ne^{i\varphi_n}a_N^\dagger a_n+{\rm H.c.}\right),
\end{equation}
where the detuning $\Delta_n(t)\equiv\omega_n-\omega_0(t)$.

Here, to clarify the underlying ideas of our theory, we would further elaborate how the framework introduced in Sec.~\ref{general} can be applied to any $N$-mode system, such as the centralized network in Fig.~\ref{modelN} or the bosonic network composed of two coupled subsystems~\cite{Jian2023Dark} if there exist more than one central node. In our framework, the dynamics of the general system can be described in the time-dependent ancillary modes $\mu_k(t)$ in Eq.~(\ref{TimeAnci}). By applying the unitary transformation $V_{N-1}(t)$ in Eq.~(\ref{unitaryV}) together with the commutation condition in Eq.~(\ref{diagEq}), the resulting constraints on the Hamiltonian~(\ref{HamNdetu}) activate the ancillary modes in Eq.~(\ref{Oper}) to be universal passages for versatile tasks, e.g., exchanging arbitrary unknown states between the desired bosonic modes in a connected network, irrespective of the network size.

As a concrete example, we consider the Hamiltonian in Eq.~(\ref{HamNdetu}) with $N=4$. Then the ancillary modes in Eq.~(\ref{TimeAnci}) is written as
\begin{equation}\label{TimeAnci4}
\begin{aligned}
\mu_1(t)&=\cos\theta_1e^{i\frac{\alpha_1}{2}}a_1-\sin\theta_1e^{-i\frac{\alpha_1}{2}}a_2,\\
\mu_2(t)&=\cos\theta_2e^{i\frac{\alpha_2}{2}}b_1(t)-\sin\theta_2e^{-i\frac{\alpha_2}{2}}a_3,\\
\mu_3(t)&=\cos\theta_3e^{i\frac{\alpha_3}{2}}b_2(t)-\sin\theta_3e^{-i\frac{\alpha_3}{2}}a_4,\\
\mu_4(t)&=\sin\theta_3e^{i\frac{\alpha_3}{2}}b_2(t)+\cos\theta_3e^{-i\frac{\alpha_3}{2}}a_4,
\end{aligned}
\end{equation}
where $b_1(t)=\sin\theta_1e^{i\alpha_1/2}a_1+\cos\theta_1e^{-i\alpha_1/2}a_2$ and $b_2(t)=\sin\theta_2e^{i\alpha_2/2}b_1(t)+\cos\theta_2e^{-i\alpha_2/2}a_3$ due to Eq.~(\ref{unitarVecBri}). Subsequently, the unitary transformation in Eq.~(\ref{unitaryV}), which connects the time-dependent and time-independent ancillary modes, is expressed as
\begin{equation}\label{unitaryV4}
V_3(t)=V_{\alpha_1}V_{\theta_1}V_{\alpha_2}V_{\theta_2}V_{\alpha_3}V_{\theta_3}=\prod_{k=1}^{3}V_{\alpha_k}V_{\theta_k}
\end{equation}
with $V_{\alpha_k}(t)$ and $V_{\theta_k}(t)$ defined in Eq.~(\ref{uniVat}).

Both ancillary modes $\mu_3(t)$ and $\mu_4(t)$ in Eq.~(\ref{TimeAnci4}) can be used to control the whole network. With no loss of generality, we here substitute $\mu_4(t)$ to the commutation condition~(\ref{diagEq}). The resulting conditions about the coupling strengths and the detunings are
\begin{equation}
\begin{aligned}\label{CondJ}
J_1&=-\big(\dot{\theta}_3\sin\theta_2\sin\theta_1+\dot{\theta}_2\tan\theta_3\cos\theta_2\sin\theta_1
+\dot{\theta}_1\tan\theta_3\\ &\times\sin\theta_2\cos\theta_1\big)/\sin(\varphi_1+\alpha_3),\\
J_2&=-\big(\dot{\theta}_3\sin\theta_2\cos\theta_1+\dot{\theta}_2\tan\theta_3\cos\theta_2\cos\theta_1
-\dot{\theta}_1\tan\theta_3\\ &\times\sin\theta_2\sin\theta_1\big)/\sin(\varphi_2-\alpha_1+\alpha_3),\\
J_3&=-\big(\dot{\theta}_3\cos\theta_2-\dot{\theta}_2\tan\theta_3\sin\theta_2\big)/\sin(\varphi_3-\alpha_2+\alpha_3),
\end{aligned}
\end{equation}
and
\begin{equation}
\begin{aligned}\label{CondDetu}
&\Delta_1(t)=-J_1\frac{\cot\theta_3}{\sin\theta_2\sin\theta_1}\cos(\varphi_1+\alpha_3),\\
&\Delta_2(t)=\dot{\alpha}_1-J_2\frac{\cot\theta_3}{\sin\theta_2\cos\theta_1}\cos(\varphi_2-\alpha_1+\alpha_3)\\
&\Delta_3(t)=\dot{\alpha}_2-J_3\frac{\cot\theta_3}{\cos\theta_2}\cos(\varphi_3-\alpha_2+\alpha_3),\\
&\Delta_4(t)=\dot{\alpha}_3-J_1\tan\theta_3\sin\theta_2\sin\theta_1\cos(\varphi_1+\alpha_3)\\
&-J_2\tan\theta_3\sin\theta_2\cos\theta_1\cos(\varphi_2-\alpha_1+\alpha_3)\\
&-J_3\tan\theta_3\times\cos\theta_2\cos(\varphi_3-\alpha_2+\alpha_3),
\end{aligned}
\end{equation}
respectively. They determine the laboratory implementation of $H(t)$.

Along the passage $\mu_4(t)$ with the constraint conditions in Eqs.~(\ref{CondJ}) and (\ref{CondDetu}), a chiral state transfer among the modes $a_1$, $a_2$, and $a_3$ can be divided into three stages of state conversion, i.e., in Stage (i) for $t\in[0, \tau]$, $a_1\leftrightarrow a_2$, in Stage (ii) for $t\in[\tau, 2\tau]$, $a_2\leftrightarrow a_3$, and in Stage (iii) for $t\in[2\tau, 3\tau]$, $a_3\leftrightarrow a_1$. In other words, an arbitrary initial state in mode $a_1$ propagates to mode $a_2$ at $t=\tau$, to mode $a_3$ at $t=2\tau$, and back to mode $a_1$ at $t=3\tau$, despite they have no direction connections. This task requires the parameters $\theta_n(t)$'s with $1\leq n\leq3$ to satisfy the boundary conditions: $\theta_1(0)=\theta_2(0)=\theta_3(0)=\pi/2$, $\theta_1(\tau)=\pi$, $\theta_2(\tau)=\theta_3(\tau)=\pi/2$, $\theta_2(2\tau)=0$, $\theta_3(2\tau)=\pi/2$, and $\theta_1(3\tau)=\theta_2(3\tau)=\theta_3(3\tau)=\pi/2$. Then one can check that  $\mu_4(0)=a_1\rightarrow\mu_4(\tau)=a_2\rightarrow\mu_4(2\tau)=a_3\rightarrow\mu_4(3\tau)=a_1$. In general, $\theta_1(t)$ and $\theta_2(t)$ during Stages (i), (ii), and (iii) of the $k$th loop, $k\geq1$, can be set as
\begin{equation}\label{CondTPC}
\begin{aligned}
\theta_1(t)&=\Phi(t)+\frac{\pi}{2},\quad \theta_2(t)=2\Phi(t)+\frac{\pi}{2},\\
\theta_1(t)&=\Phi(t)+\frac{\pi}{2},\quad \theta_2(t)=\Phi(t),\\
\theta_1(t)&=\theta_2(t)=\Phi(t),
\end{aligned}
\end{equation}
respectively, with $\Phi(t)\equiv\pi[t-3(k-1)\tau]/(2\tau)$, and $\theta_3(t)$ for the whole loop can be set as
\begin{equation}\label{CondChi}
\theta_3(t)=\frac{\pi}{2}\left[1+\frac{\sin{(\frac{\pi t}{\tau})}}{1+\sin^2(\frac{\pi t}{\tau})}\right].
\end{equation}

\begin{figure}[htbp]
\centering
\includegraphics[width=0.85\linewidth]{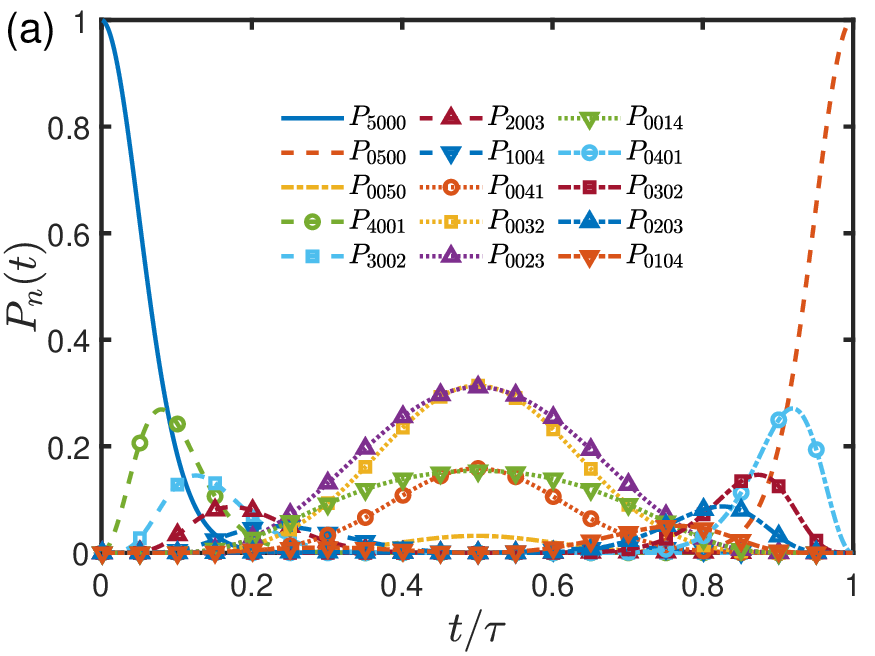}
\includegraphics[width=0.85\linewidth]{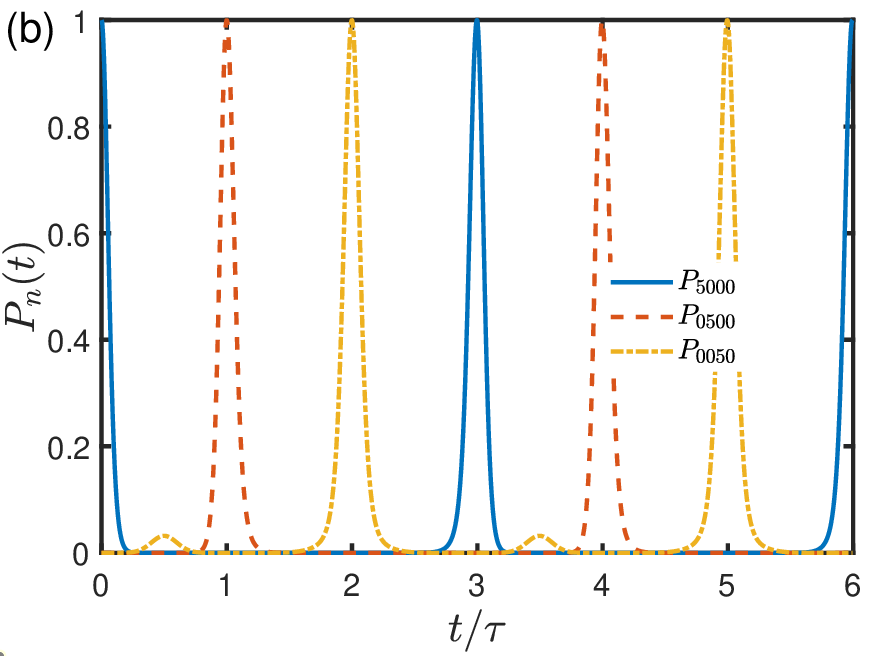}
\caption{Population dynamics $P_n(t)$ for (a) partial states with five excitations in the state transfer $|5\rangle_1|0\rangle_2|0\rangle_3|0\rangle_4\rightarrow|0\rangle_1|5\rangle_2|0\rangle_3|0\rangle_4$ and (b) the target states in chiral population transfer $|5\rangle_1|0\rangle_2|0\rangle_3|0\rangle_4
\rightarrow|0\rangle_1|5\rangle_2|0\rangle_3|0\rangle_4\rightarrow|0\rangle_1|0\rangle_2|5\rangle_3|0\rangle_4
\rightarrow|5\rangle_1|0\rangle_2|0\rangle_3|0\rangle_4$. Coupling strengths and detunings are set as Eqs.~(\ref{CondJ}) and (\ref{CondDetu}) under $\varphi_1+\alpha_3=\pi/2$, $\varphi_2-\alpha_1+\alpha_3=\pi/2$, $\varphi_3-\alpha_2+\alpha_3=\pi/2$, $\theta_1(t)$ and $\theta_2(t)$ in Eq.~(\ref{CondTPC}), and $\theta_3(t)$ in Eq.~(\ref{CondChi}).}\label{chiralfour}
\end{figure}

We assume that the whole system is initially in the Fock state $|5\rangle_1|0\rangle_2|0\rangle_3|0\rangle_4$, i.e., the mode $a_1$ is prepared in $|5\rangle$ and the other modes are in the vacuum state $|0\rangle$. Our protocol is evaluated by the population $P_n=|\langle n|\psi(t)\rangle|^2$, where $|n\rangle$'s are a group of relevant number states with conserved excitations and $|\psi(t)\rangle$ is obtained by numerical simulation. Figure~\ref{chiralfour}(a) presents the dynamics about the population on $15$ states during Stage (i). It is found that the initial population on $|5\rangle_1|0\rangle_2|0\rangle_3|0\rangle_4$ can be completely transferred to the state $|0\rangle_1|5\rangle_2|0\rangle_3|0\rangle_4$ when $t=\tau$, despite the other intermediate states can be temporally occupied. For instance, when $t=0.5\tau$, the populations are $P_{0023}(0.5\tau)=0.31$, $P_{0032}(0.5\tau)=0.31$, $P_{0014}(0.5\tau)=0.15$, $P_{0041}(0.5\tau)=0.16$, and $P_{0050}=0.032$. We plot two completed loops of chiral Fock-state transfer, i.e., $|5\rangle_1|0\rangle_2|0\rangle_3|0\rangle_4\rightarrow|0\rangle_1|5\rangle_2|0\rangle_3|0\rangle_4
\rightarrow|0\rangle_1|0\rangle_2|5\rangle_3|0\rangle_4\rightarrow|5\rangle_1|0\rangle_2|0\rangle_3|0\rangle_4$, in Fig.~\ref{chiralfour}(b). It is found that the perfect transfer can be realized as $P_{5000}(0)=1$, $P_{0500}(\tau)=1$, $P_{0050}(2\tau)=1$, and $P_{5000}(3\tau)=1$. The second loop behaves the same as the first one.

\begin{figure}[htbp]
\centering
\includegraphics[width=0.95\linewidth]{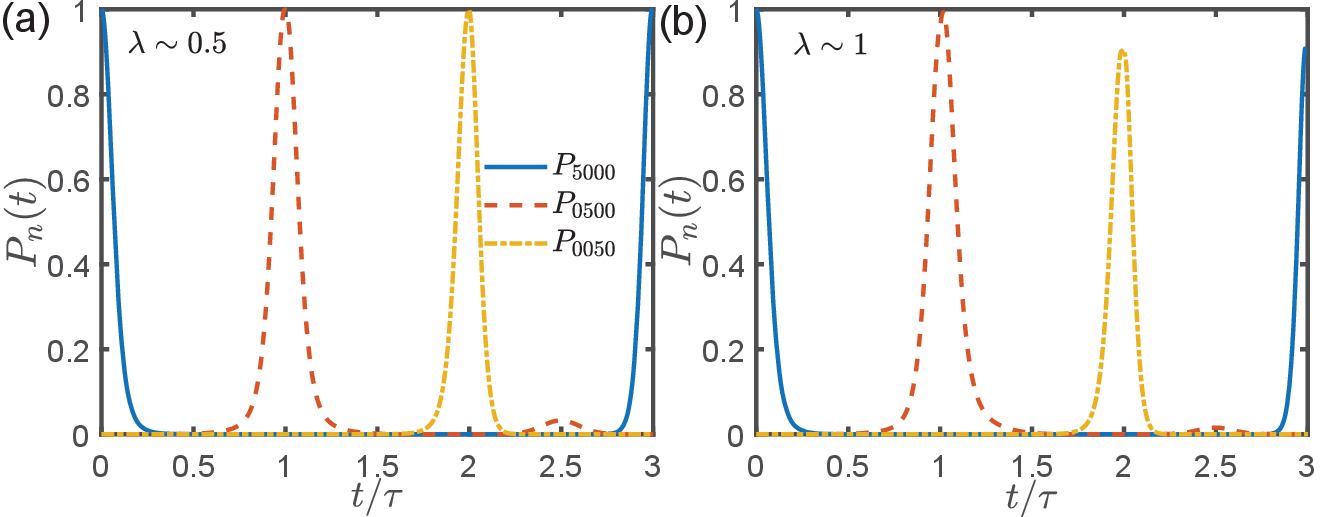}
\includegraphics[width=0.95\linewidth]{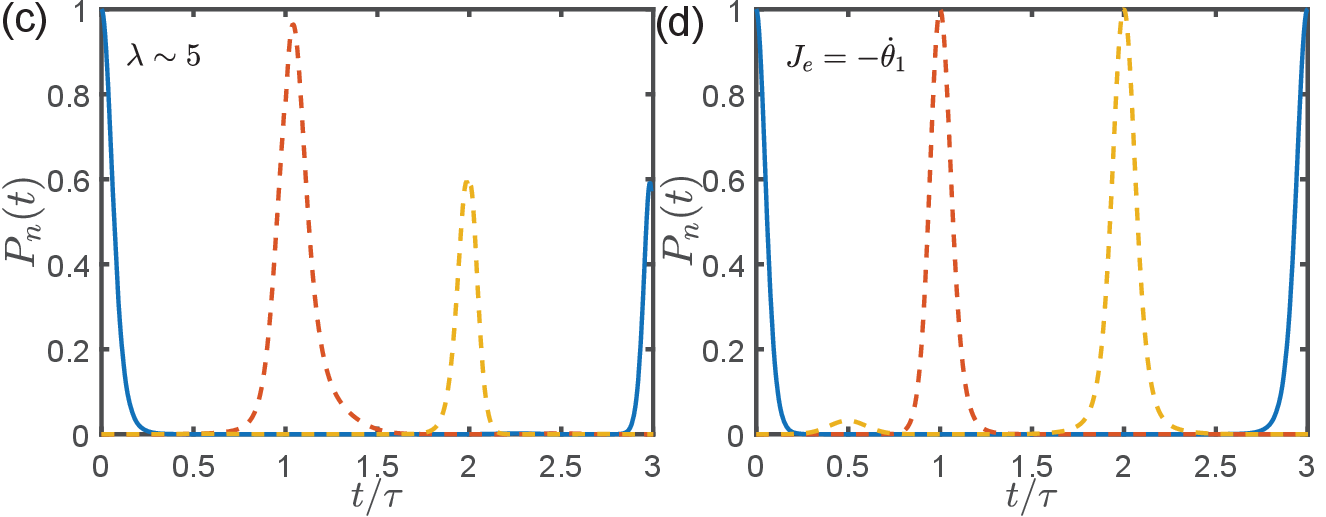}
\caption{Population dynamics $P_n(t)$ for the target states in chiral population transfer $|5\rangle_1|0\rangle_2|0\rangle_3|0\rangle_4\rightarrow|0\rangle_1|5\rangle_2|0\rangle_3|0\rangle_4
\rightarrow|0\rangle_1|0\rangle_2|5\rangle_3|0\rangle_4\rightarrow|5\rangle_1|0\rangle_2|0\rangle_3|0\rangle_4$ in the presence of unwanted coupling $J_e\sim\lambda {\rm max}\{J_n\}\times10^{-2}$, $1\leq n\leq3$ with (a) $\lambda\sim0.5$, (b) $\lambda\sim1$, and (c) $\lambda\sim5$. In comparison, the result with a known $J_e$ is shown in (d). The coupling strengths $J_n$ with $1\leq n\leq3$ and $J_e$ in (d) are constrained by Eqs.~(\ref{CondJErr}) and (\ref{CondJa}), while the other parameters in (a), (b), (c), and (d) are the same as Fig.~\ref{chiralfour}(b).}\label{chiralfourErr}
\end{figure}

In the presence of unwanted and unknown crosstalk among the network nodes, our protocol demonstrates a certain degree of robustness. For example, consider an extra coupling $J_e$ between the modes $a_1$ and $a_2$, the nonideal Hamiltonian can be written as the sum of the ideal Hamiltonian~(\ref{HamNdetu}) with $N=4$ and the extra interaction Hamiltonian:
\begin{equation}\label{HamFdetuErr}
H(t)\rightarrow H(t)+H_1(t), \quad H_1(t)=J_e\left(a_1^\dagger a_2+a_2^\dagger a_1\right).
\end{equation}
In the current cavity magnonic systems with multiple modes~\cite{Zhang2015Magnon}, it is experimentally feasible to couple two magnon modes to a common cavity mode while keeping the direct or unwanted magnon-magnon coupling negligible. Therefore $J_e$ can be about two orders of magnitude lower than the intended coupling strength, i.e., $J_e=\lambda {\rm max}\{J_n\}\times10^{-2}$, $1\leq n\leq3$, with $\lambda$ a dimensionless factor measuring the unwanted connection strength. In Figs.~\ref{chiralfourErr}(a), (b) and (c), we plot the population dynamics $P_n$ during the first loop about the chiral Fock-state transfer among the modes $a_1$, $a_2$, and $a_3$ with various $\lambda$. The network initial state remains as $|5\rangle_1|0\rangle_2|0\rangle_3|0\rangle_4$. We still have an almost perfect chiral transfer under a weak unwanted connection. In particular, in Fig.~\ref{chiralfourErr}(a) with $\lambda=0.5$, we obtain $P_{0500}(\tau)=0.999$, $P_{0050}(2\tau)=0.997$, and $P_{5000}(3\tau)=0.995$. In Fig.~\ref{chiralfourErr}(b) with $\lambda=1$, $P_{0500}(\tau)=0.957$, $P_{0050}(2\tau)=0.902$, and $P_{5000}(3\tau)=0.880$. When $\lambda$ is enhanced to $\lambda=5$ as shown in Fig.~\ref{chiralfourErr}(c), the fidelities are found to be as low as $P_{0500}(\tau)=0.772$, $P_{0050}(2\tau)=0.576$, and $P_{5000}(3\tau)=0.571$.

Nevertheless, a perfect cyclic transfer can be still attainable by the evolution along the passage $\mu_4(t)$ if $J_e$ is known. In this case, the commutation condition~(\ref{diagEq}) with the nonideal Hamiltonian~(\ref{HamFdetuErr}) gives rise to the constraint equations deviated from Eqs.~(\ref{CondJ}) and (\ref{CondDetu}). Specifically, we have $\Delta_n(t)=0$, $1\leq n\leq4$, for the detunings, $-\varphi_1=\varphi_2=\varphi_3=\pi/2$ for the phases, and Eq.~(\ref{CondJ}) for coupling strengths are modified to be
\begin{equation}
\begin{aligned}\label{CondJErr}
J_1&=J\sin\theta_1, \\
J_2&=J\cos\theta_1, \\
J_3&=-\dot{\theta}_3\cos\theta_2+\dot{\theta}_2\tan\theta_3\sin\theta_2, \\
J&=-\dot{\theta}_3\sin\theta_2-\dot{\theta}_2\tan\theta_3\cos\theta_2,
\end{aligned}
\end{equation}
where $J$ serves as a scaling coupling strength. The constraint equation for the extra coupling strength $J_e$ is found to be
\begin{equation}\label{CondJa}
J_e=-\dot{\theta}_1,
\end{equation}
which can be a constant as long as $\theta_1(t)$ is a linear function with time. Then, again with the boundary conditions given by Eqs.~(\ref{CondTPC}) and (\ref{CondChi}), any initial state in the mode $a_1$ can be perfectly transferred to the mode $a_2$ at $t=\tau$, to the mode $a_3$ at $t=2\tau$, and returns to the mode $a_1$ at $t=3\tau$, along the passage $\mu_4(t)$. The relevant population dynamics $P_n(t)$ with a known $J_e$ presents in Fig.~\ref{chiralfourErr}(d) in comparison to Figs.~\ref{chiralfourErr}(a), (b), and (c).

Moreover, the ideal constraint equations~(\ref{CondJ}) and (\ref{CondDetu}) can be slightly modified to run a four-mode chiral transfer by a four-stage passage along the ancillary mode $\mu_4(t)$. In particular, the control target becomes $a_1\rightarrow a_2$ in Stage (i) for $t\in[0, \tau]$, $a_2\rightarrow a_3$ in Stage (ii) for $t\in[\tau, 2\tau]$, $a_3\rightarrow a_4$ in Stage (iii) for $t\in[2\tau, 3\tau]$, and $a_4\rightarrow a_1$ in Stage (iv) for $t\in[3\tau, 4\tau]$. For this target, the boundary conditions for $\theta_k(t)$ with $1\leq k\leq3$ of Stages (i) and (ii) remain invariant as those for the three-mode transfer protocol. And in Stages (iii) and (iv), they are reset as $\theta_3(3\tau)=\pi$ and $\theta_1(4\tau)=\theta_2(4\tau)=\theta_3(4\tau)=\pi/2$, respectively. In practice, $\theta_1(t)$ and $\theta_2(t)$ in Stages (i-ii) and $\theta_3(t)$ in Stages (i-ii) and (iv) can be the same as in Eqs.~(\ref{CondTPC}) and (\ref{CondChi}), respectively. And one can choose $\theta_1(t)=\theta_2(t)=\Phi(t)+\pi/2$ for Stages (iii-iv) and $\theta_3(t)=(\pi/2)[1-\sin(\pi t/(2\tau))]$ for Stage (iii).

\section{Discussion and Conclusion}\label{conclusion}

In summary, we propose a theoretical framework to construct nonadiabatic passages for the general $N$-mode bosonic network that is governed by the time-dependent Hamiltonian. With a completed set of time-independent ancillary modes, we find a necessary and sufficient condition to exactly solve the time-dependent Schr\"odinger equation or equivalently to determine the dynamics of ancillary operators in the Heisenberg picture. In particular, the diagonalization of the coefficient matrix for Hamiltonian and gauge potential in the representation of the time-independent ancillary modes can be implemented by its commutation with the projection operators of the matrix bases. The commutation condition yields the parametric constraints for the network Hamiltonian, which activate the ancillary modes as versatile nonadiabatic passages. Along the activated passages, arbitrary states can be exchanged between any pair of modes in the bosonic network, which is robust against the unknown crosstalk among relevant nodes, the environmental decoherence, and the imperfect parametric setting. As illustrative examples, the feasibility of our theory is confirmed by the perfect state exchange in a two-mode network, the chiral NOON-state transfer in a three-mode system, and the chiral Fock-state transfer among three of four bosonic modes. Our work thus provides a universal approach for controlling quantum bosonic networks with arbitrary connection and size.

Our protocol for the bosonic network connected with beam-splitter coupling can be realized in a variety of physical platforms. They include and are not limited to the circuit quantum electrodynamics system~\cite{Vermersch2017Quantum,Blais2021Circuit}, which is about two cavity modes; the circuit quantum acoustodynamics system~\cite{von2024Engineering}, in which a mechanical mode is simultaneously coupled to two other mechanical modes; the cavity magnomechanical systems~\cite{Zhang2016Cavity,Li2021Quantum,Shen2022Coherent,Shen2025Cavity}, in which a cavity-magnon polariton is coupled to a phonon mode; and the cavity magnonic systems~\cite{Zhang2014Strongly,Zhang2016Optomagnonic,Lachance2019Hybrid,Xu2021Coherent,Babak2022Cavity,Xu2023Quantum,Zheng2023Tutorial}, where a cavity mode is coupled to a magnon mode. In addition, the cavity magnonic systems can be directly extended to the multi-mode case~\cite{Zhang2015Magnon}, where multiple magnon modes are coupled to a common cavity mode. In comparison to the existing methods, our protocol is free of at least three typical limitations: (1) the running time of protocol, (2) the capability to transfer arbitrary states, and (3) the scalability to a large size.

\section*{Acknowledgments}

We acknowledge grant support from the National Natural Science Foundation of China (Grant No. U25A20199) and the ``Pioneer'' and ``Leading Goose'' R\&D of Zhejiang Province (Grant No. 2025C01028).

\appendix

\section{Proof about Eq.~(\ref{unitaryV})}\label{proof}

This appendix is used to prove that the unitary transformation $V_{N-1}(t)\equiv \prod_{j=1}^{N-1}V_{\alpha_j}V_{\theta_j}$ in Eq.~(\ref{unitaryV}) can transform all the time-dependent ancillary modes $\mu_k(t)$'s in Eq.~(\ref{TimeAnci}) into a time-independent formation, i.e., $V_{N-1}^\dagger(t)\mu_k(t)V_{N-1}(t)=\mu_k(0)$ with $1\leq k\leq N$. The proof is organized by the mathematical induction method. For each $k$, we only need to prove $V_k^\dagger(t)\mu_k(t)V_k(t)=\mu_k(0)$, since $\mu_k(0)$ is invariant under the unitary transformation $V_{\alpha_{j>k}}V_{\theta_{j>k}}$.

\emph{Step one}: For $k=1$,
\begin{equation}\label{V1}
V_1(t)=V_{\alpha_1}V_{\theta_1},
\end{equation}
where
\begin{equation}
\begin{aligned}
V_{\alpha_1}(t)&=e^{-i\frac{\delta\alpha_1}{2}\left[b_0^\dagger(0)b_0(0)-a_2^\dagger a_2\right]} \\
&=e^{-i\frac{\delta\alpha_1}{2}\left(a_1^\dagger a_1-a_2^\dagger a_2\right)}, \\
V_{\theta_1}(t)&=e^{-\delta\theta_1\left[e^{i\alpha_1(0)}a_2^\dagger b_0(0)-e^{-i\alpha_1(0)}b_0^\dagger(0)a_2\right]} \\
&=e^{-\delta\theta_1\left[e^{i\alpha_1(0)}a_2^\dagger a_1-e^{-i\alpha_1(0)}a_1^\dagger a_2\right]}
\end{aligned}
\end{equation}
and $\delta\alpha_1\equiv \alpha_1(t)-\alpha_1(0)$ and $\delta\theta_1\equiv \theta_1(t)-\theta_1(0)$, according to Eq.~(\ref{uniVat}). Using Eq.~(\ref{V1}) and the Baker-Campbell-Hausdorff formula, it can be verified that $\mu_1(t)$ in Eq.~(\ref{unitaryV}) is transformed as
\begin{equation}\label{mu1trans}
\begin{aligned}
&V_1^\dagger(t)\mu_1(t)V_1(t)\\
=&V_{\theta_1}^\dagger V_{\alpha_1}^\dagger\left[\cos\theta_1(t)e^{i\frac{\alpha_1(t)}{2}}a_1-\sin\theta_1(t)e^{-i\frac{\alpha_1(t)}{2}}a_2\right]
V_{\alpha_1}V_{\theta_1}\\
=&V_{\theta_1}^\dagger\left[\cos\theta_1(t)e^{i\frac{\alpha_1(0)}{2}}a_1-\sin\theta_1(t)e^{-i\frac{\alpha_1(0)}{2}}a_2\right]
V_{\theta_1}\\
=&\cos\theta_1(0)e^{i\frac{\alpha_1(0)}{2}}a_1-\sin\theta_1(0)e^{-i\frac{\alpha_1(0)}{2}}a_2=\mu_1(0).
\end{aligned}
\end{equation}
Similarly, for the bight-mode operator $b_1(t)$, one can check that $V_1^\dagger(t)b_1(t)V_1(t)\rightarrow b_1(0)$ due to the bright vector defined in Eq.~(\ref{unitarVecBri}).

\emph{Step two}: We assume that the time-dependent ancillary modes $\mu_k(t)$, $2\leq k\leq N-2$, in Eq.~(\ref{TimeAnci}) and the bright-mode operators $b_k(t)$'s associated with Eq.~(\ref{unitarVecBri}) can be transformed to the time-independent formation, i.e.,
\begin{equation}\label{VN1Sat}
\begin{aligned}
V_k^\dagger(t)\mu_k(t)V_k(t)&=\mu_k(0),\\
V_k^\dagger(t)b_k(t)V_k(t)&=b_k(0),
\end{aligned}
\end{equation}
where the unitary transformation $V_k(t)$ is defined as
\begin{equation}\label{VN1}
V_k(t)=\prod_{j=1}^kV_{\alpha_j}V_{\theta_j}.
\end{equation}

\emph{Step three}: Using Eqs.~(\ref{VN1Sat}) and (\ref{VN1}), we can verify that $\mu_{k+1}(t)$ can be converted into $\mu_{k+1}(0)$ in Eq.~(\ref{TimeAnci}) by the transformation $V_{k+1}(t)$ in Eq.~(\ref{unitaryV}). In particular, we have
\begin{equation}\label{mu1transN}
\begin{aligned}
&V_{k+1}^\dagger(t)\mu_{k+1}(t)V_{k+1}(t)\\
&=V_{\theta_{k+1}}^\dagger V_{\alpha_{k+1}}^\dagger V_k^\dagger(t)\mu_{k+1}(t)V_k(t)V_{\alpha_{k+1}}V_{\theta_{k+1}}\\
&=V_{\theta_{k+1}}^\dagger V_{\alpha_{k+1}}^\dagger\Big[\cos\theta_{k+1}(t)e^{i\frac{\alpha_{k+1}(t)}{2}}b_k(0)\\
&-\sin\theta_{k+1}(t)e^{-i\frac{\alpha_{k+1}(t)}{2}}a_{k+2}\Big]V_{\alpha_{k+1}}V_{\theta_{k+1}}\\
&=\cos\theta_{k+1}(0)e^{i\frac{\alpha_{k+1}(0)}{2}}b_k(0)\\
&-\sin\theta_{k+1}(0)e^{-i\frac{\alpha_{k+1}(0)}{2}}a_{k+2}=\mu_{k+1}(0),
\end{aligned}
\end{equation}
Similarly, we have $V_{k+1}^\dagger(t)b_{k+1}(t)V_{k+1}(t)=b_{k+1}(0)$.

Eventually, replacing $\mu_{k+1}(t)$ in Eq.~(\ref{mu1transN}) for $k=N-2$ with $\mu_N(t)$, one can obtain $V_{N-1}^\dagger(t)\mu_N(t)V_{N-1}(t)=\mu_N(0)$.

\bibliographystyle{apsrevlong}
\bibliography{ref}

\begin{thebibliography}{99}%
\makeatletter
\providecommand \@ifxundefined [1]{%
 \@ifx{#1\undefined}
}%
\providecommand \@ifnum [1]{%
 \ifnum #1\expandafter \@firstoftwo
 \else \expandafter \@secondoftwo
 \fi
}%
\providecommand \@ifx [1]{%
 \ifx #1\expandafter \@firstoftwo
 \else \expandafter \@secondoftwo
 \fi
}%
\providecommand \natexlab [1]{#1}%
\providecommand \enquote  [1]{``#1''}%
\providecommand \bibnamefont  [1]{#1}%
\providecommand \bibfnamefont [1]{#1}%
\providecommand \citenamefont [1]{#1}%
\providecommand \href@noop [0]{\@secondoftwo}%
\providecommand \href [0]{\begingroup \@sanitize@url \@href}%
\providecommand \@href[1]{\@@startlink{#1}\@@href}%
\providecommand \@@href[1]{\endgroup#1\@@endlink}%
\providecommand \@sanitize@url [0]{\catcode `\\12\catcode `\$12\catcode
  `\&12\catcode `\#12\catcode `\^12\catcode `\_12\catcode `\%12\relax}%
\providecommand \@@startlink[1]{}%
\providecommand \@@endlink[0]{}%
\providecommand \url  [0]{\begingroup\@sanitize@url \@url }%
\providecommand \@url [1]{\endgroup\@href {#1}{\urlprefix }}%
\providecommand \urlprefix  [0]{URL }%
\providecommand \Eprint [0]{\href }%
\providecommand \doibase [0]{http://dx.doi.org/}%
\providecommand \selectlanguage [0]{\@gobble}%
\providecommand \bibinfo  [0]{\@secondoftwo}%
\providecommand \bibfield  [0]{\@secondoftwo}%
\providecommand \translation [1]{[#1]}%
\providecommand \BibitemOpen [0]{}%
\providecommand \bibitemStop [0]{}%
\providecommand \bibitemNoStop [0]{.\EOS\space}%
\providecommand \EOS [0]{\spacefactor3000\relax}%
\providecommand \BibitemShut  [1]{\csname bibitem#1\endcsname}%
\let\auto@bib@innerbib\@empty
\bibitem [{\citenamefont {Kimble}(2008)}]{Kimble2008Quantum}%
  \BibitemOpen
  \bibfield  {author} {\bibinfo {author} {\bibfnamefont {H.~J.}\ \bibnamefont
  {Kimble}},\ }\bibfield  {title} {\emph {\bibinfo {title} {The quantum
  internet},\ }}\href {\doibase 10.1038/nature07127} {\bibfield  {journal}
  {\bibinfo  {journal} {Nature}\ }\textbf {\bibinfo {volume} {453}},\ \bibinfo
  {pages} {1023} (\bibinfo {year} {2008})}\BibitemShut {NoStop}%
\bibitem [{\citenamefont {Wehner}\ \emph {et~al.}(2018)\citenamefont {Wehner},
  \citenamefont {Elkouss},\ and\ \citenamefont
  {Hanson}}]{Stephanie2018Quantum}%
  \BibitemOpen
  \bibfield  {author} {\bibinfo {author} {\bibfnamefont {S.}~\bibnamefont
  {Wehner}}, \bibinfo {author} {\bibfnamefont {D.}~\bibnamefont {Elkouss}}, \
  and\ \bibinfo {author} {\bibfnamefont {R.}~\bibnamefont {Hanson}},\
  }\bibfield  {title} {\emph {\bibinfo {title} {Quantum internet: A vision for
  the road ahead},\ }}\href {\doibase 10.1126/science.aam9288} {\bibfield
  {journal} {\bibinfo  {journal} {Science}\ }\textbf {\bibinfo {volume}
  {362}},\ \bibinfo {pages} {eaam9288} (\bibinfo {year} {2018})}\BibitemShut
  {NoStop}%
\bibitem [{\citenamefont {Xu}\ \emph {et~al.}(2020)\citenamefont {Xu},
  \citenamefont {Ma}, \citenamefont {Zhang}, \citenamefont {Lo},\ and\
  \citenamefont {Pan}}]{Xu2020Secure}%
  \BibitemOpen
  \bibfield  {author} {\bibinfo {author} {\bibfnamefont {F.}~\bibnamefont
  {Xu}}, \bibinfo {author} {\bibfnamefont {X.}~\bibnamefont {Ma}}, \bibinfo
  {author} {\bibfnamefont {Q.}~\bibnamefont {Zhang}}, \bibinfo {author}
  {\bibfnamefont {H.-K.}\ \bibnamefont {Lo}}, \ and\ \bibinfo {author}
  {\bibfnamefont {J.-W.}\ \bibnamefont {Pan}},\ }\bibfield  {title} {\emph
  {\bibinfo {title} {Secure quantum key distribution with realistic devices},\
  }}\href {\doibase 10.1103/RevModPhys.92.025002} {\bibfield  {journal}
  {\bibinfo  {journal} {Rev. Mod. Phys.}\ }\textbf {\bibinfo {volume} {92}},\
  \bibinfo {pages} {025002} (\bibinfo {year} {2020})}\BibitemShut {NoStop}%
\bibitem [{\citenamefont {Duan}\ \emph {et~al.}(2001)\citenamefont {Duan},
  \citenamefont {Lukin}, \citenamefont {Cirac},\ and\ \citenamefont
  {Zoller}}]{Duan2001Longdistance}%
  \BibitemOpen
  \bibfield  {author} {\bibinfo {author} {\bibfnamefont {L.-M.}\ \bibnamefont
  {Duan}}, \bibinfo {author} {\bibfnamefont {M.~D.}\ \bibnamefont {Lukin}},
  \bibinfo {author} {\bibfnamefont {J.~I.}\ \bibnamefont {Cirac}}, \ and\
  \bibinfo {author} {\bibfnamefont {P.}~\bibnamefont {Zoller}},\ }\bibfield
  {title} {\emph {\bibinfo {title} {Long-distance quantum communication with
  atomic ensembles and linear optics},\ }}\href {\doibase 10.1038/35106500}
  {\bibfield  {journal} {\bibinfo  {journal} {Nature}\ }\textbf {\bibinfo
  {volume} {414}},\ \bibinfo {pages} {413} (\bibinfo {year}
  {2001})}\BibitemShut {NoStop}%
\bibitem [{\citenamefont {Stannigel}\ \emph {et~al.}(2010)\citenamefont
  {Stannigel}, \citenamefont {Rabl}, \citenamefont {S\o{}rensen}, \citenamefont
  {Zoller},\ and\ \citenamefont {Lukin}}]{Stannigel2010Optomechanical}%
  \BibitemOpen
  \bibfield  {author} {\bibinfo {author} {\bibfnamefont {K.}~\bibnamefont
  {Stannigel}}, \bibinfo {author} {\bibfnamefont {P.}~\bibnamefont {Rabl}},
  \bibinfo {author} {\bibfnamefont {A.~S.}\ \bibnamefont {S\o{}rensen}},
  \bibinfo {author} {\bibfnamefont {P.}~\bibnamefont {Zoller}}, \ and\ \bibinfo
  {author} {\bibfnamefont {M.~D.}\ \bibnamefont {Lukin}},\ }\bibfield  {title}
  {\emph {\bibinfo {title} {Optomechanical transducers for long-distance
  quantum communication},\ }}\href {\doibase 10.1103/PhysRevLett.105.220501}
  {\bibfield  {journal} {\bibinfo  {journal} {Phys. Rev. Lett.}\ }\textbf
  {\bibinfo {volume} {105}},\ \bibinfo {pages} {220501} (\bibinfo {year}
  {2010})}\BibitemShut {NoStop}%
\bibitem [{\citenamefont {Muralidharan}\ \emph {et~al.}(2016)\citenamefont
  {Muralidharan}, \citenamefont {Li}, \citenamefont {Kim}, \citenamefont
  {L\"utkenhaus}, \citenamefont {Lukin},\ and\ \citenamefont
  {Jiang}}]{Muralidharan2016Optimal}%
  \BibitemOpen
  \bibfield  {author} {\bibinfo {author} {\bibfnamefont {S.}~\bibnamefont
  {Muralidharan}}, \bibinfo {author} {\bibfnamefont {L.}~\bibnamefont {Li}},
  \bibinfo {author} {\bibfnamefont {J.}~\bibnamefont {Kim}}, \bibinfo {author}
  {\bibfnamefont {N.}~\bibnamefont {L\"utkenhaus}}, \bibinfo {author}
  {\bibfnamefont {M.~D.}\ \bibnamefont {Lukin}}, \ and\ \bibinfo {author}
  {\bibfnamefont {L.}~\bibnamefont {Jiang}},\ }\bibfield  {title} {\emph
  {\bibinfo {title} {Optimal architectures for long distance quantum
  communication},\ }}\href {\doibase 10.1038/srep20463} {\bibfield  {journal}
  {\bibinfo  {journal} {Sci. Rep.}\ }\textbf {\bibinfo {volume} {6}},\ \bibinfo
  {pages} {20463} (\bibinfo {year} {2016})}\BibitemShut {NoStop}%
\bibitem [{\citenamefont {Walln\"ofer}\ \emph {et~al.}(2020)\citenamefont
  {Walln\"ofer}, \citenamefont {Melnikov}, \citenamefont {D\"ur},\ and\
  \citenamefont {Briegel}}]{Wallnofer2020Machine}%
  \BibitemOpen
  \bibfield  {author} {\bibinfo {author} {\bibfnamefont {J.}~\bibnamefont
  {Walln\"ofer}}, \bibinfo {author} {\bibfnamefont {A.~A.}\ \bibnamefont
  {Melnikov}}, \bibinfo {author} {\bibfnamefont {W.}~\bibnamefont {D\"ur}}, \
  and\ \bibinfo {author} {\bibfnamefont {H.~J.}\ \bibnamefont {Briegel}},\
  }\bibfield  {title} {\emph {\bibinfo {title} {Machine learning for
  long-distance quantum communication},\ }}\href {\doibase
  10.1103/PRXQuantum.1.010301} {\bibfield  {journal} {\bibinfo  {journal} {PRX
  Quantum}\ }\textbf {\bibinfo {volume} {1}},\ \bibinfo {pages} {010301}
  (\bibinfo {year} {2020})}\BibitemShut {NoStop}%
\bibitem [{\citenamefont {Lim}\ \emph {et~al.}(2005)\citenamefont {Lim},
  \citenamefont {Beige},\ and\ \citenamefont {Kwek}}]{Lim2005Repeat}%
  \BibitemOpen
  \bibfield  {author} {\bibinfo {author} {\bibfnamefont {Y.~L.}\ \bibnamefont
  {Lim}}, \bibinfo {author} {\bibfnamefont {A.}~\bibnamefont {Beige}}, \ and\
  \bibinfo {author} {\bibfnamefont {L.~C.}\ \bibnamefont {Kwek}},\ }\bibfield
  {title} {\emph {\bibinfo {title} {Repeat-until-success linear optics
  distributed quantum computing},\ }}\href {\doibase
  10.1103/PhysRevLett.95.030505} {\bibfield  {journal} {\bibinfo  {journal}
  {Phys. Rev. Lett.}\ }\textbf {\bibinfo {volume} {95}},\ \bibinfo {pages}
  {030505} (\bibinfo {year} {2005})}\BibitemShut {NoStop}%
\bibitem [{\citenamefont {Cohen}\ and\ \citenamefont
  {M\o{}lmer}(2018)}]{Cohen2018Deterministic}%
  \BibitemOpen
  \bibfield  {author} {\bibinfo {author} {\bibfnamefont {I.}~\bibnamefont
  {Cohen}}\ and\ \bibinfo {author} {\bibfnamefont {K.}~\bibnamefont
  {M\o{}lmer}},\ }\bibfield  {title} {\emph {\bibinfo {title} {Deterministic
  quantum network for distributed entanglement and quantum computation},\
  }}\href {\doibase 10.1103/PhysRevA.98.030302} {\bibfield  {journal} {\bibinfo
   {journal} {Phys. Rev. A}\ }\textbf {\bibinfo {volume} {98}},\ \bibinfo
  {pages} {030302} (\bibinfo {year} {2018})}\BibitemShut {NoStop}%
\bibitem [{\citenamefont {Cuomo}\ \emph {et~al.}(2020)\citenamefont {Cuomo},
  \citenamefont {Caleffi},\ and\ \citenamefont
  {Cacciapuoti}}]{Cuomo2020Towards}%
  \BibitemOpen
  \bibfield  {author} {\bibinfo {author} {\bibfnamefont {D.}~\bibnamefont
  {Cuomo}}, \bibinfo {author} {\bibfnamefont {M.}~\bibnamefont {Caleffi}}, \
  and\ \bibinfo {author} {\bibfnamefont {A.~S.}\ \bibnamefont {Cacciapuoti}},\
  }\bibfield  {title} {\emph {\bibinfo {title} {Towards a distributed quantum
  computing ecosystem},\ }}\href {\doibase 10.1049/iet-qtc.2020.0002}
  {\bibfield  {journal} {\bibinfo  {journal} {IET Quantum Commun.}\ }\textbf
  {\bibinfo {volume} {1}},\ \bibinfo {pages} {3} (\bibinfo {year}
  {2020})}\BibitemShut {NoStop}%
\bibitem [{\citenamefont {Giovannetti}\ \emph {et~al.}(2006)\citenamefont
  {Giovannetti}, \citenamefont {Lloyd},\ and\ \citenamefont
  {Maccone}}]{Glovannetti2006Quantum}%
  \BibitemOpen
  \bibfield  {author} {\bibinfo {author} {\bibfnamefont {V.}~\bibnamefont
  {Giovannetti}}, \bibinfo {author} {\bibfnamefont {S.}~\bibnamefont {Lloyd}},
  \ and\ \bibinfo {author} {\bibfnamefont {L.}~\bibnamefont {Maccone}},\
  }\bibfield  {title} {\emph {\bibinfo {title} {Quantum metrology},\ }}\href
  {\doibase 10.1103/PhysRevLett.96.010401} {\bibfield  {journal} {\bibinfo
  {journal} {Phys. Rev. Lett.}\ }\textbf {\bibinfo {volume} {96}},\ \bibinfo
  {pages} {010401} (\bibinfo {year} {2006})}\BibitemShut {NoStop}%
\bibitem [{\citenamefont {Joo}\ \emph {et~al.}(2011)\citenamefont {Joo},
  \citenamefont {Munro},\ and\ \citenamefont {Spiller}}]{Joo2011Quantum}%
  \BibitemOpen
  \bibfield  {author} {\bibinfo {author} {\bibfnamefont {J.}~\bibnamefont
  {Joo}}, \bibinfo {author} {\bibfnamefont {W.~J.}\ \bibnamefont {Munro}}, \
  and\ \bibinfo {author} {\bibfnamefont {T.~P.}\ \bibnamefont {Spiller}},\
  }\bibfield  {title} {\emph {\bibinfo {title} {Quantum metrology with
  entangled coherent states},\ }}\href {\doibase
  10.1103/PhysRevLett.107.083601} {\bibfield  {journal} {\bibinfo  {journal}
  {Phys. Rev. Lett.}\ }\textbf {\bibinfo {volume} {107}},\ \bibinfo {pages}
  {083601} (\bibinfo {year} {2011})}\BibitemShut {NoStop}%
\bibitem [{\citenamefont {Giovannetti}\ \emph {et~al.}(2011)\citenamefont
  {Giovannetti}, \citenamefont {Lloyd},\ and\ \citenamefont
  {Maccone}}]{Giovannetti2011Advances}%
  \BibitemOpen
  \bibfield  {author} {\bibinfo {author} {\bibfnamefont {V.}~\bibnamefont
  {Giovannetti}}, \bibinfo {author} {\bibfnamefont {S.}~\bibnamefont {Lloyd}},
  \ and\ \bibinfo {author} {\bibfnamefont {L.}~\bibnamefont {Maccone}},\
  }\bibfield  {title} {\emph {\bibinfo {title} {Advances in quantum
  metrology},\ }}\href {\doibase 10.1038/nphoton.2011.35} {\bibfield  {journal}
  {\bibinfo  {journal} {Nat. Photon.}\ }\textbf {\bibinfo {volume} {5}},\
  \bibinfo {pages} {222} (\bibinfo {year} {2011})}\BibitemShut {NoStop}%
\bibitem [{\citenamefont {Chin}\ \emph {et~al.}(2012)\citenamefont {Chin},
  \citenamefont {Huelga},\ and\ \citenamefont {Plenio}}]{Chin2012Quantum}%
  \BibitemOpen
  \bibfield  {author} {\bibinfo {author} {\bibfnamefont {A.~W.}\ \bibnamefont
  {Chin}}, \bibinfo {author} {\bibfnamefont {S.~F.}\ \bibnamefont {Huelga}}, \
  and\ \bibinfo {author} {\bibfnamefont {M.~B.}\ \bibnamefont {Plenio}},\
  }\bibfield  {title} {\emph {\bibinfo {title} {Quantum metrology in
  non-markovian environments},\ }}\href {\doibase
  10.1103/PhysRevLett.109.233601} {\bibfield  {journal} {\bibinfo  {journal}
  {Phys. Rev. Lett.}\ }\textbf {\bibinfo {volume} {109}},\ \bibinfo {pages}
  {233601} (\bibinfo {year} {2012})}\BibitemShut {NoStop}%
\bibitem [{\citenamefont {Zhou}\ \emph {et~al.}(2020)\citenamefont {Zhou},
  \citenamefont {Choi}, \citenamefont {Choi}, \citenamefont {Landig},
  \citenamefont {Douglas}, \citenamefont {Isoya}, \citenamefont {Jelezko},
  \citenamefont {Onoda}, \citenamefont {Sumiya}, \citenamefont {Cappellaro},
  \citenamefont {Knowles}, \citenamefont {Park},\ and\ \citenamefont
  {Lukin}}]{Zhou2020Quantum}%
  \BibitemOpen
  \bibfield  {author} {\bibinfo {author} {\bibfnamefont {H.}~\bibnamefont
  {Zhou}}, \bibinfo {author} {\bibfnamefont {J.}~\bibnamefont {Choi}}, \bibinfo
  {author} {\bibfnamefont {S.}~\bibnamefont {Choi}}, \bibinfo {author}
  {\bibfnamefont {R.}~\bibnamefont {Landig}}, \bibinfo {author} {\bibfnamefont
  {A.~M.}\ \bibnamefont {Douglas}}, \bibinfo {author} {\bibfnamefont
  {J.}~\bibnamefont {Isoya}}, \bibinfo {author} {\bibfnamefont
  {F.}~\bibnamefont {Jelezko}}, \bibinfo {author} {\bibfnamefont
  {S.}~\bibnamefont {Onoda}}, \bibinfo {author} {\bibfnamefont
  {H.}~\bibnamefont {Sumiya}}, \bibinfo {author} {\bibfnamefont
  {P.}~\bibnamefont {Cappellaro}}, \bibinfo {author} {\bibfnamefont {H.~S.}\
  \bibnamefont {Knowles}}, \bibinfo {author} {\bibfnamefont {H.}~\bibnamefont
  {Park}}, \ and\ \bibinfo {author} {\bibfnamefont {M.~D.}\ \bibnamefont
  {Lukin}},\ }\bibfield  {title} {\emph {\bibinfo {title} {Quantum metrology
  with strongly interacting spin systems},\ }}\href {\doibase
  10.1103/PhysRevX.10.031003} {\bibfield  {journal} {\bibinfo  {journal} {Phys.
  Rev. X}\ }\textbf {\bibinfo {volume} {10}},\ \bibinfo {pages} {031003}
  (\bibinfo {year} {2020})}\BibitemShut {NoStop}%
\bibitem [{\citenamefont {Reiserer}\ and\ \citenamefont
  {Rempe}(2015)}]{Reiserer2015Cavity}%
  \BibitemOpen
  \bibfield  {author} {\bibinfo {author} {\bibfnamefont {A.}~\bibnamefont
  {Reiserer}}\ and\ \bibinfo {author} {\bibfnamefont {G.}~\bibnamefont
  {Rempe}},\ }\bibfield  {title} {\emph {\bibinfo {title} {Cavity-based quantum
  networks with single atoms and optical photons},\ }}\href {\doibase
  10.1103/RevModPhys.87.1379} {\bibfield  {journal} {\bibinfo  {journal} {Rev.
  Mod. Phys.}\ }\textbf {\bibinfo {volume} {87}},\ \bibinfo {pages} {1379}
  (\bibinfo {year} {2015})}\BibitemShut {NoStop}%
\bibitem [{\citenamefont {Liu}\ \emph {et~al.}(2022)\citenamefont {Liu},
  \citenamefont {Li},\ and\ \citenamefont {Ma}}]{Liu2022Hybrid}%
  \BibitemOpen
  \bibfield  {author} {\bibinfo {author} {\bibfnamefont {Y.}~\bibnamefont
  {Liu}}, \bibinfo {author} {\bibfnamefont {L.}~\bibnamefont {Li}}, \ and\
  \bibinfo {author} {\bibfnamefont {Y.}~\bibnamefont {Ma}},\ }\bibfield
  {title} {\emph {\bibinfo {title} {Hybrid rydberg quantum gate for quantum
  network},\ }}\href {\doibase 10.1103/PhysRevResearch.4.013008} {\bibfield
  {journal} {\bibinfo  {journal} {Phys. Rev. Res.}\ }\textbf {\bibinfo {volume}
  {4}},\ \bibinfo {pages} {013008} (\bibinfo {year} {2022})}\BibitemShut
  {NoStop}%
\bibitem [{\citenamefont {Covey}\ \emph {et~al.}(2023)\citenamefont {Covey},
  \citenamefont {Weinfurter},\ and\ \citenamefont
  {Bernien}}]{Covey2023Quantum}%
  \BibitemOpen
  \bibfield  {author} {\bibinfo {author} {\bibfnamefont {J.~P.}\ \bibnamefont
  {Covey}}, \bibinfo {author} {\bibfnamefont {H.}~\bibnamefont {Weinfurter}}, \
  and\ \bibinfo {author} {\bibfnamefont {H.}~\bibnamefont {Bernien}},\
  }\bibfield  {title} {\emph {\bibinfo {title} {Quantum networks with neutral
  atom processing nodes},\ }}\href {\doibase 10.1038/s41534-023-00759-9}
  {\bibfield  {journal} {\bibinfo  {journal} {npj Quantum Inf.}\ }\textbf
  {\bibinfo {volume} {9}},\ \bibinfo {pages} {90} (\bibinfo {year}
  {2023})}\BibitemShut {NoStop}%
\bibitem [{\citenamefont {Duan}\ and\ \citenamefont
  {Monroe}(2010)}]{Duan2010Quantum}%
  \BibitemOpen
  \bibfield  {author} {\bibinfo {author} {\bibfnamefont {L.-M.}\ \bibnamefont
  {Duan}}\ and\ \bibinfo {author} {\bibfnamefont {C.}~\bibnamefont {Monroe}},\
  }\bibfield  {title} {\emph {\bibinfo {title} {Colloquium: Quantum networks
  with trapped ions},\ }}\href {\doibase 10.1103/RevModPhys.82.1209} {\bibfield
   {journal} {\bibinfo  {journal} {Rev. Mod. Phys.}\ }\textbf {\bibinfo
  {volume} {82}},\ \bibinfo {pages} {1209} (\bibinfo {year}
  {2010})}\BibitemShut {NoStop}%
\bibitem [{\citenamefont {Chen}\ \emph {et~al.}(2023)\citenamefont {Chen},
  \citenamefont {Lu}, \citenamefont {Zhang}, \citenamefont {Zhang},
  \citenamefont {Huang}, \citenamefont {Qiao}, \citenamefont {Su},
  \citenamefont {Zhang}, \citenamefont {Zhang}, \citenamefont {Banchi},
  \citenamefont {Kim},\ and\ \citenamefont {Kim}}]{Chen2023Scalable}%
  \BibitemOpen
  \bibfield  {author} {\bibinfo {author} {\bibfnamefont {W.}~\bibnamefont
  {Chen}}, \bibinfo {author} {\bibfnamefont {Y.}~\bibnamefont {Lu}}, \bibinfo
  {author} {\bibfnamefont {S.}~\bibnamefont {Zhang}}, \bibinfo {author}
  {\bibfnamefont {K.}~\bibnamefont {Zhang}}, \bibinfo {author} {\bibfnamefont
  {G.}~\bibnamefont {Huang}}, \bibinfo {author} {\bibfnamefont
  {M.}~\bibnamefont {Qiao}}, \bibinfo {author} {\bibfnamefont {X.}~\bibnamefont
  {Su}}, \bibinfo {author} {\bibfnamefont {J.}~\bibnamefont {Zhang}}, \bibinfo
  {author} {\bibfnamefont {J.-N.}\ \bibnamefont {Zhang}}, \bibinfo {author}
  {\bibfnamefont {L.}~\bibnamefont {Banchi}}, \bibinfo {author} {\bibfnamefont
  {M.~S.}\ \bibnamefont {Kim}}, \ and\ \bibinfo {author} {\bibfnamefont
  {K.}~\bibnamefont {Kim}},\ }\bibfield  {title} {\emph {\bibinfo {title}
  {Scalable and programmable phononic network with trapped ions},\ }}\href
  {\doibase 10.1038/s41567-023-01952-5} {\bibfield  {journal} {\bibinfo
  {journal} {Nat. Phys.}\ }\textbf {\bibinfo {volume} {19}},\ \bibinfo {pages}
  {877} (\bibinfo {year} {2023})}\BibitemShut {NoStop}%
\bibitem [{\citenamefont {Briegel}\ \emph {et~al.}(1998)\citenamefont
  {Briegel}, \citenamefont {D\"ur}, \citenamefont {Cirac},\ and\ \citenamefont
  {Zoller}}]{Briegel1998Quantum}%
  \BibitemOpen
  \bibfield  {author} {\bibinfo {author} {\bibfnamefont {H.-J.}\ \bibnamefont
  {Briegel}}, \bibinfo {author} {\bibfnamefont {W.}~\bibnamefont {D\"ur}},
  \bibinfo {author} {\bibfnamefont {J.~I.}\ \bibnamefont {Cirac}}, \ and\
  \bibinfo {author} {\bibfnamefont {P.}~\bibnamefont {Zoller}},\ }\bibfield
  {title} {\emph {\bibinfo {title} {Quantum repeaters: The role of imperfect
  local operations in quantum communication},\ }}\href {\doibase
  10.1103/PhysRevLett.81.5932} {\bibfield  {journal} {\bibinfo  {journal}
  {Phys. Rev. Lett.}\ }\textbf {\bibinfo {volume} {81}},\ \bibinfo {pages}
  {5932} (\bibinfo {year} {1998})}\BibitemShut {NoStop}%
\bibitem [{\citenamefont {Yung}\ and\ \citenamefont
  {Bose}(2005)}]{Yung2005Perfect}%
  \BibitemOpen
  \bibfield  {author} {\bibinfo {author} {\bibfnamefont {M.-H.}\ \bibnamefont
  {Yung}}\ and\ \bibinfo {author} {\bibfnamefont {S.}~\bibnamefont {Bose}},\
  }\bibfield  {title} {\emph {\bibinfo {title} {Perfect state transfer,
  effective gates, and entanglement generation in engineered bosonic and
  fermionic networks},\ }}\href {\doibase 10.1103/PhysRevA.71.032310}
  {\bibfield  {journal} {\bibinfo  {journal} {Phys. Rev. A}\ }\textbf {\bibinfo
  {volume} {71}},\ \bibinfo {pages} {032310} (\bibinfo {year}
  {2005})}\BibitemShut {NoStop}%
\bibitem [{\citenamefont {Ma}\ \emph {et~al.}(2021)\citenamefont {Ma},
  \citenamefont {Puri}, \citenamefont {Schoelkopf}, \citenamefont {Devoret},
  \citenamefont {Girvin},\ and\ \citenamefont {Jiang}}]{Ma2021Quantum}%
  \BibitemOpen
  \bibfield  {author} {\bibinfo {author} {\bibfnamefont {W.-L.}\ \bibnamefont
  {Ma}}, \bibinfo {author} {\bibfnamefont {S.}~\bibnamefont {Puri}}, \bibinfo
  {author} {\bibfnamefont {R.~J.}\ \bibnamefont {Schoelkopf}}, \bibinfo
  {author} {\bibfnamefont {M.~H.}\ \bibnamefont {Devoret}}, \bibinfo {author}
  {\bibfnamefont {S.}~\bibnamefont {Girvin}}, \ and\ \bibinfo {author}
  {\bibfnamefont {L.}~\bibnamefont {Jiang}},\ }\bibfield  {title} {\emph
  {\bibinfo {title} {Quantum control of bosonic modes with superconducting
  circuits},\ }}\href {\doibase https://doi.org/10.1016/j.scib.2021.05.024}
  {\bibfield  {journal} {\bibinfo  {journal} {Sci. Bull.}\ }\textbf {\bibinfo
  {volume} {66}},\ \bibinfo {pages} {1789} (\bibinfo {year}
  {2021})}\BibitemShut {NoStop}%
\bibitem [{\citenamefont {Zhou}\ \emph {et~al.}(2024)\citenamefont {Zhou},
  \citenamefont {Li}, \citenamefont {Wang}, \citenamefont {Cai}, \citenamefont
  {Hua}, \citenamefont {Xu}, \citenamefont {Pan}, \citenamefont {Xue},
  \citenamefont {Zhang}, \citenamefont {Song}, \citenamefont {Yu},
  \citenamefont {Zou},\ and\ \citenamefont {Sun}}]{Zhou2024Quantum}%
  \BibitemOpen
  \bibfield  {author} {\bibinfo {author} {\bibfnamefont {J.}~\bibnamefont
  {Zhou}}, \bibinfo {author} {\bibfnamefont {M.}~\bibnamefont {Li}}, \bibinfo
  {author} {\bibfnamefont {W.}~\bibnamefont {Wang}}, \bibinfo {author}
  {\bibfnamefont {W.}~\bibnamefont {Cai}}, \bibinfo {author} {\bibfnamefont
  {Z.}~\bibnamefont {Hua}}, \bibinfo {author} {\bibfnamefont {Y.}~\bibnamefont
  {Xu}}, \bibinfo {author} {\bibfnamefont {X.}~\bibnamefont {Pan}}, \bibinfo
  {author} {\bibfnamefont {G.}~\bibnamefont {Xue}}, \bibinfo {author}
  {\bibfnamefont {H.}~\bibnamefont {Zhang}}, \bibinfo {author} {\bibfnamefont
  {Y.}~\bibnamefont {Song}}, \bibinfo {author} {\bibfnamefont {H.}~\bibnamefont
  {Yu}}, \bibinfo {author} {\bibfnamefont {C.-L.}\ \bibnamefont {Zou}}, \ and\
  \bibinfo {author} {\bibfnamefont {L.}~\bibnamefont {Sun}},\ }\bibfield
  {title} {\emph {\bibinfo {title} {Quantum state transfer between
  superconducting cavities via exchange-free interactions},\ }}\href {\doibase
  10.1103/PhysRevLett.133.220801} {\bibfield  {journal} {\bibinfo  {journal}
  {Phys. Rev. Lett.}\ }\textbf {\bibinfo {volume} {133}},\ \bibinfo {pages}
  {220801} (\bibinfo {year} {2024})}\BibitemShut {NoStop}%
\bibitem [{\citenamefont {Cirac}\ \emph {et~al.}(1997)\citenamefont {Cirac},
  \citenamefont {Zoller}, \citenamefont {Kimble},\ and\ \citenamefont
  {Mabuchi}}]{Cirac1997Quantum}%
  \BibitemOpen
  \bibfield  {author} {\bibinfo {author} {\bibfnamefont {J.~I.}\ \bibnamefont
  {Cirac}}, \bibinfo {author} {\bibfnamefont {P.}~\bibnamefont {Zoller}},
  \bibinfo {author} {\bibfnamefont {H.~J.}\ \bibnamefont {Kimble}}, \ and\
  \bibinfo {author} {\bibfnamefont {H.}~\bibnamefont {Mabuchi}},\ }\bibfield
  {title} {\emph {\bibinfo {title} {Quantum state transfer and entanglement
  distribution among distant nodes in a quantum network},\ }}\href {\doibase
  10.1103/PhysRevLett.78.3221} {\bibfield  {journal} {\bibinfo  {journal}
  {Phys. Rev. Lett.}\ }\textbf {\bibinfo {volume} {78}},\ \bibinfo {pages}
  {3221} (\bibinfo {year} {1997})}\BibitemShut {NoStop}%
\bibitem [{\citenamefont {Chou}\ \emph {et~al.}(2007)\citenamefont {Chou},
  \citenamefont {Laurat}, \citenamefont {Deng}, \citenamefont {Choi},
  \citenamefont {de~Riedmatten}, \citenamefont {Felinto},\ and\ \citenamefont
  {Kimble}}]{Chou2007Functional}%
  \BibitemOpen
  \bibfield  {author} {\bibinfo {author} {\bibfnamefont {C.-W.}\ \bibnamefont
  {Chou}}, \bibinfo {author} {\bibfnamefont {J.}~\bibnamefont {Laurat}},
  \bibinfo {author} {\bibfnamefont {H.}~\bibnamefont {Deng}}, \bibinfo {author}
  {\bibfnamefont {K.~S.}\ \bibnamefont {Choi}}, \bibinfo {author}
  {\bibfnamefont {H.}~\bibnamefont {de~Riedmatten}}, \bibinfo {author}
  {\bibfnamefont {D.}~\bibnamefont {Felinto}}, \ and\ \bibinfo {author}
  {\bibfnamefont {H.~J.}\ \bibnamefont {Kimble}},\ }\bibfield  {title} {\emph
  {\bibinfo {title} {Functional quantum nodes for entanglement distribution
  over scalable quantum networks},\ }}\href {\doibase 10.1126/science.1140300}
  {\bibfield  {journal} {\bibinfo  {journal} {Science}\ }\textbf {\bibinfo
  {volume} {316}},\ \bibinfo {pages} {1316} (\bibinfo {year}
  {2007})}\BibitemShut {NoStop}%
\bibitem [{\citenamefont {Spring}\ \emph {et~al.}(2013)\citenamefont {Spring},
  \citenamefont {Metcalf}, \citenamefont {Humphreys}, \citenamefont
  {Kolthammer}, \citenamefont {Jin}, \citenamefont {Barbieri}, \citenamefont
  {Datta}, \citenamefont {Thomas-Peter}, \citenamefont {Langford},
  \citenamefont {Kundys}, \citenamefont {Gates}, \citenamefont {Smith},
  \citenamefont {Smith},\ and\ \citenamefont {Walmsley}}]{Justin2013Boson}%
  \BibitemOpen
  \bibfield  {author} {\bibinfo {author} {\bibfnamefont {J.~B.}\ \bibnamefont
  {Spring}}, \bibinfo {author} {\bibfnamefont {B.~J.}\ \bibnamefont {Metcalf}},
  \bibinfo {author} {\bibfnamefont {P.~C.}\ \bibnamefont {Humphreys}}, \bibinfo
  {author} {\bibfnamefont {W.~S.}\ \bibnamefont {Kolthammer}}, \bibinfo
  {author} {\bibfnamefont {X.-M.}\ \bibnamefont {Jin}}, \bibinfo {author}
  {\bibfnamefont {M.}~\bibnamefont {Barbieri}}, \bibinfo {author}
  {\bibfnamefont {A.}~\bibnamefont {Datta}}, \bibinfo {author} {\bibfnamefont
  {N.}~\bibnamefont {Thomas-Peter}}, \bibinfo {author} {\bibfnamefont {N.~K.}\
  \bibnamefont {Langford}}, \bibinfo {author} {\bibfnamefont {D.}~\bibnamefont
  {Kundys}}, \bibinfo {author} {\bibfnamefont {J.~C.}\ \bibnamefont {Gates}},
  \bibinfo {author} {\bibfnamefont {B.~J.}\ \bibnamefont {Smith}}, \bibinfo
  {author} {\bibfnamefont {P.~G.~R.}\ \bibnamefont {Smith}}, \ and\ \bibinfo
  {author} {\bibfnamefont {I.~A.}\ \bibnamefont {Walmsley}},\ }\bibfield
  {title} {\emph {\bibinfo {title} {Boson sampling on a photonic chip},\
  }}\href {\doibase 10.1126/science.1231692} {\bibfield  {journal} {\bibinfo
  {journal} {Science}\ }\textbf {\bibinfo {volume} {339}},\ \bibinfo {pages}
  {798} (\bibinfo {year} {2013})}\BibitemShut {NoStop}%
\bibitem [{\citenamefont {Broome}\ \emph {et~al.}(2013)\citenamefont {Broome},
  \citenamefont {Fedrizzi}, \citenamefont {Rahimi-Keshari}, \citenamefont
  {Dove}, \citenamefont {Aaronson}, \citenamefont {Ralph},\ and\ \citenamefont
  {White}}]{Matthew2013Photonic}%
  \BibitemOpen
  \bibfield  {author} {\bibinfo {author} {\bibfnamefont {M.~A.}\ \bibnamefont
  {Broome}}, \bibinfo {author} {\bibfnamefont {A.}~\bibnamefont {Fedrizzi}},
  \bibinfo {author} {\bibfnamefont {S.}~\bibnamefont {Rahimi-Keshari}},
  \bibinfo {author} {\bibfnamefont {J.}~\bibnamefont {Dove}}, \bibinfo {author}
  {\bibfnamefont {S.}~\bibnamefont {Aaronson}}, \bibinfo {author}
  {\bibfnamefont {T.~C.}\ \bibnamefont {Ralph}}, \ and\ \bibinfo {author}
  {\bibfnamefont {A.~G.}\ \bibnamefont {White}},\ }\bibfield  {title} {\emph
  {\bibinfo {title} {Photonic boson sampling in a tunable circuit},\ }}\href
  {\doibase 10.1126/science.1231440} {\bibfield  {journal} {\bibinfo  {journal}
  {Science}\ }\textbf {\bibinfo {volume} {339}},\ \bibinfo {pages} {794}
  (\bibinfo {year} {2013})}\BibitemShut {NoStop}%
\bibitem [{\citenamefont {Tillmann}\ \emph {et~al.}(2013)\citenamefont
  {Tillmann}, \citenamefont {Daki$\acute{c}$}, \citenamefont {Heilmann},
  \citenamefont {Nolte}, \citenamefont {Szameit},\ and\ \citenamefont
  {Walther}}]{Tillmann2013Photonic}%
  \BibitemOpen
  \bibfield  {author} {\bibinfo {author} {\bibfnamefont {M.}~\bibnamefont
  {Tillmann}}, \bibinfo {author} {\bibfnamefont {B.}~\bibnamefont
  {Daki$\acute{c}$}}, \bibinfo {author} {\bibfnamefont {R.}~\bibnamefont
  {Heilmann}}, \bibinfo {author} {\bibfnamefont {S.}~\bibnamefont {Nolte}},
  \bibinfo {author} {\bibfnamefont {A.}~\bibnamefont {Szameit}}, \ and\
  \bibinfo {author} {\bibfnamefont {P.}~\bibnamefont {Walther}},\ }\bibfield
  {title} {\emph {\bibinfo {title} {Photonic boson sampling in a tunable
  circuit},\ }}\href {\doibase 10.1038/nphoton.2013.102} {\bibfield  {journal}
  {\bibinfo  {journal} {Nat. Photon.}\ }\textbf {\bibinfo {volume} {7}},\
  \bibinfo {pages} {540} (\bibinfo {year} {2013})}\BibitemShut {NoStop}%
\bibitem [{\citenamefont {Carolan}\ \emph {et~al.}(2014)\citenamefont
  {Carolan}, \citenamefont {Meinecke}, \citenamefont {Shadbolt}, \citenamefont
  {Russell}, \citenamefont {Ismail}, \citenamefont {W\"orhoff}, \citenamefont
  {Rudolph}, \citenamefont {Thompson}, \citenamefont {O'Brien}, \citenamefont
  {Matthews},\ and\ \citenamefont {Laing}}]{Carolan2014On}%
  \BibitemOpen
  \bibfield  {author} {\bibinfo {author} {\bibfnamefont {J.}~\bibnamefont
  {Carolan}}, \bibinfo {author} {\bibfnamefont {J.~D.~A.}\ \bibnamefont
  {Meinecke}}, \bibinfo {author} {\bibfnamefont {P.~J.}\ \bibnamefont
  {Shadbolt}}, \bibinfo {author} {\bibfnamefont {N.~J.}\ \bibnamefont
  {Russell}}, \bibinfo {author} {\bibfnamefont {N.}~\bibnamefont {Ismail}},
  \bibinfo {author} {\bibfnamefont {K.}~\bibnamefont {W\"orhoff}}, \bibinfo
  {author} {\bibfnamefont {T.}~\bibnamefont {Rudolph}}, \bibinfo {author}
  {\bibfnamefont {M.~G.}\ \bibnamefont {Thompson}}, \bibinfo {author}
  {\bibfnamefont {J.~L.}\ \bibnamefont {O'Brien}}, \bibinfo {author}
  {\bibfnamefont {J.~C.~F.}\ \bibnamefont {Matthews}}, \ and\ \bibinfo {author}
  {\bibfnamefont {A.}~\bibnamefont {Laing}},\ }\bibfield  {title} {\emph
  {\bibinfo {title} {On the experimental verification of quantum complexity in
  linear optics},\ }}\href {\doibase 10.1038/nphoton.2014.152} {\bibfield
  {journal} {\bibinfo  {journal} {Nat. Photon.}\ }\textbf {\bibinfo {volume}
  {8}},\ \bibinfo {pages} {621} (\bibinfo {year} {2014})}\BibitemShut {NoStop}%
\bibitem [{\citenamefont {Spagnolo}\ \emph {et~al.}(2014)\citenamefont
  {Spagnolo}, \citenamefont {Vitelli}, \citenamefont {Bentivegna},
  \citenamefont {Brod}, \citenamefont {Crespi}, \citenamefont {Flamini},
  \citenamefont {Giacomini}, \citenamefont {Milani}, \citenamefont {Ramponi},
  \citenamefont {Mataloni}, \citenamefont {Osellame}, \citenamefont
  {Galv$\check{a}$o},\ and\ \citenamefont
  {Sciarrino}}]{Spagnolo2014Experimental}%
  \BibitemOpen
  \bibfield  {author} {\bibinfo {author} {\bibfnamefont {N.}~\bibnamefont
  {Spagnolo}}, \bibinfo {author} {\bibfnamefont {C.}~\bibnamefont {Vitelli}},
  \bibinfo {author} {\bibfnamefont {M.}~\bibnamefont {Bentivegna}}, \bibinfo
  {author} {\bibfnamefont {D.~J.}\ \bibnamefont {Brod}}, \bibinfo {author}
  {\bibfnamefont {A.}~\bibnamefont {Crespi}}, \bibinfo {author} {\bibfnamefont
  {F.}~\bibnamefont {Flamini}}, \bibinfo {author} {\bibfnamefont
  {S.}~\bibnamefont {Giacomini}}, \bibinfo {author} {\bibfnamefont
  {G.}~\bibnamefont {Milani}}, \bibinfo {author} {\bibfnamefont
  {R.}~\bibnamefont {Ramponi}}, \bibinfo {author} {\bibfnamefont
  {P.}~\bibnamefont {Mataloni}}, \bibinfo {author} {\bibfnamefont
  {R.}~\bibnamefont {Osellame}}, \bibinfo {author} {\bibfnamefont {E.~F.}\
  \bibnamefont {Galv$\check{a}$o}}, \ and\ \bibinfo {author} {\bibfnamefont
  {F.}~\bibnamefont {Sciarrino}},\ }\bibfield  {title} {\emph {\bibinfo {title}
  {Experimental validation of photonic boson sampling},\ }}\href {\doibase
  10.1038/nphoton.2014.135} {\bibfield  {journal} {\bibinfo  {journal} {Nat.
  Photon.}\ }\textbf {\bibinfo {volume} {8}},\ \bibinfo {pages} {615} (\bibinfo
  {year} {2014})}\BibitemShut {NoStop}%
\bibitem [{\citenamefont {Wang}\ \emph {et~al.}(2019)\citenamefont {Wang},
  \citenamefont {Qin}, \citenamefont {Ding}, \citenamefont {Chen},
  \citenamefont {Chen}, \citenamefont {You}, \citenamefont {He}, \citenamefont
  {Jiang}, \citenamefont {You}, \citenamefont {Wang}, \citenamefont
  {Schneider}, \citenamefont {Renema}, \citenamefont {H\"ofling}, \citenamefont
  {Lu},\ and\ \citenamefont {Pan}}]{Wang2019Boson}%
  \BibitemOpen
  \bibfield  {author} {\bibinfo {author} {\bibfnamefont {H.}~\bibnamefont
  {Wang}}, \bibinfo {author} {\bibfnamefont {J.}~\bibnamefont {Qin}}, \bibinfo
  {author} {\bibfnamefont {X.}~\bibnamefont {Ding}}, \bibinfo {author}
  {\bibfnamefont {M.-C.}\ \bibnamefont {Chen}}, \bibinfo {author}
  {\bibfnamefont {S.}~\bibnamefont {Chen}}, \bibinfo {author} {\bibfnamefont
  {X.}~\bibnamefont {You}}, \bibinfo {author} {\bibfnamefont {Y.-M.}\
  \bibnamefont {He}}, \bibinfo {author} {\bibfnamefont {X.}~\bibnamefont
  {Jiang}}, \bibinfo {author} {\bibfnamefont {L.}~\bibnamefont {You}}, \bibinfo
  {author} {\bibfnamefont {Z.}~\bibnamefont {Wang}}, \bibinfo {author}
  {\bibfnamefont {C.}~\bibnamefont {Schneider}}, \bibinfo {author}
  {\bibfnamefont {J.~J.}\ \bibnamefont {Renema}}, \bibinfo {author}
  {\bibfnamefont {S.}~\bibnamefont {H\"ofling}}, \bibinfo {author}
  {\bibfnamefont {C.-Y.}\ \bibnamefont {Lu}}, \ and\ \bibinfo {author}
  {\bibfnamefont {J.-W.}\ \bibnamefont {Pan}},\ }\bibfield  {title} {\emph
  {\bibinfo {title} {Boson sampling with 20 input photons and a 60-mode
  interferometer in a $1{0}^{14}$-dimensional hilbert space},\ }}\href
  {\doibase 10.1103/PhysRevLett.123.250503} {\bibfield  {journal} {\bibinfo
  {journal} {Phys. Rev. Lett.}\ }\textbf {\bibinfo {volume} {123}},\ \bibinfo
  {pages} {250503} (\bibinfo {year} {2019})}\BibitemShut {NoStop}%
\bibitem [{\citenamefont {Arrazola}\ \emph {et~al.}(2021)\citenamefont
  {Arrazola}, \citenamefont {Bergholm}, \citenamefont {Brádler}, \citenamefont
  {Bromley}, \citenamefont {Collins}, \citenamefont {Dhand}, \citenamefont
  {Fumagalli}, \citenamefont {Gerrits}, \citenamefont {Goussev}, \citenamefont
  {Helt}, \citenamefont {Hundal}, \citenamefont {Isacsson}, \citenamefont
  {Israel}, \citenamefont {Izaac}, \citenamefont {Jahangiri}, \citenamefont
  {Janik}, \citenamefont {Killoran}, \citenamefont {Kumar}, \citenamefont
  {Lavoie}, \citenamefont {Lita}, \citenamefont {Mahler}, \citenamefont
  {Menotti}, \citenamefont {Morrison}, \citenamefont {Nam}, \citenamefont
  {Neuhaus}, \citenamefont {Qi}, \citenamefont {Quesada}, \citenamefont
  {Repingon}, \citenamefont {Sabapathy}, \citenamefont {Schuld}, \citenamefont
  {Su}, \citenamefont {Swinarton}, \citenamefont {Sz$\acute{a}$va},
  \citenamefont {Tan}, \citenamefont {Tan}, \citenamefont {Vaidya},
  \citenamefont {Vernon}, \citenamefont {Zabaneh},\ and\ \citenamefont
  {Zhang}}]{Arrazola2021Quantum}%
  \BibitemOpen
  \bibfield  {author} {\bibinfo {author} {\bibfnamefont {J.~M.}\ \bibnamefont
  {Arrazola}}, \bibinfo {author} {\bibfnamefont {V.}~\bibnamefont {Bergholm}},
  \bibinfo {author} {\bibfnamefont {K.}~\bibnamefont {Brádler}}, \bibinfo
  {author} {\bibfnamefont {T.~R.}\ \bibnamefont {Bromley}}, \bibinfo {author}
  {\bibfnamefont {M.~J.}\ \bibnamefont {Collins}}, \bibinfo {author}
  {\bibfnamefont {I.}~\bibnamefont {Dhand}}, \bibinfo {author} {\bibfnamefont
  {A.}~\bibnamefont {Fumagalli}}, \bibinfo {author} {\bibfnamefont
  {T.}~\bibnamefont {Gerrits}}, \bibinfo {author} {\bibfnamefont
  {A.}~\bibnamefont {Goussev}}, \bibinfo {author} {\bibfnamefont {L.~G.}\
  \bibnamefont {Helt}}, \bibinfo {author} {\bibfnamefont {J.}~\bibnamefont
  {Hundal}}, \bibinfo {author} {\bibfnamefont {T.}~\bibnamefont {Isacsson}},
  \bibinfo {author} {\bibfnamefont {R.~B.}\ \bibnamefont {Israel}}, \bibinfo
  {author} {\bibfnamefont {J.}~\bibnamefont {Izaac}}, \bibinfo {author}
  {\bibfnamefont {S.}~\bibnamefont {Jahangiri}}, \bibinfo {author}
  {\bibfnamefont {R.}~\bibnamefont {Janik}}, \bibinfo {author} {\bibfnamefont
  {N.}~\bibnamefont {Killoran}}, \bibinfo {author} {\bibfnamefont {S.~P.}\
  \bibnamefont {Kumar}}, \bibinfo {author} {\bibfnamefont {J.}~\bibnamefont
  {Lavoie}}, \bibinfo {author} {\bibfnamefont {A.~E.}\ \bibnamefont {Lita}},
  \bibinfo {author} {\bibfnamefont {D.~H.}\ \bibnamefont {Mahler}}, \bibinfo
  {author} {\bibfnamefont {M.}~\bibnamefont {Menotti}}, \bibinfo {author}
  {\bibfnamefont {B.}~\bibnamefont {Morrison}}, \bibinfo {author}
  {\bibfnamefont {S.~W.}\ \bibnamefont {Nam}}, \bibinfo {author} {\bibfnamefont
  {L.}~\bibnamefont {Neuhaus}}, \bibinfo {author} {\bibfnamefont {H.~Y.}\
  \bibnamefont {Qi}}, \bibinfo {author} {\bibfnamefont {N.}~\bibnamefont
  {Quesada}}, \bibinfo {author} {\bibfnamefont {A.}~\bibnamefont {Repingon}},
  \bibinfo {author} {\bibfnamefont {K.~K.}\ \bibnamefont {Sabapathy}}, \bibinfo
  {author} {\bibfnamefont {M.}~\bibnamefont {Schuld}}, \bibinfo {author}
  {\bibfnamefont {D.}~\bibnamefont {Su}}, \bibinfo {author} {\bibfnamefont
  {J.}~\bibnamefont {Swinarton}}, \bibinfo {author} {\bibfnamefont
  {A.}~\bibnamefont {Sz$\acute{a}$va}}, \bibinfo {author} {\bibfnamefont
  {K.}~\bibnamefont {Tan}}, \bibinfo {author} {\bibfnamefont {P.}~\bibnamefont
  {Tan}}, \bibinfo {author} {\bibfnamefont {V.~D.}\ \bibnamefont {Vaidya}},
  \bibinfo {author} {\bibfnamefont {Z.}~\bibnamefont {Vernon}}, \bibinfo
  {author} {\bibfnamefont {Z.}~\bibnamefont {Zabaneh}}, \ and\ \bibinfo
  {author} {\bibfnamefont {Y.}~\bibnamefont {Zhang}},\ }\bibfield  {title}
  {\emph {\bibinfo {title} {Quantum circuits with many photons on a
  programmable nanophotonic chip},\ }}\href {\doibase
  10.1038/s41586-021-03202-1} {\bibfield  {journal} {\bibinfo  {journal}
  {Nature}\ }\textbf {\bibinfo {volume} {59}},\ \bibinfo {pages} {54} (\bibinfo
  {year} {2021})}\BibitemShut {NoStop}%
\bibitem [{\citenamefont {Aaronson}\ and\ \citenamefont
  {Arkhipov}(2013)}]{Aaronson2013Computational}%
  \BibitemOpen
  \bibfield  {author} {\bibinfo {author} {\bibfnamefont {S.}~\bibnamefont
  {Aaronson}}\ and\ \bibinfo {author} {\bibfnamefont {A.}~\bibnamefont
  {Arkhipov}},\ }\bibfield  {title} {\emph {\bibinfo {title} {The computational
  complexity of linear optics},\ }}\href {\doibase 10.4086/toc.2013.v009a004}
  {\bibfield  {journal} {\bibinfo  {journal} {Theory. Comput.}\ }\textbf
  {\bibinfo {volume} {9}},\ \bibinfo {pages} {143} (\bibinfo {year}
  {2013})}\BibitemShut {NoStop}%
\bibitem [{\citenamefont {Knill}\ \emph {et~al.}(2001)\citenamefont {Knill},
  \citenamefont {Laflamme},\ and\ \citenamefont {Milburn}}]{Knill2001Scheme}%
  \BibitemOpen
  \bibfield  {author} {\bibinfo {author} {\bibfnamefont {E.}~\bibnamefont
  {Knill}}, \bibinfo {author} {\bibfnamefont {R.}~\bibnamefont {Laflamme}}, \
  and\ \bibinfo {author} {\bibfnamefont {G.~J.}\ \bibnamefont {Milburn}},\
  }\bibfield  {title} {\emph {\bibinfo {title} {A scheme for efficient quantum
  computation with linear optics},\ }}\href {\doibase 10.1038/35051009}
  {\bibfield  {journal} {\bibinfo  {journal} {Nature}\ }\textbf {\bibinfo
  {volume} {409}},\ \bibinfo {pages} {46} (\bibinfo {year} {2001})}\BibitemShut
  {NoStop}%
\bibitem [{\citenamefont {Kok}\ \emph {et~al.}(2007)\citenamefont {Kok},
  \citenamefont {Munro}, \citenamefont {Nemoto}, \citenamefont {Ralph},
  \citenamefont {Dowling},\ and\ \citenamefont {Milburn}}]{Kok2007Linear}%
  \BibitemOpen
  \bibfield  {author} {\bibinfo {author} {\bibfnamefont {P.}~\bibnamefont
  {Kok}}, \bibinfo {author} {\bibfnamefont {W.~J.}\ \bibnamefont {Munro}},
  \bibinfo {author} {\bibfnamefont {K.}~\bibnamefont {Nemoto}}, \bibinfo
  {author} {\bibfnamefont {T.~C.}\ \bibnamefont {Ralph}}, \bibinfo {author}
  {\bibfnamefont {J.~P.}\ \bibnamefont {Dowling}}, \ and\ \bibinfo {author}
  {\bibfnamefont {G.~J.}\ \bibnamefont {Milburn}},\ }\bibfield  {title} {\emph
  {\bibinfo {title} {Linear optical quantum computing with photonic qubits},\
  }}\href {\doibase 10.1103/RevModPhys.79.135} {\bibfield  {journal} {\bibinfo
  {journal} {Rev. Mod. Phys.}\ }\textbf {\bibinfo {volume} {79}},\ \bibinfo
  {pages} {135} (\bibinfo {year} {2007})}\BibitemShut {NoStop}%
\bibitem [{\citenamefont {Chuang}\ \emph {et~al.}(1997)\citenamefont {Chuang},
  \citenamefont {Leung},\ and\ \citenamefont {Yamamoto}}]{Chuang1997Bosonic}%
  \BibitemOpen
  \bibfield  {author} {\bibinfo {author} {\bibfnamefont {I.~L.}\ \bibnamefont
  {Chuang}}, \bibinfo {author} {\bibfnamefont {D.~W.}\ \bibnamefont {Leung}}, \
  and\ \bibinfo {author} {\bibfnamefont {Y.}~\bibnamefont {Yamamoto}},\
  }\bibfield  {title} {\emph {\bibinfo {title} {Bosonic quantum codes for
  amplitude damping},\ }}\href {\doibase 10.1103/PhysRevA.56.1114} {\bibfield
  {journal} {\bibinfo  {journal} {Phys. Rev. A}\ }\textbf {\bibinfo {volume}
  {56}},\ \bibinfo {pages} {1114} (\bibinfo {year} {1997})}\BibitemShut
  {NoStop}%
\bibitem [{\citenamefont {Gottesman}\ \emph {et~al.}(2001)\citenamefont
  {Gottesman}, \citenamefont {Kitaev},\ and\ \citenamefont
  {Preskill}}]{Gottesman2001Encoding}%
  \BibitemOpen
  \bibfield  {author} {\bibinfo {author} {\bibfnamefont {D.}~\bibnamefont
  {Gottesman}}, \bibinfo {author} {\bibfnamefont {A.}~\bibnamefont {Kitaev}}, \
  and\ \bibinfo {author} {\bibfnamefont {J.}~\bibnamefont {Preskill}},\
  }\bibfield  {title} {\emph {\bibinfo {title} {Encoding a qubit in an
  oscillator},\ }}\href {\doibase 10.1103/PhysRevA.64.012310} {\bibfield
  {journal} {\bibinfo  {journal} {Phys. Rev. A}\ }\textbf {\bibinfo {volume}
  {64}},\ \bibinfo {pages} {012310} (\bibinfo {year} {2001})}\BibitemShut
  {NoStop}%
\bibitem [{\citenamefont {Mirrahimi}\ \emph {et~al.}(2014)\citenamefont
  {Mirrahimi}, \citenamefont {Leghtas}, \citenamefont {Albert}, \citenamefont
  {Touzard}, \citenamefont {Schoelkopf}, \citenamefont {Jiang},\ and\
  \citenamefont {Devoret}}]{Mirrahimi2014Dynamically}%
  \BibitemOpen
  \bibfield  {author} {\bibinfo {author} {\bibfnamefont {M.}~\bibnamefont
  {Mirrahimi}}, \bibinfo {author} {\bibfnamefont {Z.}~\bibnamefont {Leghtas}},
  \bibinfo {author} {\bibfnamefont {V.~V.}\ \bibnamefont {Albert}}, \bibinfo
  {author} {\bibfnamefont {S.}~\bibnamefont {Touzard}}, \bibinfo {author}
  {\bibfnamefont {R.~J.}\ \bibnamefont {Schoelkopf}}, \bibinfo {author}
  {\bibfnamefont {L.}~\bibnamefont {Jiang}}, \ and\ \bibinfo {author}
  {\bibfnamefont {M.~H.}\ \bibnamefont {Devoret}},\ }\bibfield  {title} {\emph
  {\bibinfo {title} {Dynamically protected cat-qubits: a new paradigm for
  universal quantum computation},\ }}\href {\doibase
  10.1088/1367-2630/16/4/045014} {\bibfield  {journal} {\bibinfo  {journal}
  {New J. Phys.}\ }\textbf {\bibinfo {volume} {16}},\ \bibinfo {pages} {045014}
  (\bibinfo {year} {2014})}\BibitemShut {NoStop}%
\bibitem [{\citenamefont {Michael}\ \emph {et~al.}(2016)\citenamefont
  {Michael}, \citenamefont {Silveri}, \citenamefont {Brierley}, \citenamefont
  {Albert}, \citenamefont {Salmilehto}, \citenamefont {Jiang},\ and\
  \citenamefont {Girvin}}]{Michael2016New}%
  \BibitemOpen
  \bibfield  {author} {\bibinfo {author} {\bibfnamefont {M.~H.}\ \bibnamefont
  {Michael}}, \bibinfo {author} {\bibfnamefont {M.}~\bibnamefont {Silveri}},
  \bibinfo {author} {\bibfnamefont {R.~T.}\ \bibnamefont {Brierley}}, \bibinfo
  {author} {\bibfnamefont {V.~V.}\ \bibnamefont {Albert}}, \bibinfo {author}
  {\bibfnamefont {J.}~\bibnamefont {Salmilehto}}, \bibinfo {author}
  {\bibfnamefont {L.}~\bibnamefont {Jiang}}, \ and\ \bibinfo {author}
  {\bibfnamefont {S.~M.}\ \bibnamefont {Girvin}},\ }\bibfield  {title} {\emph
  {\bibinfo {title} {New class of quantum error-correcting codes for a bosonic
  mode},\ }}\href {\doibase 10.1103/PhysRevX.6.031006} {\bibfield  {journal}
  {\bibinfo  {journal} {Phys. Rev. X}\ }\textbf {\bibinfo {volume} {6}},\
  \bibinfo {pages} {031006} (\bibinfo {year} {2016})}\BibitemShut {NoStop}%
\bibitem [{\citenamefont {Kimble}(1998)}]{Kimble1998strong}%
  \BibitemOpen
  \bibfield  {author} {\bibinfo {author} {\bibfnamefont {H.~J.}\ \bibnamefont
  {Kimble}},\ }\bibfield  {title} {\emph {\bibinfo {title} {Strong interactions
  of single atoms and photons in cavity qed},\ }}\href {\doibase
  10.1238/Physica.Topical.076a00127} {\bibfield  {journal} {\bibinfo  {journal}
  {Phys. Scr.}\ }\textbf {\bibinfo {volume} {T76}},\ \bibinfo {pages} {127}
  (\bibinfo {year} {1998})}\BibitemShut {NoStop}%
\bibitem [{\citenamefont {Haroche}\ and\ \citenamefont
  {Raimond}(2006)}]{Haroche2006Exploring}%
  \BibitemOpen
  \bibfield  {author} {\bibinfo {author} {\bibfnamefont {S.}~\bibnamefont
  {Haroche}}\ and\ \bibinfo {author} {\bibfnamefont {J.-M.}\ \bibnamefont
  {Raimond}},\ }\href@noop {} {\emph {\bibinfo {title} {Exploring the Quantum:
  Atoms, Cavities, and Photons}}}\ (\bibinfo  {publisher} {Oxford University
  Press, New York},\ \bibinfo {year} {2006})\BibitemShut {NoStop}%
\bibitem [{\citenamefont {Vermersch}\ \emph {et~al.}(2017)\citenamefont
  {Vermersch}, \citenamefont {Guimond}, \citenamefont {Pichler},\ and\
  \citenamefont {Zoller}}]{Vermersch2017Quantum}%
  \BibitemOpen
  \bibfield  {author} {\bibinfo {author} {\bibfnamefont {B.}~\bibnamefont
  {Vermersch}}, \bibinfo {author} {\bibfnamefont {P.-O.}\ \bibnamefont
  {Guimond}}, \bibinfo {author} {\bibfnamefont {H.}~\bibnamefont {Pichler}}, \
  and\ \bibinfo {author} {\bibfnamefont {P.}~\bibnamefont {Zoller}},\
  }\bibfield  {title} {\emph {\bibinfo {title} {Quantum state transfer via
  noisy photonic and phononic waveguides},\ }}\href {\doibase
  10.1103/PhysRevLett.118.133601} {\bibfield  {journal} {\bibinfo  {journal}
  {Phys. Rev. Lett.}\ }\textbf {\bibinfo {volume} {118}},\ \bibinfo {pages}
  {133601} (\bibinfo {year} {2017})}\BibitemShut {NoStop}%
\bibitem [{\citenamefont {Blais}\ \emph {et~al.}(2021)\citenamefont {Blais},
  \citenamefont {Grimsmo}, \citenamefont {Girvin},\ and\ \citenamefont
  {Wallraff}}]{Blais2021Circuit}%
  \BibitemOpen
  \bibfield  {author} {\bibinfo {author} {\bibfnamefont {A.}~\bibnamefont
  {Blais}}, \bibinfo {author} {\bibfnamefont {A.~L.}\ \bibnamefont {Grimsmo}},
  \bibinfo {author} {\bibfnamefont {S.~M.}\ \bibnamefont {Girvin}}, \ and\
  \bibinfo {author} {\bibfnamefont {A.}~\bibnamefont {Wallraff}},\ }\bibfield
  {title} {\emph {\bibinfo {title} {Circuit quantum electrodynamics},\ }}\href
  {\doibase 10.1103/RevModPhys.93.025005} {\bibfield  {journal} {\bibinfo
  {journal} {Rev. Mod. Phys.}\ }\textbf {\bibinfo {volume} {93}},\ \bibinfo
  {pages} {025005} (\bibinfo {year} {2021})}\BibitemShut {NoStop}%
\bibitem [{\citenamefont {Regensburger}\ \emph {et~al.}(2012)\citenamefont
  {Regensburger}, \citenamefont {Bersch}, \citenamefont {Miri}, \citenamefont
  {Onishchukov},\ and\ \citenamefont
  {Christodoulides}}]{Regensburger2012Parity}%
  \BibitemOpen
  \bibfield  {author} {\bibinfo {author} {\bibfnamefont {A.}~\bibnamefont
  {Regensburger}}, \bibinfo {author} {\bibfnamefont {C.}~\bibnamefont
  {Bersch}}, \bibinfo {author} {\bibfnamefont {M.-A.}\ \bibnamefont {Miri}},
  \bibinfo {author} {\bibfnamefont {G.}~\bibnamefont {Onishchukov}}, \ and\
  \bibinfo {author} {\bibfnamefont {D.~N.}\ \bibnamefont {Christodoulides}},\
  }\bibfield  {title} {\emph {\bibinfo {title} {Parity–time synthetic
  photonic lattices},\ }}\href {\doibase 10.1038/nature11298} {\bibfield
  {journal} {\bibinfo  {journal} {Nature}\ }\textbf {\bibinfo {volume} {488}},\
  \bibinfo {pages} {167} (\bibinfo {year} {2012})}\BibitemShut {NoStop}%
\bibitem [{\citenamefont {Celi}\ \emph {et~al.}(2014)\citenamefont {Celi},
  \citenamefont {Massignan}, \citenamefont {Ruseckas}, \citenamefont {Goldman},
  \citenamefont {Spielman}, \citenamefont {Juzeli\ifmmode~\bar{u}\else
  \={u}\fi{}nas},\ and\ \citenamefont {Lewenstein}}]{Celi2014Synthetic}%
  \BibitemOpen
  \bibfield  {author} {\bibinfo {author} {\bibfnamefont {A.}~\bibnamefont
  {Celi}}, \bibinfo {author} {\bibfnamefont {P.}~\bibnamefont {Massignan}},
  \bibinfo {author} {\bibfnamefont {J.}~\bibnamefont {Ruseckas}}, \bibinfo
  {author} {\bibfnamefont {N.}~\bibnamefont {Goldman}}, \bibinfo {author}
  {\bibfnamefont {I.~B.}\ \bibnamefont {Spielman}}, \bibinfo {author}
  {\bibfnamefont {G.}~\bibnamefont {Juzeli\ifmmode~\bar{u}\else
  \={u}\fi{}nas}}, \ and\ \bibinfo {author} {\bibfnamefont {M.}~\bibnamefont
  {Lewenstein}},\ }\bibfield  {title} {\emph {\bibinfo {title} {Synthetic gauge
  fields in synthetic dimensions},\ }}\href {\doibase
  10.1103/PhysRevLett.112.043001} {\bibfield  {journal} {\bibinfo  {journal}
  {Phys. Rev. Lett.}\ }\textbf {\bibinfo {volume} {112}},\ \bibinfo {pages}
  {043001} (\bibinfo {year} {2014})}\BibitemShut {NoStop}%
\bibitem [{\citenamefont {Lustig}\ \emph {et~al.}(2019)\citenamefont {Lustig},
  \citenamefont {Weimann}, \citenamefont {Plotnik}, \citenamefont {Lumer},
  \citenamefont {Bandres}, \citenamefont {Szameit},\ and\ \citenamefont
  {Segev}}]{Lustig2019Photonic}%
  \BibitemOpen
  \bibfield  {author} {\bibinfo {author} {\bibfnamefont {E.}~\bibnamefont
  {Lustig}}, \bibinfo {author} {\bibfnamefont {S.}~\bibnamefont {Weimann}},
  \bibinfo {author} {\bibfnamefont {Y.}~\bibnamefont {Plotnik}}, \bibinfo
  {author} {\bibfnamefont {Y.}~\bibnamefont {Lumer}}, \bibinfo {author}
  {\bibfnamefont {M.~A.}\ \bibnamefont {Bandres}}, \bibinfo {author}
  {\bibfnamefont {A.}~\bibnamefont {Szameit}}, \ and\ \bibinfo {author}
  {\bibfnamefont {M.}~\bibnamefont {Segev}},\ }\bibfield  {title} {\emph
  {\bibinfo {title} {Photonic topological insulator in synthetic dimensions},\
  }}\href {\doibase 10.1038/s41586-019-0943-7} {\bibfield  {journal} {\bibinfo
  {journal} {Nature}\ }\textbf {\bibinfo {volume} {567}},\ \bibinfo {pages}
  {356} (\bibinfo {year} {2019})}\BibitemShut {NoStop}%
\bibitem [{\citenamefont {Ozawa}\ and\ \citenamefont
  {Price}(2019)}]{Ozawa2019Topological}%
  \BibitemOpen
  \bibfield  {author} {\bibinfo {author} {\bibfnamefont {T.}~\bibnamefont
  {Ozawa}}\ and\ \bibinfo {author} {\bibfnamefont {H.~M.}\ \bibnamefont
  {Price}},\ }\bibfield  {title} {\emph {\bibinfo {title} {Topological quantum
  matter in synthetic dimensions},\ }}\href {\doibase
  10.1038/s42254-019-0045-3} {\bibfield  {journal} {\bibinfo  {journal} {Nat.
  Rev. Phys.}\ }\textbf {\bibinfo {volume} {1}},\ \bibinfo {pages} {349}
  (\bibinfo {year} {2019})}\BibitemShut {NoStop}%
\bibitem [{\citenamefont {Morsch}\ and\ \citenamefont
  {Oberthaler}(2006)}]{Morsch2006Dynamics}%
  \BibitemOpen
  \bibfield  {author} {\bibinfo {author} {\bibfnamefont {O.}~\bibnamefont
  {Morsch}}\ and\ \bibinfo {author} {\bibfnamefont {M.}~\bibnamefont
  {Oberthaler}},\ }\bibfield  {title} {\emph {\bibinfo {title} {Dynamics of
  bose-einstein condensates in optical lattices},\ }}\href {\doibase
  10.1103/RevModPhys.78.179} {\bibfield  {journal} {\bibinfo  {journal} {Rev.
  Mod. Phys.}\ }\textbf {\bibinfo {volume} {78}},\ \bibinfo {pages} {179}
  (\bibinfo {year} {2006})}\BibitemShut {NoStop}%
\bibitem [{\citenamefont {Heinrich}\ \emph {et~al.}(2011)\citenamefont
  {Heinrich}, \citenamefont {Ludwig}, \citenamefont {Qian}, \citenamefont
  {Kubala},\ and\ \citenamefont {Marquardt}}]{Heinrich2011Collective}%
  \BibitemOpen
  \bibfield  {author} {\bibinfo {author} {\bibfnamefont {G.}~\bibnamefont
  {Heinrich}}, \bibinfo {author} {\bibfnamefont {M.}~\bibnamefont {Ludwig}},
  \bibinfo {author} {\bibfnamefont {J.}~\bibnamefont {Qian}}, \bibinfo {author}
  {\bibfnamefont {B.}~\bibnamefont {Kubala}}, \ and\ \bibinfo {author}
  {\bibfnamefont {F.}~\bibnamefont {Marquardt}},\ }\bibfield  {title} {\emph
  {\bibinfo {title} {Collective dynamics in optomechanical arrays},\ }}\href
  {\doibase 10.1103/PhysRevLett.107.043603} {\bibfield  {journal} {\bibinfo
  {journal} {Phys. Rev. Lett.}\ }\textbf {\bibinfo {volume} {107}},\ \bibinfo
  {pages} {043603} (\bibinfo {year} {2011})}\BibitemShut {NoStop}%
\bibitem [{\citenamefont {Stannigel}\ \emph {et~al.}(2011)\citenamefont
  {Stannigel}, \citenamefont {Rabl}, \citenamefont {S\o{}rensen}, \citenamefont
  {Lukin},\ and\ \citenamefont {Zoller}}]{Stannigel2011Optomechanical}%
  \BibitemOpen
  \bibfield  {author} {\bibinfo {author} {\bibfnamefont {K.}~\bibnamefont
  {Stannigel}}, \bibinfo {author} {\bibfnamefont {P.}~\bibnamefont {Rabl}},
  \bibinfo {author} {\bibfnamefont {A.~S.}\ \bibnamefont {S\o{}rensen}},
  \bibinfo {author} {\bibfnamefont {M.~D.}\ \bibnamefont {Lukin}}, \ and\
  \bibinfo {author} {\bibfnamefont {P.}~\bibnamefont {Zoller}},\ }\bibfield
  {title} {\emph {\bibinfo {title} {Optomechanical transducers for
  quantum-information processing},\ }}\href {\doibase
  10.1103/PhysRevA.84.042341} {\bibfield  {journal} {\bibinfo  {journal} {Phys.
  Rev. A}\ }\textbf {\bibinfo {volume} {84}},\ \bibinfo {pages} {042341}
  (\bibinfo {year} {2011})}\BibitemShut {NoStop}%
\bibitem [{\citenamefont {Aspelmeyer}\ \emph {et~al.}(2014)\citenamefont
  {Aspelmeyer}, \citenamefont {Kippenberg},\ and\ \citenamefont
  {Marquardt}}]{Aspelmeyer2014Cavity}%
  \BibitemOpen
  \bibfield  {author} {\bibinfo {author} {\bibfnamefont {M.}~\bibnamefont
  {Aspelmeyer}}, \bibinfo {author} {\bibfnamefont {T.~J.}\ \bibnamefont
  {Kippenberg}}, \ and\ \bibinfo {author} {\bibfnamefont {F.}~\bibnamefont
  {Marquardt}},\ }\bibfield  {title} {\emph {\bibinfo {title} {Cavity
  optomechanics},\ }}\href {\doibase 10.1103/RevModPhys.86.1391} {\bibfield
  {journal} {\bibinfo  {journal} {Rev. Mod. Phys.}\ }\textbf {\bibinfo {volume}
  {86}},\ \bibinfo {pages} {1391} (\bibinfo {year} {2014})}\BibitemShut
  {NoStop}%
\bibitem [{\citenamefont {Zhang}\ \emph
  {et~al.}(2015{\natexlab{a}})\citenamefont {Zhang}, \citenamefont {Shah},
  \citenamefont {Cardenas},\ and\ \citenamefont
  {Lipson}}]{Zhang2015Synchronization}%
  \BibitemOpen
  \bibfield  {author} {\bibinfo {author} {\bibfnamefont {M.}~\bibnamefont
  {Zhang}}, \bibinfo {author} {\bibfnamefont {S.}~\bibnamefont {Shah}},
  \bibinfo {author} {\bibfnamefont {J.}~\bibnamefont {Cardenas}}, \ and\
  \bibinfo {author} {\bibfnamefont {M.}~\bibnamefont {Lipson}},\ }\bibfield
  {title} {\emph {\bibinfo {title} {Synchronization and phase noise reduction
  in micromechanical oscillator arrays coupled through light},\ }}\href
  {\doibase 10.1103/PhysRevLett.115.163902} {\bibfield  {journal} {\bibinfo
  {journal} {Phys. Rev. Lett.}\ }\textbf {\bibinfo {volume} {115}},\ \bibinfo
  {pages} {163902} (\bibinfo {year} {2015}{\natexlab{a}})}\BibitemShut
  {NoStop}%
\bibitem [{\citenamefont {Peterson}\ \emph {et~al.}(2017)\citenamefont
  {Peterson}, \citenamefont {Lecocq}, \citenamefont {Cicak}, \citenamefont
  {Simmonds}, \citenamefont {Aumentado},\ and\ \citenamefont
  {Teufel}}]{Peterson2017Demonstration}%
  \BibitemOpen
  \bibfield  {author} {\bibinfo {author} {\bibfnamefont {G.~A.}\ \bibnamefont
  {Peterson}}, \bibinfo {author} {\bibfnamefont {F.}~\bibnamefont {Lecocq}},
  \bibinfo {author} {\bibfnamefont {K.}~\bibnamefont {Cicak}}, \bibinfo
  {author} {\bibfnamefont {R.~W.}\ \bibnamefont {Simmonds}}, \bibinfo {author}
  {\bibfnamefont {J.}~\bibnamefont {Aumentado}}, \ and\ \bibinfo {author}
  {\bibfnamefont {J.~D.}\ \bibnamefont {Teufel}},\ }\bibfield  {title} {\emph
  {\bibinfo {title} {Demonstration of efficient nonreciprocity in a microwave
  optomechanical circuit},\ }}\href {\doibase 10.1103/PhysRevX.7.031001}
  {\bibfield  {journal} {\bibinfo  {journal} {Phys. Rev. X}\ }\textbf {\bibinfo
  {volume} {7}},\ \bibinfo {pages} {031001} (\bibinfo {year}
  {2017})}\BibitemShut {NoStop}%
\bibitem [{\citenamefont {Han}\ \emph {et~al.}(2020)\citenamefont {Han},
  \citenamefont {Fu}, \citenamefont {Zhong}, \citenamefont {Zou}, \citenamefont
  {Xu}, \citenamefont {Sayem}, \citenamefont {Xu}, \citenamefont {Wang},
  \citenamefont {Cheng}, \citenamefont {Jiang},\ and\ \citenamefont
  {Tang}}]{Xu2020Cavity}%
  \BibitemOpen
  \bibfield  {author} {\bibinfo {author} {\bibfnamefont {X.}~\bibnamefont
  {Han}}, \bibinfo {author} {\bibfnamefont {W.}~\bibnamefont {Fu}}, \bibinfo
  {author} {\bibfnamefont {C.}~\bibnamefont {Zhong}}, \bibinfo {author}
  {\bibfnamefont {C.-L.}\ \bibnamefont {Zou}}, \bibinfo {author} {\bibfnamefont
  {Y.}~\bibnamefont {Xu}}, \bibinfo {author} {\bibfnamefont {A.~A.}\
  \bibnamefont {Sayem}}, \bibinfo {author} {\bibfnamefont {M.}~\bibnamefont
  {Xu}}, \bibinfo {author} {\bibfnamefont {S.}~\bibnamefont {Wang}}, \bibinfo
  {author} {\bibfnamefont {R.}~\bibnamefont {Cheng}}, \bibinfo {author}
  {\bibfnamefont {L.}~\bibnamefont {Jiang}}, \ and\ \bibinfo {author}
  {\bibfnamefont {H.~X.}\ \bibnamefont {Tang}},\ }\bibfield  {title} {\emph
  {\bibinfo {title} {Cavity piezo-mechanics for superconducting-nanophotonic
  quantum interface},\ }}\href {\doibase 10.1038/s41467-020-17053-3} {\bibfield
   {journal} {\bibinfo  {journal} {Nat. Commun.}\ }\textbf {\bibinfo {volume}
  {11}},\ \bibinfo {pages} {3237} (\bibinfo {year} {2020})}\BibitemShut
  {NoStop}%
\bibitem [{\citenamefont {von L{\"u}pke}\ \emph {et~al.}(2024)\citenamefont
  {von L{\"u}pke}, \citenamefont {Rodrigues}, \citenamefont {Yang},
  \citenamefont {Fadel},\ and\ \citenamefont {Chu}}]{von2024Engineering}%
  \BibitemOpen
  \bibfield  {author} {\bibinfo {author} {\bibfnamefont {U.}~\bibnamefont {von
  L{\"u}pke}}, \bibinfo {author} {\bibfnamefont {I.~C.}\ \bibnamefont
  {Rodrigues}}, \bibinfo {author} {\bibfnamefont {Y.}~\bibnamefont {Yang}},
  \bibinfo {author} {\bibfnamefont {M.}~\bibnamefont {Fadel}}, \ and\ \bibinfo
  {author} {\bibfnamefont {Y.}~\bibnamefont {Chu}},\ }\bibfield  {title} {\emph
  {\bibinfo {title} {Engineering multimode interactions in circuit quantum
  acoustodynamics},\ }}\href {\doibase 10.1038/s41567-023-02377-w} {\bibfield
  {journal} {\bibinfo  {journal} {Nat. Phys.}\ }\textbf {\bibinfo {volume}
  {20}},\ \bibinfo {pages} {564} (\bibinfo {year} {2024})}\BibitemShut
  {NoStop}%
\bibitem [{\citenamefont {Zhang}\ \emph
  {et~al.}(2016{\natexlab{a}})\citenamefont {Zhang}, \citenamefont {Zou},
  \citenamefont {Jiang},\ and\ \citenamefont {Tang}}]{Zhang2016Cavity}%
  \BibitemOpen
  \bibfield  {author} {\bibinfo {author} {\bibfnamefont {X.}~\bibnamefont
  {Zhang}}, \bibinfo {author} {\bibfnamefont {C.-L.}\ \bibnamefont {Zou}},
  \bibinfo {author} {\bibfnamefont {L.}~\bibnamefont {Jiang}}, \ and\ \bibinfo
  {author} {\bibfnamefont {H.~X.}\ \bibnamefont {Tang}},\ }\bibfield  {title}
  {\emph {\bibinfo {title} {Cavity magnomechanics},\ }}\href {\doibase
  10.1126/sciadv.1501286} {\bibfield  {journal} {\bibinfo  {journal} {Sci.
  Adv.}\ }\textbf {\bibinfo {volume} {2}},\ \bibinfo {pages} {e1501286}
  (\bibinfo {year} {2016}{\natexlab{a}})}\BibitemShut {NoStop}%
\bibitem [{\citenamefont {Li}\ \emph {et~al.}(2021)\citenamefont {Li},
  \citenamefont {Wang}, \citenamefont {Wu}, \citenamefont {Zhu},\ and\
  \citenamefont {You}}]{Li2021Quantum}%
  \BibitemOpen
  \bibfield  {author} {\bibinfo {author} {\bibfnamefont {J.}~\bibnamefont
  {Li}}, \bibinfo {author} {\bibfnamefont {Y.-P.}\ \bibnamefont {Wang}},
  \bibinfo {author} {\bibfnamefont {W.-J.}\ \bibnamefont {Wu}}, \bibinfo
  {author} {\bibfnamefont {S.-Y.}\ \bibnamefont {Zhu}}, \ and\ \bibinfo
  {author} {\bibfnamefont {J.}~\bibnamefont {You}},\ }\bibfield  {title} {\emph
  {\bibinfo {title} {Quantum network with magnonic and mechanical nodes},\
  }}\href {\doibase 10.1103/PRXQuantum.2.040344} {\bibfield  {journal}
  {\bibinfo  {journal} {PRX Quantum}\ }\textbf {\bibinfo {volume} {2}},\
  \bibinfo {pages} {040344} (\bibinfo {year} {2021})}\BibitemShut {NoStop}%
\bibitem [{\citenamefont {Shen}\ \emph {et~al.}(2022)\citenamefont {Shen},
  \citenamefont {Xu}, \citenamefont {Zhang}, \citenamefont {Zhang},
  \citenamefont {Wang}, \citenamefont {Chai}, \citenamefont {Zou},
  \citenamefont {Guo},\ and\ \citenamefont {Dong}}]{Shen2022Coherent}%
  \BibitemOpen
  \bibfield  {author} {\bibinfo {author} {\bibfnamefont {Z.}~\bibnamefont
  {Shen}}, \bibinfo {author} {\bibfnamefont {G.-T.}\ \bibnamefont {Xu}},
  \bibinfo {author} {\bibfnamefont {M.}~\bibnamefont {Zhang}}, \bibinfo
  {author} {\bibfnamefont {Y.-L.}\ \bibnamefont {Zhang}}, \bibinfo {author}
  {\bibfnamefont {Y.}~\bibnamefont {Wang}}, \bibinfo {author} {\bibfnamefont
  {C.-Z.}\ \bibnamefont {Chai}}, \bibinfo {author} {\bibfnamefont {C.-L.}\
  \bibnamefont {Zou}}, \bibinfo {author} {\bibfnamefont {G.-C.}\ \bibnamefont
  {Guo}}, \ and\ \bibinfo {author} {\bibfnamefont {C.-H.}\ \bibnamefont
  {Dong}},\ }\bibfield  {title} {\emph {\bibinfo {title} {Coherent coupling
  between phonons, magnons, and photons},\ }}\href {\doibase
  10.1103/PhysRevLett.129.243601} {\bibfield  {journal} {\bibinfo  {journal}
  {Phys. Rev. Lett.}\ }\textbf {\bibinfo {volume} {129}},\ \bibinfo {pages}
  {243601} (\bibinfo {year} {2022})}\BibitemShut {NoStop}%
\bibitem [{\citenamefont {Shen}\ \emph {et~al.}(2025)\citenamefont {Shen},
  \citenamefont {Li}, \citenamefont {Sun}, \citenamefont {Wu}, \citenamefont
  {Zuo}, \citenamefont {Wang}, \citenamefont {Zhu},\ and\ \citenamefont
  {You}}]{Shen2025Cavity}%
  \BibitemOpen
  \bibfield  {author} {\bibinfo {author} {\bibfnamefont {R.-C.}\ \bibnamefont
  {Shen}}, \bibinfo {author} {\bibfnamefont {J.}~\bibnamefont {Li}}, \bibinfo
  {author} {\bibfnamefont {Y.-M.}\ \bibnamefont {Sun}}, \bibinfo {author}
  {\bibfnamefont {W.-J.}\ \bibnamefont {Wu}}, \bibinfo {author} {\bibfnamefont
  {X.}~\bibnamefont {Zuo}}, \bibinfo {author} {\bibfnamefont {Y.-P.}\
  \bibnamefont {Wang}}, \bibinfo {author} {\bibfnamefont {S.-Y.}\ \bibnamefont
  {Zhu}}, \ and\ \bibinfo {author} {\bibfnamefont {J.~Q.}\ \bibnamefont
  {You}},\ }\bibfield  {title} {\emph {\bibinfo {title} {Cavity-magnon
  polaritons strongly coupled to phonons},\ }}\href {\doibase
  10.1038/s41467-025-60799-x} {\bibfield  {journal} {\bibinfo  {journal} {Nat.
  Commun.}\ }\textbf {\bibinfo {volume} {16}},\ \bibinfo {pages} {5652}
  (\bibinfo {year} {2025})}\BibitemShut {NoStop}%
\bibitem [{\citenamefont {Zhang}\ \emph {et~al.}(2014)\citenamefont {Zhang},
  \citenamefont {Zou}, \citenamefont {Jiang},\ and\ \citenamefont
  {Tang}}]{Zhang2014Strongly}%
  \BibitemOpen
  \bibfield  {author} {\bibinfo {author} {\bibfnamefont {X.}~\bibnamefont
  {Zhang}}, \bibinfo {author} {\bibfnamefont {C.-L.}\ \bibnamefont {Zou}},
  \bibinfo {author} {\bibfnamefont {L.}~\bibnamefont {Jiang}}, \ and\ \bibinfo
  {author} {\bibfnamefont {H.~X.}\ \bibnamefont {Tang}},\ }\bibfield  {title}
  {\emph {\bibinfo {title} {Strongly coupled magnons and cavity microwave
  photons},\ }}\href {\doibase 10.1103/PhysRevLett.113.156401} {\bibfield
  {journal} {\bibinfo  {journal} {Phys. Rev. Lett.}\ }\textbf {\bibinfo
  {volume} {113}},\ \bibinfo {pages} {156401} (\bibinfo {year}
  {2014})}\BibitemShut {NoStop}%
\bibitem [{\citenamefont {Zhang}\ \emph
  {et~al.}(2016{\natexlab{b}})\citenamefont {Zhang}, \citenamefont {Zhu},
  \citenamefont {Zou},\ and\ \citenamefont {Tang}}]{Zhang2016Optomagnonic}%
  \BibitemOpen
  \bibfield  {author} {\bibinfo {author} {\bibfnamefont {X.}~\bibnamefont
  {Zhang}}, \bibinfo {author} {\bibfnamefont {N.}~\bibnamefont {Zhu}}, \bibinfo
  {author} {\bibfnamefont {C.-L.}\ \bibnamefont {Zou}}, \ and\ \bibinfo
  {author} {\bibfnamefont {H.~X.}\ \bibnamefont {Tang}},\ }\bibfield  {title}
  {\emph {\bibinfo {title} {Optomagnonic whispering gallery microresonators},\
  }}\href {\doibase 10.1103/PhysRevLett.117.123605} {\bibfield  {journal}
  {\bibinfo  {journal} {Phys. Rev. Lett.}\ }\textbf {\bibinfo {volume} {117}},\
  \bibinfo {pages} {123605} (\bibinfo {year} {2016}{\natexlab{b}})}\BibitemShut
  {NoStop}%
\bibitem [{\citenamefont {Lachance-Quirion}\ \emph {et~al.}(2019)\citenamefont
  {Lachance-Quirion}, \citenamefont {Tabuchi}, \citenamefont {Gloppe},
  \citenamefont {Usami},\ and\ \citenamefont {Nakamura}}]{Lachance2019Hybrid}%
  \BibitemOpen
  \bibfield  {author} {\bibinfo {author} {\bibfnamefont {D.}~\bibnamefont
  {Lachance-Quirion}}, \bibinfo {author} {\bibfnamefont {Y.}~\bibnamefont
  {Tabuchi}}, \bibinfo {author} {\bibfnamefont {A.}~\bibnamefont {Gloppe}},
  \bibinfo {author} {\bibfnamefont {K.}~\bibnamefont {Usami}}, \ and\ \bibinfo
  {author} {\bibfnamefont {Y.}~\bibnamefont {Nakamura}},\ }\bibfield  {title}
  {\emph {\bibinfo {title} {Hybrid quantum systems based on magnonics},\
  }}\href {\doibase 10.7567/1882-0786/ab248d} {\bibfield  {journal} {\bibinfo
  {journal} {Appl. Phys. Express}\ }\textbf {\bibinfo {volume} {12}},\ \bibinfo
  {pages} {070101} (\bibinfo {year} {2019})}\BibitemShut {NoStop}%
\bibitem [{\citenamefont {Xu}\ \emph {et~al.}(2021)\citenamefont {Xu},
  \citenamefont {Zhong}, \citenamefont {Han}, \citenamefont {Jin},
  \citenamefont {Jiang},\ and\ \citenamefont {Zhang}}]{Xu2021Coherent}%
  \BibitemOpen
  \bibfield  {author} {\bibinfo {author} {\bibfnamefont {J.}~\bibnamefont
  {Xu}}, \bibinfo {author} {\bibfnamefont {C.}~\bibnamefont {Zhong}}, \bibinfo
  {author} {\bibfnamefont {X.}~\bibnamefont {Han}}, \bibinfo {author}
  {\bibfnamefont {D.}~\bibnamefont {Jin}}, \bibinfo {author} {\bibfnamefont
  {L.}~\bibnamefont {Jiang}}, \ and\ \bibinfo {author} {\bibfnamefont
  {X.}~\bibnamefont {Zhang}},\ }\bibfield  {title} {\emph {\bibinfo {title}
  {Coherent gate operations in hybrid magnonics},\ }}\href {\doibase
  10.1103/PhysRevLett.126.207202} {\bibfield  {journal} {\bibinfo  {journal}
  {Phys. Rev. Lett.}\ }\textbf {\bibinfo {volume} {126}},\ \bibinfo {pages}
  {207202} (\bibinfo {year} {2021})}\BibitemShut {NoStop}%
\bibitem [{\citenamefont {{Zare Rameshti}}\ \emph {et~al.}(2022)\citenamefont
  {{Zare Rameshti}}, \citenamefont {{Viola Kusminskiy}}, \citenamefont {Haigh},
  \citenamefont {Usami}, \citenamefont {Lachance-Quirion}, \citenamefont
  {Nakamura}, \citenamefont {Hu}, \citenamefont {Tang}, \citenamefont {Bauer},\
  and\ \citenamefont {Blanter}}]{Babak2022Cavity}%
  \BibitemOpen
  \bibfield  {author} {\bibinfo {author} {\bibfnamefont {B.}~\bibnamefont
  {{Zare Rameshti}}}, \bibinfo {author} {\bibfnamefont {S.}~\bibnamefont
  {{Viola Kusminskiy}}}, \bibinfo {author} {\bibfnamefont {J.~A.}\ \bibnamefont
  {Haigh}}, \bibinfo {author} {\bibfnamefont {K.}~\bibnamefont {Usami}},
  \bibinfo {author} {\bibfnamefont {D.}~\bibnamefont {Lachance-Quirion}},
  \bibinfo {author} {\bibfnamefont {Y.}~\bibnamefont {Nakamura}}, \bibinfo
  {author} {\bibfnamefont {C.-M.}\ \bibnamefont {Hu}}, \bibinfo {author}
  {\bibfnamefont {H.~X.}\ \bibnamefont {Tang}}, \bibinfo {author}
  {\bibfnamefont {G.~E.}\ \bibnamefont {Bauer}}, \ and\ \bibinfo {author}
  {\bibfnamefont {Y.~M.}\ \bibnamefont {Blanter}},\ }\bibfield  {title} {\emph
  {\bibinfo {title} {Cavity magnonics},\ }}\href {\doibase
  https://doi.org/10.1016/j.physrep.2022.06.001} {\bibfield  {journal}
  {\bibinfo  {journal} {Phys. Rep.}\ }\textbf {\bibinfo {volume} {979}},\
  \bibinfo {pages} {1} (\bibinfo {year} {2022})}\BibitemShut {NoStop}%
\bibitem [{\citenamefont {Xu}\ \emph {et~al.}(2023)\citenamefont {Xu},
  \citenamefont {Gu}, \citenamefont {Li}, \citenamefont {Weng}, \citenamefont
  {Wang}, \citenamefont {Li}, \citenamefont {Wang}, \citenamefont {Zhu},\ and\
  \citenamefont {You}}]{Xu2023Quantum}%
  \BibitemOpen
  \bibfield  {author} {\bibinfo {author} {\bibfnamefont {D.}~\bibnamefont
  {Xu}}, \bibinfo {author} {\bibfnamefont {X.-K.}\ \bibnamefont {Gu}}, \bibinfo
  {author} {\bibfnamefont {H.-K.}\ \bibnamefont {Li}}, \bibinfo {author}
  {\bibfnamefont {Y.-C.}\ \bibnamefont {Weng}}, \bibinfo {author}
  {\bibfnamefont {Y.-P.}\ \bibnamefont {Wang}}, \bibinfo {author}
  {\bibfnamefont {J.}~\bibnamefont {Li}}, \bibinfo {author} {\bibfnamefont
  {H.}~\bibnamefont {Wang}}, \bibinfo {author} {\bibfnamefont {S.-Y.}\
  \bibnamefont {Zhu}}, \ and\ \bibinfo {author} {\bibfnamefont {J.~Q.}\
  \bibnamefont {You}},\ }\bibfield  {title} {\emph {\bibinfo {title} {Quantum
  control of a single magnon in a macroscopic spin system},\ }}\href {\doibase
  10.1103/PhysRevLett.130.193603} {\bibfield  {journal} {\bibinfo  {journal}
  {Phys. Rev. Lett.}\ }\textbf {\bibinfo {volume} {130}},\ \bibinfo {pages}
  {193603} (\bibinfo {year} {2023})}\BibitemShut {NoStop}%
\bibitem [{\citenamefont {Zheng}\ \emph {et~al.}(2023)\citenamefont {Zheng},
  \citenamefont {Wang}, \citenamefont {Wang}, \citenamefont {Sun},
  \citenamefont {He}, \citenamefont {Yan},\ and\ \citenamefont
  {Yuan}}]{Zheng2023Tutorial}%
  \BibitemOpen
  \bibfield  {author} {\bibinfo {author} {\bibfnamefont {S.}~\bibnamefont
  {Zheng}}, \bibinfo {author} {\bibfnamefont {Z.}~\bibnamefont {Wang}},
  \bibinfo {author} {\bibfnamefont {Y.}~\bibnamefont {Wang}}, \bibinfo {author}
  {\bibfnamefont {F.}~\bibnamefont {Sun}}, \bibinfo {author} {\bibfnamefont
  {Q.}~\bibnamefont {He}}, \bibinfo {author} {\bibfnamefont {P.}~\bibnamefont
  {Yan}}, \ and\ \bibinfo {author} {\bibfnamefont {H.~Y.}\ \bibnamefont
  {Yuan}},\ }\bibfield  {title} {\emph {\bibinfo {title} {Tutorial: Nonlinear
  magnonics},\ }}\href {\doibase 10.1063/5.0152543} {\bibfield  {journal}
  {\bibinfo  {journal} {J. Appl. Phys.}\ }\textbf {\bibinfo {volume} {134}},\
  \bibinfo {pages} {151101} (\bibinfo {year} {2023})}\BibitemShut {NoStop}%
\bibitem [{\citenamefont {Claridge}(2009)}]{Claridge2009high}%
  \BibitemOpen
  \bibfield  {author} {\bibinfo {author} {\bibfnamefont {T.~D.~W.}\
  \bibnamefont {Claridge}},\ }\href@noop {} {\emph {\bibinfo {title}
  {High-Resolution {NMR} Techniques in Organic Chemistry}}}\ (\bibinfo
  {publisher} {Elsevier New York},\ \bibinfo {year} {2009})\BibitemShut
  {NoStop}%
\bibitem [{\citenamefont {Dong}\ \emph {et~al.}(2012)\citenamefont {Dong},
  \citenamefont {Fiore}, \citenamefont {Kuzyk},\ and\ \citenamefont
  {Wang}}]{Dong2012Optomechanical}%
  \BibitemOpen
  \bibfield  {author} {\bibinfo {author} {\bibfnamefont {C.}~\bibnamefont
  {Dong}}, \bibinfo {author} {\bibfnamefont {V.}~\bibnamefont {Fiore}},
  \bibinfo {author} {\bibfnamefont {M.~C.}\ \bibnamefont {Kuzyk}}, \ and\
  \bibinfo {author} {\bibfnamefont {H.}~\bibnamefont {Wang}},\ }\bibfield
  {title} {\emph {\bibinfo {title} {Optomechanical dark mode},\ }}\href
  {\doibase 10.1126/science.1228370} {\bibfield  {journal} {\bibinfo  {journal}
  {Science}\ }\textbf {\bibinfo {volume} {338}},\ \bibinfo {pages} {1609}
  (\bibinfo {year} {2012})}\BibitemShut {NoStop}%
\bibitem [{\citenamefont {Wang}\ and\ \citenamefont
  {Clerk}(2012)}]{Wang2012Using}%
  \BibitemOpen
  \bibfield  {author} {\bibinfo {author} {\bibfnamefont {Y.-D.}\ \bibnamefont
  {Wang}}\ and\ \bibinfo {author} {\bibfnamefont {A.~A.}\ \bibnamefont
  {Clerk}},\ }\bibfield  {title} {\emph {\bibinfo {title} {Using interference
  for high fidelity quantum state transfer in optomechanics},\ }}\href
  {\doibase 10.1103/PhysRevLett.108.153603} {\bibfield  {journal} {\bibinfo
  {journal} {Phys. Rev. Lett.}\ }\textbf {\bibinfo {volume} {108}},\ \bibinfo
  {pages} {153603} (\bibinfo {year} {2012})}\BibitemShut {NoStop}%
\bibitem [{\citenamefont {Tian}(2012)}]{Tian2012Adiabatic}%
  \BibitemOpen
  \bibfield  {author} {\bibinfo {author} {\bibfnamefont {L.}~\bibnamefont
  {Tian}},\ }\bibfield  {title} {\emph {\bibinfo {title} {Adiabatic state
  conversion and pulse transmission in optomechanical systems},\ }}\href
  {\doibase 10.1103/PhysRevLett.108.153604} {\bibfield  {journal} {\bibinfo
  {journal} {Phys. Rev. Lett.}\ }\textbf {\bibinfo {volume} {108}},\ \bibinfo
  {pages} {153604} (\bibinfo {year} {2012})}\BibitemShut {NoStop}%
\bibitem [{\citenamefont {Fedoseev}\ \emph {et~al.}(2021)\citenamefont
  {Fedoseev}, \citenamefont {Luna}, \citenamefont {Hedgepeth}, \citenamefont
  {L\"offler},\ and\ \citenamefont {Bouwmeester}}]{Fedoseev2021Stimulated}%
  \BibitemOpen
  \bibfield  {author} {\bibinfo {author} {\bibfnamefont {V.}~\bibnamefont
  {Fedoseev}}, \bibinfo {author} {\bibfnamefont {F.}~\bibnamefont {Luna}},
  \bibinfo {author} {\bibfnamefont {I.}~\bibnamefont {Hedgepeth}}, \bibinfo
  {author} {\bibfnamefont {W.}~\bibnamefont {L\"offler}}, \ and\ \bibinfo
  {author} {\bibfnamefont {D.}~\bibnamefont {Bouwmeester}},\ }\bibfield
  {title} {\emph {\bibinfo {title} {Stimulated raman adiabatic passage in
  optomechanics},\ }}\href {\doibase 10.1103/PhysRevLett.126.113601} {\bibfield
   {journal} {\bibinfo  {journal} {Phys. Rev. Lett.}\ }\textbf {\bibinfo
  {volume} {126}},\ \bibinfo {pages} {113601} (\bibinfo {year}
  {2021})}\BibitemShut {NoStop}%
\bibitem [{\citenamefont {Vitanov}\ \emph {et~al.}(2017)\citenamefont
  {Vitanov}, \citenamefont {Rangelov}, \citenamefont {Shore},\ and\
  \citenamefont {Bergmann}}]{Vitanov2017Stimulated}%
  \BibitemOpen
  \bibfield  {author} {\bibinfo {author} {\bibfnamefont {N.~V.}\ \bibnamefont
  {Vitanov}}, \bibinfo {author} {\bibfnamefont {A.~A.}\ \bibnamefont
  {Rangelov}}, \bibinfo {author} {\bibfnamefont {B.~W.}\ \bibnamefont {Shore}},
  \ and\ \bibinfo {author} {\bibfnamefont {K.}~\bibnamefont {Bergmann}},\
  }\bibfield  {title} {\emph {\bibinfo {title} {Stimulated raman adiabatic
  passage in physics, chemistry, and beyond},\ }}\href {\doibase
  10.1103/RevModPhys.89.015006} {\bibfield  {journal} {\bibinfo  {journal}
  {Rev. Mod. Phys.}\ }\textbf {\bibinfo {volume} {89}},\ \bibinfo {pages}
  {015006} (\bibinfo {year} {2017})}\BibitemShut {NoStop}%
\bibitem [{\citenamefont {Huang}\ \emph {et~al.}(2023)\citenamefont {Huang},
  \citenamefont {Liu}, \citenamefont {Xu},\ and\ \citenamefont
  {Liao}}]{Jian2023Dark}%
  \BibitemOpen
  \bibfield  {author} {\bibinfo {author} {\bibfnamefont {J.}~\bibnamefont
  {Huang}}, \bibinfo {author} {\bibfnamefont {C.}~\bibnamefont {Liu}}, \bibinfo
  {author} {\bibfnamefont {X.-W.}\ \bibnamefont {Xu}}, \ and\ \bibinfo {author}
  {\bibfnamefont {J.-Q.}\ \bibnamefont {Liao}},\ }\bibfield  {title} {\emph
  {\bibinfo {title} {Dark-mode theorems for quantum networks},\ }}\href
  {https://doi.org/10.48550/arXiv.2312.06274} {\bibfield  {journal} {\bibinfo
  {journal} {arXiv:2312.06274}\ } (\bibinfo {year} {2023})}\BibitemShut
  {NoStop}%
\bibitem [{\citenamefont {Jing}\ \emph {et~al.}(2016)\citenamefont {Jing},
  \citenamefont {Sarandy}, \citenamefont {Lidar}, \citenamefont {Luo},\ and\
  \citenamefont {Wu}}]{Jing2016Eigenstate}%
  \BibitemOpen
  \bibfield  {author} {\bibinfo {author} {\bibfnamefont {J.}~\bibnamefont
  {Jing}}, \bibinfo {author} {\bibfnamefont {M.~S.}\ \bibnamefont {Sarandy}},
  \bibinfo {author} {\bibfnamefont {D.~A.}\ \bibnamefont {Lidar}}, \bibinfo
  {author} {\bibfnamefont {D.-W.}\ \bibnamefont {Luo}}, \ and\ \bibinfo
  {author} {\bibfnamefont {L.-A.}\ \bibnamefont {Wu}},\ }\bibfield  {title}
  {\emph {\bibinfo {title} {Eigenstate tracking in open quantum systems},\
  }}\href {\doibase 10.1103/PhysRevA.94.042131} {\bibfield  {journal} {\bibinfo
   {journal} {Phys. Rev. A}\ }\textbf {\bibinfo {volume} {94}},\ \bibinfo
  {pages} {042131} (\bibinfo {year} {2016})}\BibitemShut {NoStop}%
\bibitem [{\citenamefont {Jing}\ and\ \citenamefont
  {Wu}(2022)}]{Jing2022Onecomponent}%
  \BibitemOpen
  \bibfield  {author} {\bibinfo {author} {\bibfnamefont {J.}~\bibnamefont
  {Jing}}\ and\ \bibinfo {author} {\bibfnamefont {L.-A.}\ \bibnamefont {Wu}},\
  }\bibfield  {title} {\emph {\bibinfo {title} {One-component quantum mechanics
  and dynamical leakage-free paths},\ }}\href {\doibase
  10.1038/s41598-022-13130-3} {\bibfield  {journal} {\bibinfo  {journal} {Sci.
  Rep.}\ }\textbf {\bibinfo {volume} {12}},\ \bibinfo {pages} {9247} (\bibinfo
  {year} {2022})}\BibitemShut {NoStop}%
\bibitem [{\citenamefont {Jing}\ \emph {et~al.}(2013)\citenamefont {Jing},
  \citenamefont {Wu}, \citenamefont {You},\ and\ \citenamefont
  {Yu}}]{Jing2013Nonperturbative}%
  \BibitemOpen
  \bibfield  {author} {\bibinfo {author} {\bibfnamefont {J.}~\bibnamefont
  {Jing}}, \bibinfo {author} {\bibfnamefont {L.-A.}\ \bibnamefont {Wu}},
  \bibinfo {author} {\bibfnamefont {J.~Q.}\ \bibnamefont {You}}, \ and\
  \bibinfo {author} {\bibfnamefont {T.}~\bibnamefont {Yu}},\ }\bibfield
  {title} {\emph {\bibinfo {title} {Nonperturbative quantum dynamical
  decoupling},\ }}\href {\doibase 10.1103/PhysRevA.88.022333} {\bibfield
  {journal} {\bibinfo  {journal} {Phys. Rev. A}\ }\textbf {\bibinfo {volume}
  {88}},\ \bibinfo {pages} {022333} (\bibinfo {year} {2013})}\BibitemShut
  {NoStop}%
\bibitem [{\citenamefont {Jing}\ \emph {et~al.}(2015)\citenamefont {Jing},
  \citenamefont {Wu}, \citenamefont {Byrd}, \citenamefont {You}, \citenamefont
  {Yu},\ and\ \citenamefont {Wang}}]{Jing2015Nonperturbative}%
  \BibitemOpen
  \bibfield  {author} {\bibinfo {author} {\bibfnamefont {J.}~\bibnamefont
  {Jing}}, \bibinfo {author} {\bibfnamefont {L.-A.}\ \bibnamefont {Wu}},
  \bibinfo {author} {\bibfnamefont {M.}~\bibnamefont {Byrd}}, \bibinfo {author}
  {\bibfnamefont {J.~Q.}\ \bibnamefont {You}}, \bibinfo {author} {\bibfnamefont
  {T.}~\bibnamefont {Yu}}, \ and\ \bibinfo {author} {\bibfnamefont {Z.-M.}\
  \bibnamefont {Wang}},\ }\bibfield  {title} {\emph {\bibinfo {title}
  {Nonperturbative leakage elimination operators and control of a three-level
  system},\ }}\href {\doibase 10.1103/PhysRevLett.114.190502} {\bibfield
  {journal} {\bibinfo  {journal} {Phys. Rev. Lett.}\ }\textbf {\bibinfo
  {volume} {114}},\ \bibinfo {pages} {190502} (\bibinfo {year}
  {2015})}\BibitemShut {NoStop}%
\bibitem [{\citenamefont {Zhang}\ \emph {et~al.}(2019)\citenamefont {Zhang},
  \citenamefont {Song}, \citenamefont {Ai}, \citenamefont {Wang}, \citenamefont
  {Yang},\ and\ \citenamefont {Deng}}]{Zhang2019Fast}%
  \BibitemOpen
  \bibfield  {author} {\bibinfo {author} {\bibfnamefont {H.}~\bibnamefont
  {Zhang}}, \bibinfo {author} {\bibfnamefont {X.-K.}\ \bibnamefont {Song}},
  \bibinfo {author} {\bibfnamefont {Q.}~\bibnamefont {Ai}}, \bibinfo {author}
  {\bibfnamefont {H.}~\bibnamefont {Wang}}, \bibinfo {author} {\bibfnamefont
  {G.-J.}\ \bibnamefont {Yang}}, \ and\ \bibinfo {author} {\bibfnamefont
  {F.-G.}\ \bibnamefont {Deng}},\ }\bibfield  {title} {\emph {\bibinfo {title}
  {Fast and robust quantum control for multimode interactions using shortcuts
  to adiabaticity},\ }}\href {\doibase 10.1364/OE.27.007384} {\bibfield
  {journal} {\bibinfo  {journal} {Opt. Express}\ }\textbf {\bibinfo {volume}
  {27}},\ \bibinfo {pages} {7384} (\bibinfo {year} {2019})}\BibitemShut
  {NoStop}%
\bibitem [{\citenamefont {Qi}\ and\ \citenamefont
  {Jing}(2022{\natexlab{a}})}]{Qi2022Accelerated}%
  \BibitemOpen
  \bibfield  {author} {\bibinfo {author} {\bibfnamefont {S.-f.}\ \bibnamefont
  {Qi}}\ and\ \bibinfo {author} {\bibfnamefont {J.}~\bibnamefont {Jing}},\
  }\bibfield  {title} {\emph {\bibinfo {title} {Accelerated adiabatic passage
  in cavity magnomechanics},\ }}\href {\doibase 10.1103/PhysRevA.105.053710}
  {\bibfield  {journal} {\bibinfo  {journal} {Phys. Rev. A}\ }\textbf {\bibinfo
  {volume} {105}},\ \bibinfo {pages} {053710} (\bibinfo {year}
  {2022}{\natexlab{a}})}\BibitemShut {NoStop}%
\bibitem [{\citenamefont {Chen}\ \emph {et~al.}(2018)\citenamefont {Chen},
  \citenamefont {Shi}, \citenamefont {Song},\ and\ \citenamefont
  {Xia}}]{Chen2018Invariant}%
  \BibitemOpen
  \bibfield  {author} {\bibinfo {author} {\bibfnamefont {Y.-H.}\ \bibnamefont
  {Chen}}, \bibinfo {author} {\bibfnamefont {Z.-C.}\ \bibnamefont {Shi}},
  \bibinfo {author} {\bibfnamefont {J.}~\bibnamefont {Song}}, \ and\ \bibinfo
  {author} {\bibfnamefont {Y.}~\bibnamefont {Xia}},\ }\bibfield  {title} {\emph
  {\bibinfo {title} {Invariant-based inverse engineering for fluctuation
  transfer between membranes in an optomechanical cavity system},\ }}\href
  {\doibase 10.1103/PhysRevA.97.023841} {\bibfield  {journal} {\bibinfo
  {journal} {Phys. Rev. A}\ }\textbf {\bibinfo {volume} {97}},\ \bibinfo
  {pages} {023841} (\bibinfo {year} {2018})}\BibitemShut {NoStop}%
\bibitem [{\citenamefont {Xiang}\ \emph {et~al.}(2023)\citenamefont {Xiang},
  \citenamefont {Olivares}, \citenamefont {Garc\'{\i}a-Ripoll},\ and\
  \citenamefont {Rabl}}]{Xiang2023Universal}%
  \BibitemOpen
  \bibfield  {author} {\bibinfo {author} {\bibfnamefont {Z.-L.}\ \bibnamefont
  {Xiang}}, \bibinfo {author} {\bibfnamefont {D.~G.}\ \bibnamefont {Olivares}},
  \bibinfo {author} {\bibfnamefont {J.~J.}\ \bibnamefont {Garc\'{\i}a-Ripoll}},
  \ and\ \bibinfo {author} {\bibfnamefont {P.}~\bibnamefont {Rabl}},\
  }\bibfield  {title} {\emph {\bibinfo {title} {Universal time-dependent
  control scheme for realizing arbitrary linear bosonic transformations},\
  }}\href {\doibase 10.1103/PhysRevLett.130.050801} {\bibfield  {journal}
  {\bibinfo  {journal} {Phys. Rev. Lett.}\ }\textbf {\bibinfo {volume} {130}},\
  \bibinfo {pages} {050801} (\bibinfo {year} {2023})}\BibitemShut {NoStop}%
\bibitem [{\citenamefont {Lu}\ \emph {et~al.}(2023)\citenamefont {Lu},
  \citenamefont {Maiti}, \citenamefont {Garmon}, \citenamefont {Ganjam},
  \citenamefont {Zhang}, \citenamefont {Claes}, \citenamefont {Frunzio},
  \citenamefont {Girvin},\ and\ \citenamefont {Schoelkopf}}]{Lu2023High}%
  \BibitemOpen
  \bibfield  {author} {\bibinfo {author} {\bibfnamefont {Y.}~\bibnamefont
  {Lu}}, \bibinfo {author} {\bibfnamefont {A.}~\bibnamefont {Maiti}}, \bibinfo
  {author} {\bibfnamefont {J.~W.~O.}\ \bibnamefont {Garmon}}, \bibinfo {author}
  {\bibfnamefont {S.}~\bibnamefont {Ganjam}}, \bibinfo {author} {\bibfnamefont
  {Y.}~\bibnamefont {Zhang}}, \bibinfo {author} {\bibfnamefont
  {J.}~\bibnamefont {Claes}}, \bibinfo {author} {\bibfnamefont
  {L.}~\bibnamefont {Frunzio}}, \bibinfo {author} {\bibfnamefont {S.~M.}\
  \bibnamefont {Girvin}}, \ and\ \bibinfo {author} {\bibfnamefont {R.~J.}\
  \bibnamefont {Schoelkopf}},\ }\bibfield  {title} {\emph {\bibinfo {title}
  {High-fidelity parametric beamsplitting with a parity-protected converter},\
  }}\href {\doibase 10.1038/s41467-023-41104-0} {\bibfield  {journal} {\bibinfo
   {journal} {Nat. Commun.}\ }\textbf {\bibinfo {volume} {14}},\ \bibinfo
  {pages} {5767} (\bibinfo {year} {2023})}\BibitemShut {NoStop}%
\bibitem [{\citenamefont {Jin}\ and\ \citenamefont
  {Jing}(2025{\natexlab{a}})}]{Jin2025Universal}%
  \BibitemOpen
  \bibfield  {author} {\bibinfo {author} {\bibfnamefont {Z.-y.}\ \bibnamefont
  {Jin}}\ and\ \bibinfo {author} {\bibfnamefont {J.}~\bibnamefont {Jing}},\
  }\bibfield  {title} {\emph {\bibinfo {title} {Universal perspective on
  nonadiabatic quantum control},\ }}\href {\doibase
  10.1103/PhysRevA.111.012406} {\bibfield  {journal} {\bibinfo  {journal}
  {Phys. Rev. A}\ }\textbf {\bibinfo {volume} {111}},\ \bibinfo {pages}
  {012406} (\bibinfo {year} {2025}{\natexlab{a}})}\BibitemShut {NoStop}%
\bibitem [{\citenamefont {Jin}\ and\ \citenamefont
  {Jing}(2025{\natexlab{b}})}]{Jin2025Entangling}%
  \BibitemOpen
  \bibfield  {author} {\bibinfo {author} {\bibfnamefont {Z.-y.}\ \bibnamefont
  {Jin}}\ and\ \bibinfo {author} {\bibfnamefont {J.}~\bibnamefont {Jing}},\
  }\bibfield  {title} {\emph {\bibinfo {title} {Entangling distant systems via
  universal nonadiabatic passage},\ }}\href {\doibase
  10.1103/PhysRevA.111.022628} {\bibfield  {journal} {\bibinfo  {journal}
  {Phys. Rev. A}\ }\textbf {\bibinfo {volume} {111}},\ \bibinfo {pages}
  {022628} (\bibinfo {year} {2025}{\natexlab{b}})}\BibitemShut {NoStop}%
\bibitem [{\citenamefont {Jin}\ and\ \citenamefont
  {Jing}(2025{\natexlab{c}})}]{Jin2025ErrCorr}%
  \BibitemOpen
  \bibfield  {author} {\bibinfo {author} {\bibfnamefont {Z.-y.}\ \bibnamefont
  {Jin}}\ and\ \bibinfo {author} {\bibfnamefont {J.}~\bibnamefont {Jing}},\
  }\bibfield  {title} {\emph {\bibinfo {title} {Universal quantum control with
  dynamical correction},\ }}\href {\doibase 10.1103/vfrs-fyzw} {\bibfield
  {journal} {\bibinfo  {journal} {Phys. Rev. A}\ }\textbf {\bibinfo {volume}
  {112}},\ \bibinfo {pages} {022427} (\bibinfo {year}
  {2025}{\natexlab{c}})}\BibitemShut {NoStop}%
\bibitem [{\citenamefont {Jin}\ and\ \citenamefont
  {Jing}(2025{\natexlab{d}})}]{Jin2025Rydberg}%
  \BibitemOpen
  \bibfield  {author} {\bibinfo {author} {\bibfnamefont {Z.-y.}\ \bibnamefont
  {Jin}}\ and\ \bibinfo {author} {\bibfnamefont {J.}~\bibnamefont {Jing}},\
  }\bibfield  {title} {\emph {\bibinfo {title} {Preparing
  greenberger-horne-zeilinger states on ground levels of neutral atoms},\
  }}\href {\doibase 10.1103/c47j-cw46} {\bibfield  {journal} {\bibinfo
  {journal} {Phys. Rev. A}\ }\textbf {\bibinfo {volume} {112}},\ \bibinfo
  {pages} {022602} (\bibinfo {year} {2025}{\natexlab{d}})}\BibitemShut
  {NoStop}%
\bibitem [{\citenamefont {Jin}\ and\ \citenamefont
  {Jing}(2025{\natexlab{e}})}]{Jin2025NonHerm}%
  \BibitemOpen
  \bibfield  {author} {\bibinfo {author} {\bibfnamefont {Z.-y.}\ \bibnamefont
  {Jin}}\ and\ \bibinfo {author} {\bibfnamefont {J.}~\bibnamefont {Jing}},\
  }\bibfield  {title} {\emph {\bibinfo {title} {Universal quantum control by
  non-hermitian hamiltonian},\ }}\href {\doibase 10.1103/7h6g-1z57} {\bibfield
  {journal} {\bibinfo  {journal} {Phys. Rev. A}\ }\textbf {\bibinfo {volume}
  {112}},\ \bibinfo {pages} {032605} (\bibinfo {year}
  {2025}{\natexlab{e}})}\BibitemShut {NoStop}%
\bibitem [{\citenamefont {Jin}\ and\ \citenamefont
  {Jing}(2025{\natexlab{f}})}]{Jin2025Majorana}%
  \BibitemOpen
  \bibfield  {author} {\bibinfo {author} {\bibfnamefont {Z.-y.}\ \bibnamefont
  {Jin}}\ and\ \bibinfo {author} {\bibfnamefont {J.}~\bibnamefont {Jing}},\
  }\bibfield  {title} {\emph {\bibinfo {title} {Universal quantum control over
  majorana zero modes},\ }}\href {\doibase 10.1103/sf19-3dc4} {\bibfield
  {journal} {\bibinfo  {journal} {Phys. Rev. A}\ }\textbf {\bibinfo {volume}
  {112}},\ \bibinfo {pages} {052614} (\bibinfo {year}
  {2025}{\natexlab{f}})}\BibitemShut {NoStop}%
\bibitem [{\citenamefont {Kolodrubetz}\ \emph {et~al.}(2017)\citenamefont
  {Kolodrubetz}, \citenamefont {Sels}, \citenamefont {Mehta},\ and\
  \citenamefont {Polkovnikov}}]{Michael2017Geometry}%
  \BibitemOpen
  \bibfield  {author} {\bibinfo {author} {\bibfnamefont {M.}~\bibnamefont
  {Kolodrubetz}}, \bibinfo {author} {\bibfnamefont {D.}~\bibnamefont {Sels}},
  \bibinfo {author} {\bibfnamefont {P.}~\bibnamefont {Mehta}}, \ and\ \bibinfo
  {author} {\bibfnamefont {A.}~\bibnamefont {Polkovnikov}},\ }\bibfield
  {title} {\emph {\bibinfo {title} {Geometry and non-adiabatic response in
  quantum and classical systems},\ }}\href {\doibase
  https://doi.org/10.1016/j.physrep.2017.07.001} {\bibfield  {journal}
  {\bibinfo  {journal} {Phys. Rep.}\ }\textbf {\bibinfo {volume} {697}},\
  \bibinfo {pages} {1} (\bibinfo {year} {2017})}\BibitemShut {NoStop}%
\bibitem [{\citenamefont {Claeys}\ \emph {et~al.}(2019)\citenamefont {Claeys},
  \citenamefont {Pandey}, \citenamefont {Sels},\ and\ \citenamefont
  {Polkovnikov}}]{Claeys2019Floquet}%
  \BibitemOpen
  \bibfield  {author} {\bibinfo {author} {\bibfnamefont {P.~W.}\ \bibnamefont
  {Claeys}}, \bibinfo {author} {\bibfnamefont {M.}~\bibnamefont {Pandey}},
  \bibinfo {author} {\bibfnamefont {D.}~\bibnamefont {Sels}}, \ and\ \bibinfo
  {author} {\bibfnamefont {A.}~\bibnamefont {Polkovnikov}},\ }\bibfield
  {title} {\emph {\bibinfo {title} {Floquet-engineering counterdiabatic
  protocols in quantum many-body systems},\ }}\href {\doibase
  10.1103/PhysRevLett.123.090602} {\bibfield  {journal} {\bibinfo  {journal}
  {Phys. Rev. Lett.}\ }\textbf {\bibinfo {volume} {123}},\ \bibinfo {pages}
  {090602} (\bibinfo {year} {2019})}\BibitemShut {NoStop}%
\bibitem [{\citenamefont {Takahashi}\ and\ \citenamefont {del
  Campo}(2024)}]{Takahashi2024Shortcuts}%
  \BibitemOpen
  \bibfield  {author} {\bibinfo {author} {\bibfnamefont {K.}~\bibnamefont
  {Takahashi}}\ and\ \bibinfo {author} {\bibfnamefont {A.}~\bibnamefont {del
  Campo}},\ }\bibfield  {title} {\emph {\bibinfo {title} {Shortcuts to
  adiabaticity in krylov space},\ }}\href {\doibase 10.1103/PhysRevX.14.011032}
  {\bibfield  {journal} {\bibinfo  {journal} {Phys. Rev. X}\ }\textbf {\bibinfo
  {volume} {14}},\ \bibinfo {pages} {011032} (\bibinfo {year}
  {2024})}\BibitemShut {NoStop}%
\bibitem [{\citenamefont {Dyson}(1949)}]{Dyson1949TheRadiation}%
  \BibitemOpen
  \bibfield  {author} {\bibinfo {author} {\bibfnamefont {F.~J.}\ \bibnamefont
  {Dyson}},\ }\bibfield  {title} {\emph {\bibinfo {title} {The radiation
  theories of tomonaga, schwinger, and feynman},\ }}\href {\doibase
  10.1103/PhysRev.75.486} {\bibfield  {journal} {\bibinfo  {journal} {Phys.
  Rev.}\ }\textbf {\bibinfo {volume} {75}},\ \bibinfo {pages} {486} (\bibinfo
  {year} {1949})}\BibitemShut {NoStop}%
\bibitem [{\citenamefont {Sj\"oqvist}\ \emph {et~al.}(2012)\citenamefont
  {Sj\"oqvist}, \citenamefont {Tong}, \citenamefont {Andersson}, \citenamefont
  {Hessmo}, \citenamefont {Johansson},\ and\ \citenamefont
  {Singh}}]{Sjoqvist2012Nonadiabatic}%
  \BibitemOpen
  \bibfield  {author} {\bibinfo {author} {\bibfnamefont {E.}~\bibnamefont
  {Sj\"oqvist}}, \bibinfo {author} {\bibfnamefont {D.}~\bibnamefont {Tong}},
  \bibinfo {author} {\bibfnamefont {L.~M.}\ \bibnamefont {Andersson}}, \bibinfo
  {author} {\bibfnamefont {B.}~\bibnamefont {Hessmo}}, \bibinfo {author}
  {\bibfnamefont {M.}~\bibnamefont {Johansson}}, \ and\ \bibinfo {author}
  {\bibfnamefont {K.}~\bibnamefont {Singh}},\ }\bibfield  {title} {\emph
  {\bibinfo {title} {Non-adiabatic holonomic quantum computation},\ }}\href
  {\doibase 10.1088/1367-2630/14/10/103035} {\bibfield  {journal} {\bibinfo
  {journal} {New J. Phys.}\ }\textbf {\bibinfo {volume} {14}},\ \bibinfo
  {pages} {103035} (\bibinfo {year} {2012})}\BibitemShut {NoStop}%
\bibitem [{\citenamefont {Carmichael}(1999)}]{Carmichael1999statistical}%
  \BibitemOpen
  \bibfield  {author} {\bibinfo {author} {\bibfnamefont {H.}~\bibnamefont
  {Carmichael}},\ }\href@noop {} {\emph {\bibinfo {title} {Statistical Methods
  in Quantum Optics}}}\ (\bibinfo  {publisher} {Springer, Berlin},\ \bibinfo
  {year} {1999})\BibitemShut {NoStop}%
\bibitem [{\citenamefont {Lee}\ \emph {et~al.}(2002)\citenamefont {Lee},
  \citenamefont {Kok},\ and\ \citenamefont {Dowling}}]{Lee2002Quantum}%
  \BibitemOpen
  \bibfield  {author} {\bibinfo {author} {\bibfnamefont {H.}~\bibnamefont
  {Lee}}, \bibinfo {author} {\bibfnamefont {P.}~\bibnamefont {Kok}}, \ and\
  \bibinfo {author} {\bibfnamefont {J.~P.}\ \bibnamefont {Dowling}},\
  }\bibfield  {title} {\emph {\bibinfo {title} {A quantum rosetta stone for
  interferometry},\ }}\href {\doibase 10.1080/0950034021000011536} {\bibfield
  {journal} {\bibinfo  {journal} {J. Mod. Opt.}\ }\textbf {\bibinfo {volume}
  {49}},\ \bibinfo {pages} {2325} (\bibinfo {year} {2002})}\BibitemShut
  {NoStop}%
\bibitem [{\citenamefont {Pezz\`e}\ \emph {et~al.}(2018)\citenamefont
  {Pezz\`e}, \citenamefont {Smerzi}, \citenamefont {Oberthaler}, \citenamefont
  {Schmied},\ and\ \citenamefont {Treutlein}}]{Pezze2018Quantum}%
  \BibitemOpen
  \bibfield  {author} {\bibinfo {author} {\bibfnamefont {L.}~\bibnamefont
  {Pezz\`e}}, \bibinfo {author} {\bibfnamefont {A.}~\bibnamefont {Smerzi}},
  \bibinfo {author} {\bibfnamefont {M.~K.}\ \bibnamefont {Oberthaler}},
  \bibinfo {author} {\bibfnamefont {R.}~\bibnamefont {Schmied}}, \ and\
  \bibinfo {author} {\bibfnamefont {P.}~\bibnamefont {Treutlein}},\ }\bibfield
  {title} {\emph {\bibinfo {title} {Quantum metrology with nonclassical states
  of atomic ensembles},\ }}\href {\doibase 10.1103/RevModPhys.90.035005}
  {\bibfield  {journal} {\bibinfo  {journal} {Rev. Mod. Phys.}\ }\textbf
  {\bibinfo {volume} {90}},\ \bibinfo {pages} {035005} (\bibinfo {year}
  {2018})}\BibitemShut {NoStop}%
\bibitem [{\citenamefont {Qi}\ and\ \citenamefont
  {Jing}(2022{\natexlab{b}})}]{Qi2022Floquet}%
  \BibitemOpen
  \bibfield  {author} {\bibinfo {author} {\bibfnamefont {S.-f.}\ \bibnamefont
  {Qi}}\ and\ \bibinfo {author} {\bibfnamefont {J.}~\bibnamefont {Jing}},\
  }\bibfield  {title} {\emph {\bibinfo {title} {Chiral current in floquet
  cavity magnonics},\ }}\href {\doibase 10.1103/PhysRevA.106.033711} {\bibfield
   {journal} {\bibinfo  {journal} {Phys. Rev. A}\ }\textbf {\bibinfo {volume}
  {106}},\ \bibinfo {pages} {033711} (\bibinfo {year}
  {2022}{\natexlab{b}})}\BibitemShut {NoStop}%
\bibitem [{\citenamefont {Zhang}\ \emph
  {et~al.}(2015{\natexlab{b}})\citenamefont {Zhang}, \citenamefont {Zou},
  \citenamefont {Zhu}, \citenamefont {Marquardt}, \citenamefont {Jiang},\ and\
  \citenamefont {Tang}}]{Zhang2015Magnon}%
  \BibitemOpen
  \bibfield  {author} {\bibinfo {author} {\bibfnamefont {X.}~\bibnamefont
  {Zhang}}, \bibinfo {author} {\bibfnamefont {C.-L.}\ \bibnamefont {Zou}},
  \bibinfo {author} {\bibfnamefont {N.}~\bibnamefont {Zhu}}, \bibinfo {author}
  {\bibfnamefont {F.}~\bibnamefont {Marquardt}}, \bibinfo {author}
  {\bibfnamefont {L.}~\bibnamefont {Jiang}}, \ and\ \bibinfo {author}
  {\bibfnamefont {H.~X.}\ \bibnamefont {Tang}},\ }\bibfield  {title} {\emph
  {\bibinfo {title} {Magnon dark modes and gradient memory},\ }}\href {\doibase
  10.1038/ncomms9914} {\bibfield  {journal} {\bibinfo  {journal} {Nat.
  Commun.}\ }\textbf {\bibinfo {volume} {6}},\ \bibinfo {pages} {8914}
  (\bibinfo {year} {2015}{\natexlab{b}})}\BibitemShut {NoStop}%
\end{thebibliography}%

\end{document}